\documentclass[
 aip,
 amsmath,amssymb,
 reprint,
]{revtex4-1}

\usepackage{graphicx}
\usepackage{dcolumn}
\usepackage{bm}

\usepackage[utf8]{inputenc}
\usepackage[T1]{fontenc}
\usepackage{mathptmx}
\usepackage{etoolbox}
\usepackage{float}
\usepackage{verbatim}

\usepackage{xcolor,colortbl}
\definecolor{table_highlight}{RGB}{200,200,200}
\usepackage{dcolumn}

\makeatletter
\def\@email#1#2{
 \endgroup
 \patchcmd{\titleblock@produce}
  {\frontmatter@RRAPformat}
  {\frontmatter@RRAPformat{\produce@RRAP{*#1\href{mailto:#2}{#2}}}\frontmatter@RRAPformat}
  {}{}
}%
\makeatother

\def\tableStretch{1.15}

\newcommand{\TzzOne}{\parbox[c]{4em}{\includegraphics[width=0.6cm]{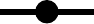}}}

\newcommand{\TzzTwo}{\parbox[c]{6em}{\includegraphics[width=0.9cm]{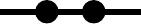}}}

\newcommand{\TzzThree}{\parbox[c]{6em}{\includegraphics[width=1.2cm]{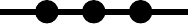}}}

\newcommand{\QzzzOne}{\parbox[c]{4em}{\includegraphics[width=0.5cm]{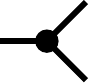}}}

\newcommand{\SzzzzOne}{\parbox[c]{4em}{\includegraphics[width=0.5cm]{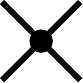}}}

\newcommand{\QuantumCorrection}{\parbox[c]{4em}{\includegraphics[width=0.7cm]{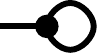}}}

\newcommand{\QzzzOneTzzOne}{\parbox[c]{4em}{\includegraphics[width=0.75cm]{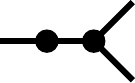}}}

\begin{document}

\preprint{AIP/123-QED}

\title{Feynman Diagrams for Matter Wave Interferometry}

\author{Jonah Glick}
\email{jonah.glick@gmail.com}
\affiliation{Department of Physics and Astronomy and Center for Fundamental Physics, Northwestern University, Evanston, Illinois, USA}
\affiliation{Fermi National Accelerator Laboratory, Batavia, Illinois, USA}
\author{Tim Kovachy}
\affiliation{Department of Physics and Astronomy and Center for Fundamental Physics, Northwestern University, Evanston, Illinois, USA}

\date{\today}

\begin{abstract}
We introduce a new theoretical framework based on Feynman diagrams to compute phase shifts in matter wave interferometry. The method allows for analytic computation of higher order quantum corrections, beyond the traditional semi-classical approximation. These additional terms depend on the finite size of the initial matter wavefunction and/or have higher order dependence on $\hbar$.
We apply the method to compute the response of matter wave interferometers to power law potentials and potentials with an arbitrary spatial dependence. The analytic expressions are validated by comparing to numerical simulations, and estimates are provided for the scale of the quantum corrections to the phase shift response to the gravitational field of the earth, anharmonic trapping potentials, and gravitational fields from local proof masses.
We also find that for certain experimentally feasible parameters,  these corrections are large enough to be measured, and could lead to systematic errors if they are not mitigated. 
We find that to first order in a spatially dependent potential, quantum corrections vanish when the initial matter wavepacket has spherical symmetry and the potential satisfies Laplace's equation.
We anticipate these quantum corrections will be especially important for trapped matter wave interferometers and for free-space matter wave interferometers in the presence of proof masses. These interferometers are becoming increasingly sensitive tools for mobile inertial sensing, gravity surveying, tests of gravity and its interplay with quantum mechanics, and searches for dark energy.
\end{abstract}

\maketitle

\section{\label{sec:introduction}Introduction}

Matter wave interferometry is a powerful tool for precision sensing and tests of fundamental physics. It is employed in a wide range of applications, including
inertial navigation, \cite{Cronin_2009, A_Peters_2001, McGuirk_2002, Durfee_2006}
gravity surveying, \cite{Wu2019_mobile, bongs2019taking, Bidel2020, Canuel2018_MIGA}
geodesy and geophysical studies,  \cite{Bongs2006, Stockton_2011}
tests of the equivalence principle and searches for fifth forces, \cite{Biedermann_2015, Fray_2004, Schlippert_2014, Zhou_2015, Kuhn_2014, Tarallo_2014, Bonnin_2013, Hartwig_2015, Asenbaum_2020, Williams_2016, Battelier_2021, Rosi2017, Barrett2016,wacker2010using}
measurements of Newton's gravitational constant \cite{Fixler_2007, Rosi2014, Lamporesi_2008}
and the fine structure constant, \cite{Parker_2018, Bouchendira2011_finestructure}
searches for dark energy, \cite{Hamilton_2015, chiow2018multiloop}
searches for dark matter, \cite{Arvanitaki_2018, el-neaj_aedge_2020,Graham_2016,abend2023terrestrial, Graham2018axion, di2024optimal,abe2021matter,Badurina_2020,Zhan_2020}
gravitational wave detection, \cite{Graham2016_GW, torres2023detecting, el-neaj_aedge_2020, baum2024gravitational, schubert2019scalable, Dimopoulos_2008, Yu2011, Chaibi_2016, Zhan_2020, Hogan2011, Graham_2013, Canuel2018_MIGA, Canuel_2020_ELGAR, abdalla_terrestrial_2025, Badurina_2020, Graham_2016, graham2017midband}
and tests of quantum mechanics and its interplay with gravity. \cite{Bassi_2013, Nimmrichter_2013, Bassi_2017, Altamirano_2018, Kovachy2015, Asenbaum_2017, Xu_2019, Arndt2014, Zych2011, Roura_2020,carney2021using, Overstreet_2022}

The calculation of the differential phase accumulated between two arms of a matter wave interferometer due to spatially varying external potentials has traditionally relied on a semi-classical approximation. \cite{Storey_1994, hogan2008lightpulse, tino2014atom,Bongs2006,Antoine_2003_quantum, Antoine_2003_exact, Dimopoulos_2008,dubetsky2006atom} This approximation neglects the finite spatial extent of an individual arm of the matter wavefunction and excludes contributions to the phase shift at orders of $\hbar$ higher than $\hbar^{-1}$. \footnote{Here we consider the recoil velocity kicks $v_r$ the matter-wave receives from absorbing/emitting a photon to not explicitly include its dependence on $\hbar$ through the photon momentum, see Tab. \ref{tab:power_law_phase_shifts}}
While numerical simulation tools have been developed that do not rely on semi-classical approximations, \cite{Fitzek2020UniversalAI} analytical expressions have been highly valuable for understanding scalings of various contributions to the phase shift in order to guide experimental design and analysis. \cite{hogan2008lightpulse, tino2014atom, Bongs2006}

We adapt the mathematical tools of quantum field theory (QFT) to introduce a Feynman diagram based approach to analytically computing the phase shift in a matter wave interferometer which incorporates higher-order quantum corrections from the finite size of a matter wavefunction, and higher order terms in $\hbar$. Our approach is complementary to analyses based on Wigner function \cite{dubetsky2016atom} or Magnus/Dyson series \cite{bertoldi2019phase,ufrecht2020perturbative, larow_beyond_2025} formalisms,
and it provides a structured framework for higher-order corrections,
similar to the way Feynman diagrams are used in perturbative quantum field theory.
In field-theoretic treatments, Feynman diagrams have been used to estimate phase shifts from hypothetical scalar fields. \cite{burrage_open_2019, kading_new_2023, kading_dilaton-induced_2023} Here, we instead apply diagrammatic methods within a purely quantum-mechanical framework.

We find higher order quantum corrections to be especially relevant in trapped matter wave interferometers with small anharmonicities, where external potentials exhibit strong position dependence. This new diagrammatic tool can be used to generate analytic estimates of these corrections in interferometers confined to magnetic \cite{
Sapiro_2009, Hansel_2001, Schumm_2006, Beydler_2024, Horikoshi_2006}
and optical \cite{Raithel_2023, Dash_2024, McDonald_2013_80hk, McDonald_2013_optically, Krzyzanowska_2023, Hyosub_2022,Charriere_2012, Panda_2024_measuring, Panda_2024_coherence, Kovachy_2010, premawardhana_feasibility_2024, nemirovsky_atomic_2023}
potentials, which commonly consist of anharmonicities that can induce beyond-semi-classical phase shifts.
Trapped interferometers are emerging as powerful tools for applications including mobile inertial sensing and gravity surveying,  \cite{Beydler_2024,Raithel_2023,McDonald_2013_80hk, panda2023atomic, Dash_2024,Krzyzanowska_2023,Charriere_2012,McDonald_2013_optically,Hyosub_2022} searches for dark energy and tests of gravity at short distances,  \cite{Panda_2024_measuring} and tests of quantum gravity. \cite{carney2021using}
Higher order quantum corrections will also be relevant to free-space interferometers that measure gravitational effects from local proof masses, like those aimed at obtaining precision measurements of Newton's gravitational constant \cite{Fixler_2007, Rosi2014, Lamporesi_2008,dubetsky2016atom} and testing the gravitational inverse square law to search for new fundamental interactions. \cite{Biedermann_2015,wacker2010using}

As illustrative case studies, we identify examples in which these quantum corrections would yield measurable effects in experimentally realistic settings. 
The first case study is a trapped atom interferometer, where we compute quantum corrections using parameters from a previously demonstrated experimental setup (see Sec. \ref{sec:anharmonic_traps}).
The second case study is a proof-mass-style matter wave interferometer (see Sec. \ref{sec:calculation_of_phase_response_3D}), which could be realized as a future upgrade to the MAGIS-100 experiment. \cite{abe2021matter}

The paper is organized as follows:
In Sec. \ref{sec:the_framework}, we introduce the framework by describing one way to compute phase shifts in matter wave interferometry (Sec. \ref{sec:phase_shift_calculation_background}), and how to compute phase shifts with Feynman diagrams (Sec. \ref{sec:phase_shifts_with_feynman_diagrams}), then we make connections to the traditional semi-classical approach to computing phase shifts (Sec. \ref{sec:connection_to_traditional_semi_classical}), and to calculations of scattering amplitudes in QFT (Sec. \ref{sec:QFT_background}).
In Sec. \ref{sec:treating_harmonic_term_perturbatively} and Sec. \ref{sec:harmonic_potential}, we apply the diagrammatic method to compute phase shifts under power law potentials.
We investigate the validity of these expression by comparing to numerical evaluations of the phase shift in a certain parameter space, propagating wavepackets with the split step method. \cite{FEIT1982412}
We then take the analytic terms which emerge from the diagrammatic framework and use them to evaluate phase shifts from gravity gradients of the earth (Sec. \ref{sec:gravity_gradients}), and for interferometers confined to anharmonic traps (Sec. \ref{sec:anharmonic_traps}). In Sec. \ref{sec:phase_shift_from_arbitary_potential}, we sum over an infinite number of diagrams to produce an analytic expression for the phase shift response to an external potential with arbitrary spatial dependence, and apply the expression to estimate quantum corrections to the phase shift for a proof-mass style experiment.
Detailed derivations of the key results are presented in the Appendices, with an organizational summary provided at the beginning of Appendix A.

\section{Framework for Phase Shift Calculations}\label{sec:the_framework}

\subsection{\label{sec:phase_shift_calculation_background}Phase Shift Calculation Background}

\begin{figure}
\centering
\includegraphics[width=3.25in]{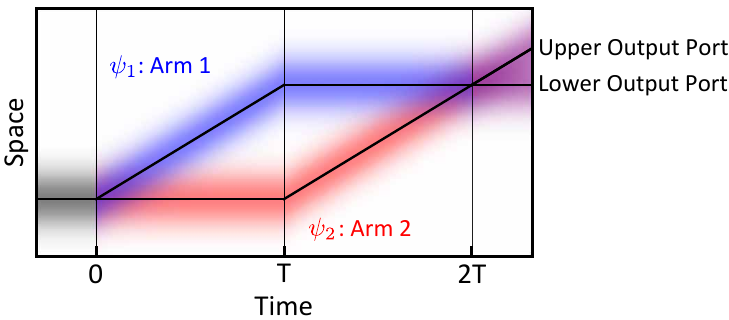}
\caption{A Mach-Zehnder interferometer sequence. A beamsplitter pulse is performed at time $t=0$, which splits the initial matter wavefunction into a superposition of two momentum states. A mirror pulse at time $t=T$ redirects the two arms back toward one another and a final beamsplitter pulse at time $t=2T$ interferes the two arms with one another. Black solid lines denote classical trajectories, the blue corresponds to the wavepacket of interferometer arm 1, and the red corresponds to the interferometer arm 2. These arms are represented by wavefunctions $\psi_1$ and $\psi_2$, respectively. The gray density  plot corresponds to the wavepacket before the first beamsplitter and the purple denotes the interference of the blue and red interferometer arms. The vertical lines denote the times of the instantaneous beamsplitter and mirror operations.}\label{fig:ai_introduction_figure}
\end{figure}

Matter-wave interferometer sequences typically begin and end with a beamsplitter pulse. The first beamsplitter pulse creates the initial superposition that forms the two interferometer arms, and the final beamsplitter pulse produces interference between them.
Figure \ref{fig:ai_introduction_figure} shows a Mach-Zehnder interferometer sequence consisting of three pulses: an initial beamsplitter pulse at $t=0$ which creates the superposition in the matter wavefunction, a mirror pulse at $t=T$ which redirects the interferometer arms back toward one another, and a final beamsplitter pulse at $t=2T$ which interferes the two arms.  For simplicity, we assume throughout this work that the atom-optics operations occur instantaneously.
When the wavefunctions interfere at $t=2T$, the population in the upper and lower output ports encodes the phase shift between the two interferometer arms.
The components of the atom wavefunction corresponding to the upper and lower output ports, $|\psi_{\text{upper}}\rangle$ and $|\psi_{\text{lower}}\rangle$, can be written as

\begin{equation}
\begin{split}
|\psi_{\text{upper}}\rangle &= |\psi_{1, \text{upper}}\rangle + |\psi_{2, \text{upper}}\rangle
\\
|\psi_{\text{lower}}\rangle &= |\psi_{1, \text{lower}}\rangle + |\psi_{2, \text{lower}}\rangle
\end{split}
\end{equation}

\noindent
where $|\psi_{j, \text{upper}}\rangle$ ($|\psi_{j, \text{lower}}\rangle$) represents the component of the matter wavefunction that traversed arm $j$ and ended up in the upper (lower) output port.
Matter-wave interferometer phase shifts are traditionally measured via these output port populations $P$, which are expressed in terms of the interferometer contrast $C$ and phases shift $\Delta\phi$ as

\begin{equation}
\begin{split}
P_{\text{upper}} &= \langle \psi_{\text{upper}} | \psi_{\text{upper}} \rangle = \frac{1}{2}
\left(
1 - C \cos\left[\Delta \phi\right]
\right)
\\
P_{\text{lower}} &= \langle \psi_{\text{lower}} | \psi_{\text{lower}} \rangle = \frac{1}{2}
\left(
1 + C \cos\left[\Delta \phi\right]
\right)
\end{split}
\end{equation}

In this paper, we adopt the convention that square brackets denote functional dependence (i.e., $f[x]$ means $f$ is a function of $x$), in contrast to round brackets $f(x)$ which could be confused for multiplication (i.e. $f$ multiplied by $x$). This notation becomes particularly useful as the mathematical expressions grow more complex, helping to clarify whether terms are arguments or multiplicative factors.

Figure \ref{fig:ai_introduction_figure} depicts the interferometer in one spatial dimension $x$ and one temporal dimension $t$, where the population is evenly distributed between the two output ports. Focusing on the lower output port, we define the wavefunctions $\psi_j[x, t] = \langle x | \psi_{j, \text{lower}}\rangle$ and define $t_i$ to be an \textit{initial} time before the first beamsplitter ($t_i<0$) and $t_f$ to be a \textit{final} time after the final beamsplitter operation ($t_f>2T$). The interferometer contrast $C$ and phase shift $\Delta\phi$ under spatial averaging  can be expressed in terms of the position space representations of $| \psi_{j, \text{lower}}\rangle$ as

\begin{equation}
\begin{split}
P_{\text{lower}} &= \int_{-\infty}^{\infty} dx_f ~ \langle \psi_{\text{lower}}|x_f\rangle \langle x_f | \psi_{\text{lower}}\rangle
\\
\frac{1}{2}\Big(1 + C \cos[\Delta\phi]\Big) &= \frac{1}{4}\int_{-\infty}^{\infty}dx_f
\big|
\psi_{1}[x_f,t_f] + \psi_{2}[x_f, t_f]
\big|^2
\end{split}
\end{equation}

\noindent
where we normalize the wavefunctions $\psi_1$ and $\psi_2$ so that the total probability associated with each is $1$.
The factor of 1/4 comes from two factors of $1/\sqrt{2}$ from the first and final 50:50 beamsplitters, which are then squared under the modulus squared operation. Expanding out this equation, we have

\begin{equation}
\begin{split}
&\frac{1}{4} + \frac{1}{4} + \frac{C}{4}e^{i\Delta \phi} + \frac{C}{4}e^{-i\Delta\phi}
\\
&= \int_{-\infty}^{\infty}dx_f
\bigg(
\frac{1}{4} \big| \psi_{1} \big|^2
+\frac{1}{4} \big| \psi_{2} \big|^2
+\frac{1}{4} \psi_{1}\psi_{2}^{*}
+\frac{1}{4} \psi_{1}^{*}\psi_{2}
\bigg)
\\
&= \frac{1}{4} + \frac{1}{4} + 
\int_{-\infty}^{\infty}dx_f 
\bigg(
\frac{1}{4} \psi_{1}\psi_{2}^{*} + \frac{1}{4} \psi_{1}^{*}\psi_{2}
\bigg)
\end{split}
\end{equation}

\noindent
Then comparing cross terms in a manner consistent with the sign conventions of Ref. \citenum{hogan2008lightpulse} results in

\begin{equation}\label{eq:integral_over_psi}
C e^{i\Delta \phi} 
= \int_{-\infty}^{\infty}dx_f \;
\psi_{1}[x_f,t_f]\psi_{2}^{*}[x_f,t_f]
\end{equation}

Equation \eqref{eq:integral_over_psi} is the main result of this section, and will form the basis for the diagrammatic computation of interferometer phase shifts in the following section.

\subsection{Computing Phase Shifts with Feynman Diagrams}\label{sec:phase_shifts_with_feynman_diagrams}

We now derive an expression for the right hand side of Eq. \eqref{eq:integral_over_psi}
which will enable the diagrammatic computation of phase shifts in matter wave interferometry.
The amplitude associated with a trajectory which is located at position $x_i$ at time $t_i$ and located at position $x_f$ at a later time $t_f$ can be written following the Feynman path integral approach \cite{feynman1965quantum} as

\begin{equation}
K[x_f, t_f ; x_i, t_i] = \int \mathcal{D}x[t] e^{\frac{i}{\hbar}S[\dot{x}, x,t]}
\end{equation}

\noindent
where $S$ is the action associated with a classical path $x[t]$, and the path integral is performed over all paths which start at position $x_i$ and time $t_i$ and arrive at position $x_f$ and time $t_f$.
The amplitude $K$ is sometimes referred to as the kernel propagator.

Consider a system with Lagrangian 
\begin{equation}
L = L_0[\dot{x}, x, t] + \epsilon L_1[x] + J[t]x
\end{equation}

\noindent
where $L_0$ corresponds to the ‘unperturbed’ Lagrangian
which consists of a kinetic energy term (with second order dependence on $\dot{x}$) and potential energy terms (containing only second order or lower dependence on $x$), 
$L_1$ consists of terms of any order in $x$, and $\epsilon$ is a perturbative parameter to keep track of the orders of the expansion.
We use $J[t]$ to represent the effective force from the laser kicks. While these kicks could in principle be included in $L_0$, separating them in this way facilitates the perturbative calculation.
We perform the perturbative expansion by taking functional derivatives with respect to $J[t]$, analogous to scattering amplitude calculations in QFT. However, unlike in QFT where source terms are set to zero after differentiation, here $J[t]$ represents the physical laser kicks and remains nonzero.
We treat the laser pulses as applying a momentum kick that is uniform in space (i.e., every spatial part of the matter wavefunction receives a uniform kick), corresponding to the plane wave approximation for the laser pulse, and instantaneous in time so that $J[t]$ is composed of Dirac delta functions.

The kernel $K$ can be written as

\begin{subequations}\label{eq:kernel_equation}
\begin{align}
&K[x_f, t_f ; x_i, t_i ; J[t]] = \int \mathcal{D}x[t] e^{\frac{i}{\hbar}
\int_{t_i}^{t_f} dt (L_0 + \epsilon L_1[x] + J[t]x)
}\label{eq:kernel_equation_a}
\\
&
\;\;\;\;\;\;\;\;
= e^{\frac{i}{\hbar}\int_{t_i}^{t_f} dt' \; \epsilon L_1\big[\frac{\hbar}{i}\frac{\delta}{\delta J[t']}\big]}
\int \mathcal{D}x[t] e^{\frac{i}{\hbar}
\int_{t_i}^{t_f} dt (L_0 + J[t]x)
}\label{eq:kernel_equation_b}
\\
&
\;\;\;\;\;\;\;\;
= 
e^{\frac{i}{\hbar}\int_{t_i}^{t_f}dt' \;\epsilon L_1\big[\frac{\hbar}{i}\frac{\delta}{\delta J[t']}\big]}
K^{(0)}[x_f, t_f ; x_i, t_i ; J[t]]
\label{eq:kernel_equation_c}
\end{align}
\end{subequations}

\noindent
where for notational convenience, the explicit dependence of $x$ and $\dot{x}$ on $t$, and the explicit dependence of $L_0$ on $\dot{x}$, $x$, and $t$ is not written.
We arrive at Eq. \eqref{eq:kernel_equation_b} by using an integral trick similar to one used in QFT\cite{zee2010quantum} to take the nonlinear component of the Lagrangian outside the path integral by replacing its argument in $x$ with functional derivatives with respect to $J[t]$, and we arrive at Eq. \eqref{eq:kernel_equation_c}
by solving the path integral in closed form,
where $K^{(0)}$ is the propagator kernel for the unperturbed Lagrangian (i.e., the Lagrangian excluding $L_1$). This represents the amplitude for free propagation under $L = L_0 + J x$ for a wavefunction which at time $t_i$ is a delta function in space centered at position $x_i$ (see Ref. \citenum{feynman1965quantum} for details).
We have explicitly written the dependence of the kernel propagator on the external force term $J[t]$. If the wave function of a matter wave at a time $t_i$ just before the first beamsplitter of a Mach-Zehnder sequence is written as $\psi_0[x_i, t_i]$, then we can write the wave function that corresponds to the $k^{\text{th}}$ arm of the interferometer, $\psi_k$,($k=1 , 2$) as

\begin{subequations}\label{eq:integral_over_K}
\begin{align}
&\psi_{k}[x_f, t_f] = \int_{-\infty}^{\infty}dx_i \; K\big[x_f, t_f ; x_i, t_i ; J_{k}[t]\big] \psi_0[x_i, t_i]
\label{eq:integral_over_K_a}
\\
\begin{split}
&=
e^{\frac{i}{\hbar}\int_{t_i}^{t_f}dt' \; \epsilon L_1\big[\frac{\hbar}{i}\frac{\delta}{\delta J_{k}[t']}\big]}
\label{eq:integral_over_K_b}
\\
&\;\;\;\;\; \times
\int_{-\infty}^{\infty}dx_i \; K^{(0)}[x_f, t_f ; x_i, t_i ; J_{k}[t]] \psi_0[x_i, t_i]
\end{split}
\\
&=
e^{\frac{i}{\hbar}\int_{t_i}^{t_f}dt' \; \epsilon L_1\big[\frac{\hbar}{i}\frac{\delta}{\delta J_{k}[t']}\big]}
\psi^{(0)}[x_f, t_f ; J_{k}[t]]
\label{eq:integral_over_K_c}
\end{align}
\end{subequations}

\noindent
where $\psi^{(0)}$ is the wavepacket which at time $t_i$ is equal to $\psi_0$ and evolves in time under the unperturbed Lagrangian (i.e. under $L = L_0 + J x$ alone), and $J_{k}[t]$ is the effective force on by the $k^{\text{th}}$ arm due to the recoil kicks of the laser on the matter waves.
To get the first equality, we replaced $K$ with its definition in Eq. \eqref{eq:kernel_equation_c} and took the functional derivative outside of the integral over $x_i$.
In this work, we will take the initial wavefunction $\psi_0[x_i, t_i]$ to have a Gaussian form, for which $\psi^{(0)}$ also has Gaussian form, since for Lagrangians $L_0$ with terms that are second order in $\dot{x}$ and no more than second order in $x$, the kernel solution $K^{(0)}$ will have a Gaussian form, \cite{feynman1965quantum} and an integral over a Gaussian is a Gaussian.  \footnote{This follows from the fact that $\psi^{(0)}$ is a convolution of $\psi_0$ and $K^{(0)}$, which are both Gaussian. Convolving two Gaussians is equivalent to taking the inverse Fourier transform of the product of the Fourier transform of the two Gaussians. Since the Fourier transform of a Gaussian is Gaussian, and the product of two Gaussians is Gaussian, $\psi^{(0)}$ must be Gaussian.} Extensions of the diagrammatic method to arbitrary (potentially non-Gaussian) initial wavepackets are explored in Appendix \ref{sec:non_gaussian_initial_conditions}.
Example forms for $J$, for a Mach-Zehnder interferometer sequence and a sequence where the two arms of the interferometer are kicked symmetrically, are given in Secs. \ref{sec:free_potential_diagram_values} and \ref{sec:harmonic_potential}, respectively. Since the two arms are kicked differently, $J_{1} \neq J_{2}$.
As shown in Appendix \ref{sec:appendix_edge_values}, We can now write Eq. \eqref{eq:integral_over_psi} using Eq. \eqref{eq:integral_over_K} as

\begin{subequations}\label{eq:phase_shift_expression}
\begin{align}
\begin{split}
C & e^{i \Delta \phi}
\\
& =
e^{
\int_{t_i}^{t_f} dt' \epsilon
\Big(
\frac{i}{\hbar}
L_1\big[\frac{\hbar}{i}\frac{\delta}{\delta J_{1}[t']}\big]
+ \frac{-i}{\hbar}
L_1\big[\frac{\hbar}{-i}\frac{\delta}{\delta J_{2}[t']}\big]
\Big)
}
\\
& \;\;\;\;
\times
\int_{-\infty}^{\infty}dx_f \;
\psi^{(0)}[x_f, t_f ; J_{1}[t]]
\;
\psi^{(0)*}[x_f, t_f ; J_{2}[t]]
\end{split}
\label{eq:phase_shift_expression_a}
\\
\begin{split}
& =
e^{
\int_{t_i}^{t_f} dt' \epsilon
\Big(
\frac{i}{\hbar}
L_1\big[\frac{\hbar}{i}\frac{\delta}{\delta J_{1}[t']}\big]
+ \frac{-i}{\hbar}
L_1\big[\frac{\hbar}{-i}\frac{\delta}{\delta J_{2}[t']}\big]
\Big)
}
\\
& \;\;\;\;
\times
\exp\bigg[
\frac{1}{2}\int_{t_i}^{t_f}dt_1\int_{t_i}^{t_f}dt_2 \Big(
J_{1}[t_1]G_{11}[t_1, t_2]J_{1}[t_2]
\\
& \;\;\;\;\;\;\;\;\;\;\;\;\;\;\;\;\;\;\;\;\;\;\;\;\;\;\;\;\;\;\;\;
\;\;\;\;\;\;\;\;\;\;
+J_{1}[t_1]G_{12}[t_1, t_2]J_{2}[t_2]
\\
& \;\;\;\;\;\;\;\;\;\;\;\;\;\;\;\;\;\;\;\;\;\;\;\;\;\;\;\;\;\;\;\;
\;\;\;\;\;\;\;\;\;\;
+J_{2}[t_1]G_{21}[t_1, t_2]J_{1}[t_2]
\\
& \;\;\;\;\;\;\;\;\;\;\;\;\;\;\;\;\;\;\;\;\;\;\;\;\;\;\;\;\;\;\;\;
\;\;\;\;\;\;\;\;\;\;
+J_{2}[t_1]G_{22}[t_1, t_2]J_{2}[t_2]
\Big)
\bigg]
\end{split}
\label{eq:phase_shift_expression_b}
\\
\begin{split}
& =
\underbrace{
e^{
\int_{t_i}^{t_f} dt'
\epsilon
\Big(
\frac{i}{\hbar}
\delta_{a1}
L_1\big[\frac{\hbar}{i}\frac{\delta}{\delta J_a[t']}\big]
+ \frac{-i}{\hbar}
\delta_{a2}
L_1\big[\frac{\hbar}{-i}\frac{\delta}{\delta J_{a}[t']}\big]
\Big)
}
}_{\text{Factor 1}}
\\
& \;\;\;\;
\times
\underbrace{
e^{
\frac{1}{2}
\int_{t_i}^{t_f}dt_1
\int_{t_i}^{t_f}dt_2
\;
J_b[t_1]
G_{bc}[t_1,t_2]
J_c[t_2]
}
}_{\text{Factor 2}}
\end{split}
\label{eq:phase_shift_expression_c}
\end{align}
\end{subequations}

\noindent
where summation over repeated Latin indices $a, b, c$ is assumed, Latin indices run from $1$ to $2$, and $\delta_{ij}$ is the Kronecker delta function.
The function $G_{ab}[t_1, t_2]$ encodes how the interferometer contrast and phase respond to laser kicks at different times on different arms of the interferometer, where the arms are indexed by Latin indices. The function $G$ is symmetric under simultaneous swapping of time arguments and Latin indices $G_{ab}[t_1, t_2] = G_{ba}[t_2, t_1]$, and the analytic form for $G_{ab}[t_1, t_2]$ for a harmonic potential is derived in Appendix \ref{sec:appendix_edge_values}.

Equation \eqref{eq:phase_shift_expression_c} is exactly the kind of expression that emerges when computing scattering amplitudes in QFT (see Sec. \ref{sec:QFT_background}). One can solve for $C e^{i\Delta\phi}$ perturbatively in $\varepsilon$ by applying the functional derivatives of Factor 1 to the functions in Factor 2 in Eq. \eqref{eq:phase_shift_expression_c} which would require Taylor expanding both factors, applying the functional derivatives and evaluating the resulting integral expressions.
While such a calculation would be computationally intensive, the Feynman diagram formalism naturally organizes these terms and eliminates the need to manually evaluate functional derivatives.
The philosophy of Feynman diagrams is to ascribe value to the edges and vertices which compose a diagram so that a single diagram corresponds to a single term in the Taylor expansion of Eq. \eqref{eq:phase_shift_expression_c}, after the functional derivatives have been applied. Consider a perturbative term in the Lagrangian of the form $L_1[x] = c_n x^n/n!$. The four step process of diagrammatically computing the $j^{\text{th}}$ order expansion of $C e^{i \Delta\phi}$ in $\epsilon$ is to (1) compute the value of the edges and vertices, (2) compose all diagrams which consist of $j$ vertices, where each vertex has $n$ edges connected to it, (3) evaluate the integrals associated with the diagrams, and (4) sum the diagrams together.
When organized in this way, the interferometer contrast and phase shift can be written as

\begin{equation}\label{eq:ai_phase_shift}
C e^{i \Delta \phi} = e^{\text{sum of all connected diagrams}}
\end{equation}

\noindent
where the value of each diagram is weighted by the inverse of the number of automorphisms of the diagram (also called the symmetry factor). As part of step 1 of the diagrammatic computation, we write the value of diagram components as

\begin{equation}\label{eq:diagram_component_values_1D}
\centering
\vcenter{\hbox{\includegraphics[width=2.25in]{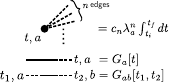}}}
\end{equation}

\noindent
where

\begin{equation}\label{eq:varepsilon_definition}
\lambda_a^{n}= \epsilon \,
\bigg(
\Big(\frac{\hbar}{i}\Big)^{n-1} 
\delta_{a1}
+
\Big(\frac{\hbar}{-i}\Big)^{n-1}
\delta_{a2}
\bigg)
\end{equation}

\noindent
and $G_a[t] = \int_{t_i}^{t_f}dt' J_{b}[t']G_{ba}[t',t]$. Determining the explicit form for $G_{ab}[t_1, t_2]$ and $G_{a}[t]$ completes step 1 of the diagrammatic computation. Explicit forms for $G_a[t]$ for some example interferometer sequences are provided in Appendix \ref{sec:appendix_edge_values}.

The focus of this paper will be on using Eq. \eqref{eq:ai_phase_shift} to solve Eq. \eqref{eq:phase_shift_expression_c} perturbatively $\varepsilon$.
In Eq. \eqref{eq:diagram_component_values_1D}, dashed ends indicate attachment points where diagram components join to one another, while solid ends are unconnected.
An external edge, with value $G_a[t]$, has one solid and one dashed end, while an internal edge, with value $G_{ab}[t_1, t_2]$, has two dashed ends, indicating that it connects to a vertex on both sides. This notation also captures the dependence of the vertex value on the number of attached edges $n$, indicated by the $n$ dashed lines emanating from it.
As an example, consider a system with a nonlinear potential $L_1 = \frac{c_3}{3!} x^3$. To compute a second order correction to the phase shift under this nonlinearity in the external potential, we must write down and sum all diagrams with two 3-point vertices. One such diagram is composed of two external edges and two internal edges and can be written as

\begin{figure}[H]
\centering
\includegraphics[width=2.25in]{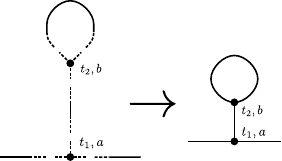}
\end{figure}

\noindent
where on the left we draw the diagram components including the dashed lines to better make the connection to Eq. \eqref{eq:diagram_component_values_1D}, and on the right we glue the diagram components together. Following the rules of Eq. \eqref{eq:diagram_component_values_1D}, the value of this diagram can be written as

\[
\begin{alignedat}{2}
& \frac{1}{4}\lambda_{a}^3 \lambda_{b}^3  (c_3)^2
\int_{t_i}^{t_f} \!\! dt_1 \int_{t_i}^{t_f}  \!\! dt_2 && \;
G_a[ t_1]^2 G_{ab}[t_1, t_2]G_{bb}[t_2, t_2] \\
&= \frac{(\epsilon c_3)^2 \hbar^4}{4}\!\!\!
\int_{t_i}^{t_f} \!\!\! dt_1 \int_{t_i}^{t_f} \!\!\! dt_2 \Big( &&
G_1[t_1]^2 G_{11}[t_1, t_2]G_{11}[t_2, t_2]
\\
& &&
+G_1[t_1]^2 G_{12}[t_1, t_2]G_{22}[t_2, t_2]
\\
& &&
+G_2[t_1]^2 G_{21}[t_1, t_2]G_{11}[t_2, t_2]
\\
& &&
+G_2[t_1]^2 G_{22}[t_1, t_2]G_{22}[t_2, t_2] \Big)
\end{alignedat}
\]

\noindent
where the factor of $1/4$ comes from the inverse of the symmetry factor of the diagram, and where we have used Eq. \eqref{eq:varepsilon_definition} and summed over Latin indices to arrive at the expression on the right hand side. Referencing Eq. \eqref{eq:ai_phase_shift}, the contribution to the phase is then given by the imaginary part of this term, and the quantum corrections to the interferometer contrast are determined by exponentiating the real part of this term. 

The rules described above are valid for doing calculations in one spatial dimension, and we now generalize the diagrammatic phase shift rules to three spatial dimensions. In 1D, vertices are labeled by a time variable $t_i$ and an interferometer arm index $a = 1,2$. In 3D, diagram components pick up a label 
that indexes the cartesian dimension. One notationally convenient way to write the diagram rules in 3D is to denote these new labels with Greek indices, which run from 1 to 3 for each of the three spatial dimensions. An alternate way to write out the diagram rules in 3D is to explicitly label the diagram edges with the cartesian coordinate, and write out all the permutations over the possible cartesian coordinate labels as independent diagrams. This second approach is explored in more detail in Appendix \ref{sec:appendix_vertex_values_3D}. 
For the purpose of this paper, we assume a separable form for the unperturbed Lagrangian $L_0$ (i.e. $L_0[\dot{x},x, \dot{y},y, \dot{z}, z] = L_x[\dot{x},x] + L_y[\dot{y},y] + L_z[\dot{z}, z]$). In three spatial dimensions, the value of the different components of the diagrams can be written as

\begin{equation}\label{eq:diagram_component_values_3D}
\centering
\vcenter{\hbox{\includegraphics[width=2.9in]{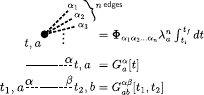}}}
\end{equation}

\noindent
where $G_a^{\alpha}[t] = \int_{t_i}^{t_f}dt' J_{b}^{\beta}[t']G_{ba}^{\beta \alpha}[t',t]$
, and
$\Phi_{\alpha_1 \alpha_2 ... \alpha_n}$
are the coefficients of the 3D Taylor series expansion of $L_1$:

\begin{equation}
L_1[x, y, z] = \sum_{n=0}^{\infty}\frac{1}{n!}\Phi_{\alpha_1\alpha_2...\alpha_n}r_{\alpha_1} r_{\alpha_2} ... r_{\alpha_n}
\end{equation}

\noindent
where the summation of repeated Greek and Latin indices is assumed ($\alpha = 1,2,3$), $r_1 = x, r_2 = y, r_3 = z$, and the coefficients $\Phi_{\alpha_1\alpha_2...\alpha_n}$ are symmetric about an exchange of indices (e.g. $\Phi_{1,1,2} = \Phi_{1,2,1} = \Phi_{2,1,1}$). The value of the functions $G^{\alpha}_a[t]$ and $G_{ab}^{\alpha\beta}[t_1, t_2]$ depend on the interferometer laser sequence, as well as the form of the unperturbed Lagrangian. Specific values for these diagram values are given in example calculations in Secs. \ref{sec:free_potential_diagram_values} and \ref{sec:harmonic_potential_diagram_values}.

\subsection{Connection to Traditional Semi-classical Framework}\label{sec:connection_to_traditional_semi_classical}

The traditional semi-classical approach to computing phase shifts involves separating the contributions of the phase into three parts : (1) a laser phase $\Delta\phi_{\text{laser}}$, which has to do with the imprint of the laser phase onto the matter waves, (2) a propagation phase 
$\Delta\phi_{\text{prop}}$, which has to do with the differential phase accumulated during the free propagation of the two arms under the spatially varying potential, and (3) a separation phase 
$\Delta\phi_{\text{sep}}$, which has to do with the two arms of the interferometer within the same output port not being perfectly overlapped with one another under spatial averaging. The total phase shift $\Delta\phi$ is expressed as a sum of these three components, $\Delta\phi = \Delta\phi_{\text{laser}} + \Delta\phi_{\text{prop}} + \Delta\phi_{\text{sep}}$. \cite{hogan2008lightpulse,Bongs2006}

In the diagram formalism, the propagation phase is associated with the convolution of wavepackets with kernels $K$, the separation phase is associated with the spatial averaging integrals over $x_f$ in Eq. \eqref{eq:phase_shift_expression}, and the laser phase is encoded in the external force term $J[t]$. In the semi-classical formalism, the laser phase associated with an individual interferometer arm can be written as 
$
\sum_{j} \pm k_x x_{\text{cl}}[t_j]
$,
where the sum is over all kicks to the classical trajectory, $x_{\text{cl}}[t]$, indexed by $j$. Kick $j$ occurs at time $t_j$, and $k_x$ is the wavenumber of the light causing the kick. The sign of the phase imprint depends on the direction of the kick.
In the limit of instantaneous matter-optics pulses, where $J[t]$ consists of delta functions in time (see for example Eqs. \eqref{eq:Mach-Zehnder_J} and \eqref{eq:symmetric_J}), the laser kick can be thought of as applying an instantaneous phase shear across the wavepacket, akin to applying the  operator $e^{\pm i \frac{m}{\hbar}v_r \hat{x}}$ to the wavepacket at the time of the pulse, where $v_r = \hbar k_x / m$ is the recoil velocity associated with the photon kick.
The sign of the phase gradient depends on the direction of the kick.
The phase gradient across the wavepacket imparts a momentum kick, and the mean phase imprinted onto the wavepacket on top of the gradient has the form $\pm k_x x_{\text{cl}}[t_j]$, as in the semi-classical formalism

It has been shown using Ehrenfest's theorem that when the external potential is at most quadratic in $x$, the interferometer phase shift is fully captured by the semiclassical approximation: there are no higher-order $\hbar$ corrections, and no contributions from the finite spatial extent of the wavepacket. \cite{hogan2008lightpulse}
There is a natural way to see this feature within the diagram formalism.
If there is no harmonic component included in $L_0$, so that harmonic contributions to the Lagrangian are treated perturbatively, the one-vertex tree-level (loop-free) diagrams reproduce the semi-classical solution to first order in the perturbative potential. Higher-order quantum corrections for these one-vertex diagrams arise from loops (see Table \ref{tab:power_law_phase_shifts}, where the leading order quantum correction comes from a diagram with a loop).
Vertices and internal edges carry no dependence on $J[t]$, only external edges encode information about the laser kicks.
"Bubble" diagrams, which have no external edges, therefore carry no information about the interaction between the matter wave and the interferometer laser. If there is no interaction with the interferometer laser, there can be no phase shift, so these bubble diagrams always sum to zero.
\footnote{In other words, the laser produces the superposition of interferometer arms. Without the laser, there is no superposition, and the two `arms' of the interferometer will correspond to exactly the same trajectory. In this case, the phase shift between the two `arms' is zero. The value of a bubble diagram is completely independent of the parameters of the laser pulse, and so has the same value as when the laser performs no atom-optic operations. In this case, the phase shift is zero, so it must be that the bubble diagrams sum to zero for all possible interferometer sequences}
The fact that higher order than quadratic terms are required to induce beyond-semi-classical corrections to the phase shift arises in the diagram formalism by the fact that it is not possible to construct a diagram with one loop and at least one external edge using only $2$-point vertices. Only with a $3$-point vertex (a cubic term in the Lagrangian) can a diagram with at least one external edge and one loop be constructed, leading to higher order quantum corrections. In this context, an $n$-vertex diagram is a diagram with $n$ vertices, and an $n$-point vertex is a vertex with $n$ edges connected to it.

\subsection{\label{sec:QFT_background}Connection to Quantum Field Theory}

Here we make explicit the correspondence between the diagrammatic framework for phase shifts developed in Sec. \ref{sec:phase_shifts_with_feynman_diagrams} and the standard diagrammatic treatment of scattering amplitudes in QFT.
Consider a scalar field theory with Lagrangian density $\mathcal{L} = \mathcal{L}_0+ \epsilon \mathcal{L}_1 + J \varphi$, where $\mathcal{L}_0 = -\frac{1}{2}\partial^\mu \varphi \partial_{\mu}\varphi -\frac{1}{2} m^2 \varphi^2$ is the "free" component of the Lagrangian density, $\mathcal{L}_1$ is a contribution that is third order or higher in $\varphi$, $J$ is a source term, and $\epsilon$ is a perturbative parameter.
The ground-to-ground state amplitude in the presence of a source $J$, often written as $\langle 0 | 0 \rangle_J$ or $Z[J]$, serves as a generating functional for correlation functions (also called Green's functions or n-point functions). Functional derivatives of $\langle 0 | 0 \rangle_J$ with respect to $J$, evaluated at $J=0$, yield time-ordered correlation functions which can be related to scattering amplitudes via the LSZ reduction formula (see Ref. \citenum{Srednicki:2007qs}). 
The ground state to ground state amplitude in the scalar field theory described by $\mathcal{L}$ can be written as

\begin{subequations}\label{eq:ground_to_ground_amplitude}
\begin{align}
&\langle 0 | 0 \rangle_J = \int \mathcal{D}\varphi e^{i \int d^4 x \big( \mathcal{L}_0 + \epsilon \mathcal{L}_1 + J \varphi \big)}
\label{eq:ground_to_ground_amplitude_a}
\\
&= e^{i \int d^4 x \; \epsilon \mathcal{L}_1\big[\frac{1}{i}\frac{\delta}{\delta J[x]}\big]}
\int \mathcal{D}\varphi e^{i \int d^4 x \big( \mathcal{L}_0 + J \varphi \big)}
\label{eq:ground_to_ground_amplitude_b}
\\
&= e^{i\int d^4 x \; \epsilon \mathcal{L}_1\big[\frac{1}{i}\frac{\delta}{\delta J[x]}\big]}
e^{
\frac{i}{2}
\int d^4 x_1 \int d^4 x_2 \;
J[x_1]G[x_1, x_2]J[x_2]
}
\label{eq:ground_to_ground_amplitude_c}
\end{align}
\end{subequations}

\noindent
where for the second equality, Eq. \eqref{eq:ground_to_ground_amplitude_b}, we have made use of a well-established trick in QFT to take the nonlinear component of the Lagrangian density outside of the path integral by replacing its argument in $\varphi$ with functional derivatives with respect to the source term $J$.\cite{zee2010quantum} This is the same trick we use to arrive at Eqs. \eqref{eq:kernel_equation_b} and \eqref{eq:integral_over_K_b}.
The last equality, Eq. \eqref{eq:ground_to_ground_amplitude_c}, highlights that after taking the nonlinear component of the Lagrangian density outside of the integrand, the path integral can be solved in closed form.
Notice that the equalities of Eqs. \eqref{eq:phase_shift_expression_c} and \eqref{eq:ground_to_ground_amplitude_c} have similar forms, with the former including Latin indices.
QFT employs second quantization, in which fields become operators that create and annihilate particles. In contrast, our matter-wave interferometer formalism uses first quantization, where only position and momentum are operators while wavefunctions remain functions (not operators). Therefore, the analog of the field $\varphi$ in Eq. \eqref{eq:ground_to_ground_amplitude} is the position coordinate $x$ in Eq. \eqref{eq:phase_shift_expression}.
In QFT, the \textit{propagator} $G$ is an amplitude associated with a particle propagating from space-time point $x_1$ to space-time point $x_2$, and is symmetric under an exchange of its arguments, $G[x_1, x_2] = G[x_2, x_1]$ \cite{Srednicki:2007qs, zee2010quantum, peskin2018introduction}.
For simplicity, we take $\hbar = 1$ in this section.
One can solve for $\langle 0 | 0 \rangle_J$ perturbatively in $\epsilon$ in exactly the same way as we outline in Sec. \ref{sec:phase_shifts_with_feynman_diagrams}, organizing terms in the expansion with diagrams. 
The values of edges and vertices in this expansion are

\begin{equation}
\centering
\vcenter{\hbox{\includegraphics[width=2.25in]{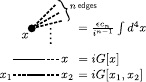}}}
\end{equation}

\noindent
where $G[x] = \int d^4 x' J[x']G[x', x]$, $c_n$ is the $n$th order Taylor series coefficient in the expansion of $\mathcal{L}_1$, $\mathcal{L}_1 = \sum_{n=0}^{\infty} \frac{1}{n!}c_n \varphi^n$, dashed lines correspond to connections of an edge and vertex
,and $x_i$ is the label of the vertex.
The key identity that we leverage when we apply this formalism to phase shift calculations in matter wave interferometry is that the ground state to ground state amplitude can then be written as \cite{Srednicki:2007qs, zee2010quantum}

\begin{equation}\label{eq:numbered_equation_1}
\langle 0 | 0 \rangle_J = 
e^{\text{sum of all connected diagrams}}
\end{equation}

\noindent
where the value of each diagram is weighted by the inverse of the symmetry factor of the diagram. Further details can be found in Refs. \citenum{Srednicki:2007qs}, \citenum{peskin2018introduction}, and \citenum{zee2010quantum}. Equation \eqref{eq:numbered_equation_1} is the QFT analog of Eq. \eqref{eq:ai_phase_shift}, and both expressions arise from the same underlying mathematics.

\section{Treating Harmonic Terms Perturbatively}\label{sec:treating_harmonic_term_perturbatively}

One can choose whether or not to include the harmonic component as part of the unperturbed component of the Lagrangian $L_0$, or to include it as part of the perturbed component $L_1$. In this section, we explore the latter option, and in Sec. \ref{sec:harmonic_potential}, we explore the former. When the harmonic component is small, it can be more convenient to treat it perturbatively. Neglecting it in $L_0$ leads to simpler integrands and therefore time integrals that are easier to evaluate. The trade-off is that one must then explicitly compute diagrams with two-point vertices associated with the perturbative treatment of the harmonic term.
Conversely, when the interferometer is operated in a harmonic trap, or more generally when the harmonic term is not small, and many orders of the harmonic component are required, a perturbative treatment of the harmonic term can require summing a large number of diagrams to account for all relevant two-point vertices.
In these cases, it can be simpler to absorb the harmonic component into the unperturbed Lagrangian and treat only the $n$-point vertices with $n\geq 3$ as perturbations. This avoids the need for a high-order diagrammatic expansion in the trap frequency at the cost of a slightly more complicated unperturbed solution.

\subsection{Diagram Values}\label{sec:free_potential_diagram_values}

In this section we use the rules outlined in Sec. \ref{sec:phase_shifts_with_feynman_diagrams} to evaluate the value of diagram elements for a free unperturbed potential, and use the diagrams to analytically compute phase shifts for matter wave interferometers subject to an external potential expressed as a power series in position. 
Consider a matter wave interferometer sensing Earth's gravitational field, where the interferometer arms evolve under a Lagrangian of the form $L = L_0 + \epsilon L_1 + \mathbf{J}[t]\cdot\mathbf{r}$, where

\begin{equation}\label{eq:power_law_lagrangian}
\begin{split}
L_0 &= \frac{m}{2}\left(\dot{x}^2 + \dot{y}^2 + \dot{z}^2 \right) - m g z\\ 
L_1 &= -\bigg(\frac{m}{2} 
\left(
T_{xx}x^2 + T_{yy}y^2  + T_{zz}z^2 
\right)
\\
&
\;\;\;\;\;\;\;\;\;\;\;
+ \frac{m}{3!}
\left(
3Q_{xxz}x^2 z + 3Q_{yyz}y^2 z + Q_{zzz}z^3
\right) 
\\
&
\;\;\;\;\;\;\;\;\;\;\;
+ \frac{m}{4!}S_{zzzz}z^4 \bigg)
\end{split}
\end{equation}

\noindent
and $g$, $T_{ij}$, $Q_{ijk}$, and $S_{ijkl}$ are the coefficients of the Taylor series expansion of Earth's gravitational field in space, evaluated about the location of the interferometer on the surface of the Earth.
We choose the $z$ axis to be the axis of the interferometer.
More generally, these coefficients could correspond to the Taylor expansion of any spatially varying potential, such as that generated by a dipole beam, or nonuniform magnetic fields.
The spatial integrals that are part of 
Eqs. \eqref{eq:integral_over_K} and \eqref{eq:phase_shift_expression}
are performed over one spatial dimension $x$.
In Appendix \ref{sec:appendix_edge_values}, we derive the values of the internal edges, $G_{ab}[t_1, t_2]$ in 1D.
In Appendix \ref{sec:appendix_edge_values_3D}, we extend these calculations to three spatial dimensions, where we take the initial matter wavepacket (the wavepacket at time $t_i = 0$ when the first beamsplitter operation occurs) to have the following form:

\begin{equation}
\psi_0[x,y,z]
=
\prod_{\alpha=1}^{3}
\left(\frac{2}{\pi w_{0,\alpha}^2}\right)^{1/4}
\exp\!\left(-\frac{r_{\alpha}^2}{w_{0,\alpha}^2}\right)
\end{equation}

\noindent
where $w_{0,1}, w_{0,2}, w_{0,3}$ are the initial wavepacket waists in the $r_1= x$, $r_2=y$, and $r_3 = z$ dimensions, respectively.
This calculation 
yields the 3D internal diagram edges $G_{ab}^{\alpha\alpha}[t_1, t_2]$, which for the Lagrangian of Eq. \eqref{eq:power_law_lagrangian}, can be expressed as

\begin{equation}\label{eq:Gab_free_potential}
\begin{split}
G_{11}^{\alpha\alpha}[t_j, t_k] &= \frac{1}{2m\hbar}\Big(-\beta_{\alpha} t_j t_k  - \frac{1}{\beta_{\alpha}} + i |t_j-t_k|\Big) \\
G_{12}^{\alpha\alpha}[t_j, t_k] &= \frac{1}{2m\hbar}\Big( \beta_{\alpha} t_j t_k + \frac{1}{\beta_{\alpha}} + i (t_j-t_k) \Big) \\
G_{21}^{\alpha\alpha}[t_j, t_k] &= \frac{1}{2m\hbar}\Big( \beta_{\alpha} t_j t_k + \frac{1}{\beta_{\alpha}} - i (t_j-t_k) \Big) \\
G_{22}^{\alpha\alpha}[t_j, t_k] &= \frac{1}{2m\hbar}\Big( -\beta_{\alpha} t_j t_k - \frac{1}{\beta_{\alpha}} - i |t_j - t_k| \Big)
\end{split}
\end{equation}

\noindent
where $\beta_{\alpha} = 2\hbar/m (w_{0,\alpha})^2$.
For notational convenience, we absorb the $m g z$ term in $L_0$ of Eq. \eqref{eq:power_law_lagrangian} into $\mathbf{J}[t]$, which simplifies the form of $G_{ab}^{\alpha\alpha}[t_1, t_2]$.
For a Mach-Zehnder interferometer, with an interferometer beam oriented along the $z$-axis, the external force term has the following form:

\begin{equation}\label{eq:Mach-Zehnder_J}
\frac{J_a^3[t]}{m} = 
-g + 
\begin{cases} 
      v_r\delta[t-0]-v_r\delta[t-T], & a = 1 \\
      v_r\delta[t-T]-v_r\delta[t-2T], & a = 2 \\
\end{cases}
\end{equation}

\noindent
corresponding to interference between the two arms in the "lower" output port of the interferometer referencing Fig. \ref{fig:ai_introduction_figure}.
$J_1^3$ corresponds to the kicks on the "upper" arm of the interferometer, and $J_2^3$ corresponds to the kicks on the lower arm. The superscript 3 denotes the $z$ axis, and laser kicks along the $x$ and $y$ axes can be captured by modifying the values of $J_a^{1}[t]$ and $J_a^{2}[t]$ respectively. For the calculations we will do in this section, we will take $J_a^{1}[t]$ and $J_a^{2}[t] = 0$.
Alternate pulse sequences can be computed by adding/modifying terms that are proportional to the recoil velocity $v_r$.
In this system, the value of the external edges can be expressed in terms of the semi-classical trajectories of the interferometer arms as follows:

\begin{equation}\label{eq:free_potential_external_edge_value}
\begin{split}
G_a^1[t] &= (-1)^{a+1} \frac{i}{\hbar} x_{\text{cl},a}^{(0)} [t] \\
G_a^2[t] &= (-1)^{a+1} \frac{i}{\hbar} y_{\text{cl},a}^{(0)} [t] \\
G_a^3[t] &= (-1)^{a+1} \frac{i}{\hbar} z_{\text{cl},a}^{(0)} [t]
\end{split}
\end{equation}

\noindent
where $z_{\text{cl}, 1}^{(0)} [t]$ and $z_{\text{cl}, 2}^{(0)} [t]$ are the unperturbed classical trajectories of the two interferometer arms (indexed by $a$) along $z$, which evolve under the Lagrangian $L_0$. For the calculations we will perform in this section, the deflections of the semiclassical trajectories transverse to the interferometer axis will be zero, such that $x_{\text{cl},a}^{(0)} [t] = y_{\text{cl},a}^{(0)} [t] = 0$.
Equation \eqref{eq:free_potential_external_edge_value} holds even when we account for a nonzero initial position $z_0$ and velocity $v_z$ of the classical trajectory. See Appendix \ref{sec:appendix_edge_values} for more details. We now use these rules to compute phase shifts under power law potentials.

\subsection{Validating with Numerics in 1D}\label{sec:power_law_potential_split_step}

Here we demonstrate agreement between the diagrammatic approach to computing phase shifts and an \textit{ab-initio} numerical evaluation of the phase shift which accounts for the contributions that emerge from the finite-size nature of the matter wavepacket.
Consider the Lagrangian of Eq. \eqref{eq:power_law_lagrangian}, with motion restricted to the $z$-axis and $L_1$ containing only the $Q_{zzz}$ term.
We leverage the split-step \cite{FEIT1982412} approach in one spatial dimension ($z$) to numerically time evolve matter wavepackets in this power-law potential. We independently time evolve the two wavepackets associated with the two interferometer arms. The instantaneous laser kicks are performed by applying linear phase gradients across the two wavefunctions between discrete time steps.
The phase is extracted by numerically evaluating $P_{\text{lower}}[\phi_b]$, the population in the lower output port as a function of the final beamsplitter phase $\phi_b$, via a discrete sum over spatial coordinates. This population can be expressed as
$
P_{\text{lower}}[\phi_b] = 
\int_{\infty}^{\infty}dz_f \big| \psi_1[z_f, t_f]+ \psi_2[z_f, t_f]e^{-i\phi_b} \big|^2
$.
$P_{\text{lower}}[\phi_b]$ is evaluated for 63 evenly spaced values of $\phi_b$ between $0$ and $2 \pi$, at a time $t_f$ immediately following the final beamsplitter. 
Here, $\psi_1$ is the wavefunction which corresponds to the `upper' interferometer arm in Fig. \ref{fig:ai_introduction_figure}. 
The interferometer contrast $C$ and phase shift $\Delta\phi$ are then extracted by fitting $P_{\text{lower}}[\phi_b]$ to $\frac{1}{2}\big(1 + C \cos[\Delta\phi + \phi_b]\big)$.

\begin{figure}
\includegraphics[width=3in]{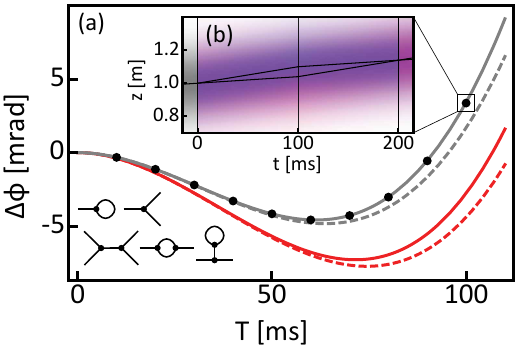}
\caption{\label{fig:power_law_split_step} 
(a) Phase shifts under the Lagrangian $L = m \dot{z}^2/2 - m g z - m Q_{zzz} z^3/3!$ as computed numerically (black dots), via the traditional semi-classical formalism to first order in $Q_{zzz}$ (red dashed line), to second order in $Q_{zzz}$ (red solid line), and computed diagrammatically to first order in $Q_{zzz}$ (gray dashed line), and to second order in $Q_{zzz}$ (gray solid line). The diagrams which contribute to the first and second order corrections in $Q_{zzz}$ are indicated in the bottom left hand side. The simulation parameters are as follows, with each parameter expressed in SI units: $\hbar/m = 10^{-2}$, $w_0 = 0.3$, $v_r = 0.6$, $v_z = 0.4$, $z_0 = 1$, $g = 0.52$, $Q_{zzz} = -0.9$, $T_{zz} = S_{zzzz} = 0$, and no higher order potential terms. The split step numerics were done with spatial resolution of $\Delta z = 2.5\text{ mm}$, a temporal resolution of $\Delta t = 0.5 \text{ ms}$, and a grid centered at $z=0$ with a size of $4 \text{ m}$.
(b) The span of the two interferometer arms for $T = 100 \text{ ms}$. The gray density plot indicates the width of the wavefunction before the first beamsplitter. The first beamsplitter operation splits the wavefunction into an upper arm (blue) and lower arm (red). The final beamsplitter then interferes the two arms (purple). The classical trajectories are indicated with solid black lines.
}
\end{figure}

We run these simulations to verify the mathematical approach and use unphysical parameters (e.g. larger values of $\hbar$) to amplify quantum corrections and reduce computational complexity. Realistic physical parameters require high spatial-temporal resolution for accuracy, which increases computation time.
These simulations do not represent expected phase shifts in physically realistic conditions, but serve to validate the analytic terms which come out of the diagrammatic formalism.
The analytic terms are then used to estimate realistic phase shifts (see Secs. \ref{sec:gravity_gradients}, \ref{sec:anharmonic_traps}, and \ref{sec:calculation_of_phase_response_3D}).
Figure \ref{fig:power_law_split_step} plots the interferometer phase shift as a function of interrogation time under this power-law potential computed in three different ways: (1) numerically (black dots), (2) using the traditional semi-classical approach to first order in $Q_{zzz}$ (dashed red line) and to second order in $Q_{zzz}$ (solid red line), and (3) using the diagrammatic approach, to first order in $Q_{zzz}$ (dashed gray line) and to second order in $Q_{zzz}$ (solid gray line).
The two contributing diagrams to first order in $Q_{zzz}$, along with the three diagrams that contribute at second order in $Q_{zzz}$ are indicated in the bottom left corner of panel (a). 
Simulation parameters are indicated in the caption. In this parameter space, the terms that emerge from the diagrammatic calculator have strong agreement with the output of the split-step numerics, whereas the semi-classical approach deviates more strongly from the numerical results. 
Panel (b) of Fig. \ref{fig:power_law_split_step} displays the classical trajectories of the two interferometer arms (black lines) for a $T=100\text{ ms}$ interferometer sequence. The gray density plot indicates the wavepacket waist before the first beamsplitter operation at time $t=0$. The blue density plot indicates the wavepacket associated with the upper interferometer arm, and the red wavepacket corresponds to the lower interferometer arm.
The diagrammatic approach re-produces the terms which emerge from the semi-classical solver, and produce new terms associated with higher order quantum corrections.
Let us define the interferometer phase shift $\Delta\phi$ as a sum of the semi-classically accumulated phase and quantum corrections, $\Delta\phi = \Delta\phi_{\text{cl}} + \Delta\phi_{\text{qu}}$.
The terms which emerge from the diagrammatic calculation but which do not emerge from the semi-classical calculation to second order in $Q_{zzz}$ are

\begin{equation}
\begin{split}
\Delta\phi_{\text{qu}} = 
& -Q_{zzz} v_r \left(\frac{1}{4\beta_3} T^2 + \frac{7 \beta_3}{24} T^4\right)
\\
&+ Q_{zzz}^2 v_r \bigg(
\frac{7}{16 \beta_3} z_0 T^4 + \frac{5}{32 \beta_3}v_r T^5 + \frac{5}{16 \beta_3} v_z T^5
\\
&
\;\;\;\;\;\;\;\;\;\;\;\;\;\;\;\
-\frac{31}{180 \beta_3} g T^6 + \frac{31 \beta_3}{144}z_0 T^6 + \frac{3 \beta_3}{32}v_r T^7 
\\
&
\;\;\;\;\;\;\;\;\;\;\;\;\;\;\;\
+ \frac{3 \beta_3}{16} v_z T^7 - \frac{1397 \beta_3}{13440} g T^8
\bigg)
\\
&+ \mathcal{O}[Q_{zzz}^3]
\end{split}
\end{equation}

\noindent
These quantum correction terms are the source of the disagreement between the semi-classically calculated phase shift and the diagrammatically calculated phase shift in Fig. \ref{fig:power_law_split_step}.
In order to emphasize the utility of the diagrammatic method for computing higher order corrections to atom interferometer phase shifts, we chose a numerical parameters in Fig. \ref{fig:power_law_split_step} so that the second order $Q_{zzz}$ contributions can be numerically resolved.

\subsection{3D Calculation of Higher Order Quantum Corrections under Gravity Gradients of the Earth} \label{sec:gravity_gradients}

To explore the extent to which these higher order quantum corrections under power law potentials associated with the gravitational field of the Earth could be measured with a state-of-the-art matter wave interferometer, we take the analytic terms that come out of the diagrammatic formalism, and plug in typical operating parameters (listed in the caption of Table \ref{tab:power_law_phase_shifts}).
The terms are ranked in Table \ref{tab:power_law_phase_shifts} by the size of their contribution to the interferometer phase shift, with the higher-order quantum correction terms highlighted in gray.

The numerical values of the Taylor series coefficients $T_{ij}$, $Q_{ijk}$, and $S_{ijkl}$ ultimately reflect the gravitational potential local to the interferometer and therefore may differ, for a given site and a given time, from the values considered here. The values of the coefficients used in Table \ref{tab:power_law_phase_shifts} are obtained by first assuming a spherical Earth with homogeneous mass density. We then add corrections (also called gravitational anomalies) $T_a$, $Q_a$, and $S_a$ to the coefficients $T_{zz}$, $Q_{zzz}$, and $S_{zzzz}$.
Under this prescription, the corrected coefficients become $T_{zz} = -2g/R - T_a$, $Q_{zzz} = 6g/R^2 + Q_a$, and $S_{zzzz} = -24g/R^3 - S_a$.
We take the following example values based on the estimates in Ref. \citenum{dubetsky_comment_2020}: $T_a = 10^{-7}~\text{s}^{-2}$, $Q_a = 3\times 10^{-11}~\text{m}^{-1} \text{s}^{-2}$, $S_a = 10^{-14}~\text{m}^{-2} \text{s}^{-2}$, $g=9.8~\text{m}/\text{s}^2$, and $R = 6.38\times10^6 ~\text{m}$.
The coefficients governing the transverse dependence of the gravitational field (e.g. $Q_{xxz}$) are chosen such that the resulting potential satisfies Laplace’s equation.
The numerical value of the Taylor series coefficients at a specific interferometer site will differ due to local contributions of the non-spherical nature of the Earth, the inhomogeneous mass density of the Earth, nearby structures, sea level variations, and other site-dependent sources of gravitational potentials. \cite{ufrecht_reply_2020}
We emphasize that the calculation presented here is intended as a demonstration of the method for computing quantum corrections to power-law potentials in general. In practice, the true coefficients ($T_{ij}$, $Q_{ijk}$, $...$) would need to be measured at the site of the interferometer.

It is important to carry out calculations of quantum corrections to interferometer phase shifts in three dimensions, even if the semiclassical trajectories are confined to one spatial dimension. For the calculation we perform here, even though the semiclassical trajectories do not deflect in the directions transverse to the interferometer axis (i.e. the classical paths lie entirely along $z$), quantum corrections still depend on the transverse gravity gradients. For the potentials considered in Table \ref{tab:power_law_phase_shifts}, if the initial wavepacket is spherically symmetric ($w_{0,1} = w_{0,2} =  w_{0,3}$), then the value of the 1-loop 3-point vertex diagram of Table \ref{tab:power_law_phase_shifts} vanishes: the terms proportional to $Q_{xxz}$ (term 27) and $Q_{yyz}$ (term 28) cancel with the corresponding $Q_{zzz}$ contribution (term 24). This cancellation was noted in Ref. \citenum{ufrecht2020perturbative}. 
As summarized in Sec. \ref{sec:conditions_for_quantum_corrections_to_be_zero}, 
and discussed in greater detail in Appendix \ref{sec:appendix_summing_1_vertex_diagrams_in_3D}, 
the diagrammatic formalism can be used to prove that, more generally, if the wavepacket is prepared with spherical symmetry, then the first-order quantum corrections to the semiclassical approximation vanish for any external potential $V[\mathbf{r}]$ that satisfies Laplace's equation, $\nabla^2 V[\mathbf{r}] = 0$.
Since the gravitational potential in mass-free regions satisfies Laplace's equation, it follows that if the initial wavepacket is spherically symmetric, there are no quantum corrections to the interferometer phase shift to first order in $V[\mathbf{r}]$.
For the numerical values of Table \ref{tab:power_law_phase_shifts}, we take the initial wavepacket to be twice as large along the interferometer axis than in the transverse dimensions, which gives rise to nonzero quantum corrections at first order in $Q_{ijk}$. The asymmetry in the initial wavepacket can be interpreted as arising from asymmetric cooling. For the parameters of Table \ref{tab:power_law_phase_shifts}, the momentum-space width is smaller along the interferometer axis than in the transverse directions, corresponding to a lower temperature along that axis. 
In Ref. \citenum{ufrecht2020perturbative}, the authors compute the quantum corrections arising from a first-order expansion of the interferometer phase shift in $Q_{ijk}$. Using the same choice of initial wavepacket waists, $\beta_{1} = \beta_{2} = 2 \beta_3 = \omega$, our first order result in $Q_{ijk}$ agrees with their first-order expression.

Notably, some quantum correction terms share the same scaling as experimentally adjustable parameters (e.g. scaling with $T$, $v_z$, and $z_0$) as the semiclassical terms, and so would therefore manifest as `anomalous' corrections to these terms. The quantum correction terms considered here can be distinguished by their dependence on the wavepacket waist. For example, term 24 in Table I scales as $T^2$, identically to term 1, and would therefore manifest as an anomalous acceleration in experimental measurements. However, this correction depends on both the cubic nonlinearity $Q_{zzz}$ and the wavepacket size along the $z$ axis $w_{0,z}$, which could provide an experimental signature to distinguish it from the semiclassical contribution.

\begin{table}
\centering
\caption{\label{tab:power_law_phase_shifts}
Phase shift terms for a Mach-Zehnder matter wave interferometer in 3D subject to the spatially varying gravitational field of the Earth (including gravitational anomalies, see main text for details), along with the diagrams which produce the term, ranked in order of the magnitude of their contribution to the total phase shift. The phase shifts are computed with the following parameters: $k_z = 2 \pi / (698 \text{ nm})$, $T = 1 \text{ s}$, $g = 9.8 \text{ m/s}^2$, $T_{zz} \approx -3.17\times10^{-6}~\text{s}^{-2}$, $T_{xx} = T_{yy} = -T_{zz}/2$, $Q_{zzz} \approx 3.15\times10^{-11}~\text{m}^{-1}\text{s}^{-2}$, $Q_{xxz} = Q_{yyz} = -Q_{zzz}/2$, $S_{zzzz} \approx -1.00 \times 10^{-14} ~ \text{m}^{-2}\text{s}^{-2}$, $x_0 = y_0 = z_0 = 0$, $v_z = 3 \text{ m/s}$, $v_x = v_y = 0$,
$m = 86.91\text{u}$, $v_r = \hbar k_z / m$, $\beta_{\ell} = 2\hbar/(m w_{0,\ell}^2)$ with $\ell = x,y,z$. $w_{0,z} = 2 \text{ mm}$, $w_{0,x} = w_{0,y} = 1 \text{ mm}$. The quantum correction terms are highlighted in light gray.
}
\renewcommand{\arraystretch}{\tableStretch}
\begin{tabular}{cccc}
\hline \hline
& Phase Shift & Size [rad] & Diagram \\
\hline
1 & $-\frac{m}{\hbar} gv_rT^2$ & $-8.82\times 10^{7}$ & \\
2 & $\frac{7}{12}\frac{m}{\hbar }gv_rT^4T_{\text{zz}}$ & $-1.63\times 10^{2}$ & \TzzOne \\
3 & $-\frac{m}{\hbar }v_rv_zT^3T_{\text{zz}}$ & $8.57\times 10^{1}$ & \TzzOne \\
4 & $-\frac{1}{2}\frac{m}{\hbar }v_r^2T^3T_{\text{zz}}$ & $9.39\times 10^{-2}$ & \TzzOne \\
5 & $-\frac{31}{120}\frac{m}{\hbar }g^2v_rT^6Q_{\text{zzz}}$ & $-7.02\times 10^{-3}$ & \QzzzOne \\
6 & $\frac{3}{4}\frac{m}{\hbar }gv_rv_zT^5Q_{\text{zzz}}$ & $6.24\times 10^{-3}$ & \QzzzOne \\
7 & $-\frac{7}{12}\frac{m}{\hbar }v_rv_z^2T^4Q_{\text{zzz}}$ & $-1.49\times 10^{-3}$ & \QzzzOne \\
8 & $-\frac{31}{360}\frac{m}{\hbar }gv_rT^6T_{\text{zz}}^2$ & $-7.64\times 10^{-5}$ & \TzzTwo \\
9 & $\frac{1}{4}\frac{m}{\hbar }v_rv_zT^5T_{\text{zz}}^2$ & $6.79\times 10^{-5}$ & \TzzTwo \\
10 & $-\frac{3}{8}\frac{m}{\hbar }g^2v_rv_zT^7S_{\text{zzzz}}$ & $9.73\times 10^{-6}$ & \SzzzzOne \\
11 & $\frac{127}{1344}\frac{m}{\hbar }g^3v_rT^8S_{\text{zzzz}}$ & $-8.01\times 10^{-6}$ & \SzzzzOne \\
12 & $\frac{3}{8}\frac{m}{\hbar }gv_r^2T^5Q_{\text{zzz}}$ & $6.84\times 10^{-6}$ & \QzzzOne \\
13 & $\frac{31}{60}\frac{m}{\hbar }gv_rv_z^2T^6S_{\text{zzzz}}$ & $-4.10\times 10^{-6}$ & \SzzzzOne \\
14 & $-\frac{7}{12}\frac{m}{\hbar }v_r^2v_zT^4Q_{\text{zzz}}$ & $-3.26\times 10^{-6}$ & \QzzzOne \\
15 & $-\frac{1}{4}\frac{m}{\hbar }v_rv_z^3T^5S_{\text{zzzz}}$ & $6.08\times 10^{-7}$ &\SzzzzOne  \\
16 & $\frac{1}{8}\frac{m}{\hbar }v_r^2T^5T_{\text{zz}}^2$ & $7.45\times 10^{-8}$ & \TzzTwo \\
17 & $-\frac{9}{20}\frac{m}{\hbar }gv_rv_zT^7Q_{\text{zzz}}T_{\text{zz}}$ & $1.19\times 10^{-8}$ & \QzzzOneTzzOne \\
18 & $-\frac{3}{16}\frac{m}{\hbar }g^2v_r^2T^7S_{\text{zzzz}}$ & $1.07\times 10^{-8}$ & \SzzzzOne \\
19 & $\frac{127}{1120}\frac{m}{\hbar }g^2v_rT^8Q_{\text{zzz}}T_{\text{zz}}$ & $-9.78\times 10^{-9}$ & \QzzzOneTzzOne \\
20 & $\frac{31}{60}\frac{m}{\hbar }gv_r^2v_zT^6S_{\text{zzzz}}$ & $-9.00\times 10^{-9}$ & \SzzzzOne \\
21 & $\frac{31}{72}\frac{m}{\hbar }v_rv_z^2T^6Q_{\text{zzz}}T_{\text{zz}}$ & $-3.48\times 10^{-9}$ & \QzzzOneTzzOne \\
22 & $-\frac{1}{6}\frac{m}{\hbar }v_r^3T^4Q_{\text{zzz}}$ & $-2.04\times 10^{-9}$ & \QzzzOne \\
23 & $-\frac{3}{8}\frac{m}{\hbar }v_r^2v_z^2T^5S_{\text{zzzz}}$ & $2.00\times 10^{-9}$ & \SzzzzOne \\
\cellcolor{gray!25}24 & \cellcolor{gray!25}$-\frac{1}{4}\frac{1}{\beta_z}v_rT^2Q_{\text{zzz}}$ & \cellcolor{gray!25}$-1.42\times 10^{-10}$ & \QuantumCorrection \\
25 & $-\frac{1}{40}\frac{m}{\hbar }v_rv_zT^7T_{\text{zz}}^3$ & $2.15\times 10^{-11}$ & \TzzThree \\
26 & $\frac{127}{20160}\frac{m}{\hbar }gv_rT^8T_{\text{zz}}^3$ & $-1.77\times 10^{-11}$ & \TzzThree \\
\cellcolor{gray!25}27 & \cellcolor{gray!25}$-\frac{1}{4}\frac{1}{\beta_x}v_rT^2Q_{\text{xxz}}$ & \cellcolor{gray!25}$1.77\times 10^{-11}$ & \QuantumCorrection \\
\cellcolor{gray!25}28 & \cellcolor{gray!25}$-\frac{1}{4}\frac{1}{\beta_y}v_rT^2Q_{\text{yyz}}$ & \cellcolor{gray!25}$1.77\times 10^{-11}$ & \QuantumCorrection \\
29 & $-\frac{9}{40}\frac{m}{\hbar }gv_r^2T^7Q_{\text{zzz}}T_{\text{zz}}$ & $1.30\times 10^{-11}$ & \QzzzOneTzzOne \\
30 & $\frac{31}{72}\frac{m}{\hbar }v_r^2v_zT^6Q_{\text{zzz}}T_{\text{zz}}$ & $-7.63\times 10^{-12}$ & \QzzzOneTzzOne \\

\hline \hline
\end{tabular}
\end{table}

\section{Including a Harmonic Component as Part of the Unperturbed Lagrangian}\label{sec:harmonic_potential}

In Sec. \ref{sec:treating_harmonic_term_perturbatively}, we considered treating the harmonic component of the potential (with strength $T_{ij}$) perturbatively. Now we consider how the value of the elements of a diagram change when these harmonic terms are considered as part of $L_0$ rather than $L_1$.

\subsection{Diagram Values}\label{sec:harmonic_potential_diagram_values}
Consider a Lagrangian describing a matter wave confined inside of an anharmonic potential with a harmonic component (e.g. $\omega_z = \sqrt{T_{zz}}$) and a small cubic component $Q_{zzz}$, so that the Lagrangian can be described by

\begin{equation}
\begin{split}
L_0 &= \frac{m}{2}\left(
\dot{x}^2 + \dot{y}^2 + \dot{z}^2
\right)
-
\frac{m}{2}\left(
\omega_x^2 x^2 + \omega_y^2 y^2 + \omega_z^2 z^2
\right)
\\
L_1 &= - \frac{m}{3!} Q_{zzz} z^3
\end{split}
\end{equation}

\noindent
As shown in Appendix \ref{sec:appendix_edge_values_3D}, in the case that we absorb the harmonic component of the potential as part of the unperturbed Lagrangian, the value of the internal edges of the diagram can be expressed as

\begin{equation}\label{eq:3D_internal_edge_values}
\begin{split}
&
G_{ab}^{\alpha\alpha}[t_1, t_2] =
\frac{1}{2m\hbar}
\bigg(
i
\left(\boldsymbol{\sigma}_3\right)_{ab}
\frac{\sin[\omega_{\alpha}|t_1-t_2|]}{\omega_{\alpha}}
\\
&
\;\;\;\;\;\;\;\;\;\;\;\;\;\;\;
-
\left(\boldsymbol{\sigma}_2\right)_{ab}
\frac{\sin[\omega_{\alpha}(t_1-t_2)]}{\omega_{\alpha}}
\\
&
\;\;\;\;\;
+\left(
\boldsymbol{\sigma}_1 - \boldsymbol{1}
\right)_{ab}
\bigg(
\frac{\beta_{\alpha}\sin[\omega_{\alpha}(t_1-t_i)]\sin[\omega_{\alpha}(t_2-t_i)]}{\omega_{\alpha}^2}
\\
&
\;\;\;\;\;\;\;\;\;\;\;\;\;\;\;\;\;\;\;\;\;\;\;\;
+\frac{\cos[\omega_{\alpha}(t_1-t_i)]\cos[\omega_{\alpha}(t_2-t_i)]}{\beta_{\alpha}}
\bigg)
\bigg)
\end{split}
\end{equation}

\noindent
where $\boldsymbol{\sigma}_1$, $\boldsymbol{\sigma}_2$, and $\boldsymbol{\sigma}_3$ are the Pauli matrices and $\boldsymbol{1}$ is the identity matrix.
Note that this value for propagators $G_{ab}^{\alpha\alpha}[t_1, t_2]$ asymptote as $\omega_{\alpha} \rightarrow 0$ to the propagator value of Sec. \ref{sec:free_potential_diagram_values}, where we did not include a harmonic component of the potential in $L_0$.
The value of the vertex is the same as in Eq. \eqref{eq:diagram_component_values_1D} and the value of the external edges $G_a^{\alpha}[t]$ depends on $J_a^{\alpha}[t]$, but is fully written out in Appendix \ref{sec:appendix_edge_values} for the case of a traditional Mach Zehnder interferometer sequence (Eq. \eqref{eq:Mach-Zehnder_J}) and symmetrically kicked interferometer arms (see Eq. \eqref{eq:symmetric_J}). In the case that we have a harmonic term as part of $L_0$, $G_a^{\alpha}[t]$
has a more complicated form than just being proportional to the unperturbed classical trajectory of the arm (Eq. \eqref{eq:free_potential_external_edge_value}).
Consider an interferometer sequence where the two arms are kicked symmetrically, and the external force on the two arms which emerge in the same output port is expressed by

\begin{equation}\label{eq:symmetric_J}
\frac{J_a^3[t]}{m v_r} = 
\begin{cases} 
      2\delta[t-0] -4 \delta[t-T] + 2\delta[t-2T] & a = 1 \\
    -2\delta[t-0]+4\delta[t-T] -2\delta[t-2T] & a = 2 \\
\end{cases}
\end{equation}

\noindent
Here one arm of the interferometer ($a=1$) is kicked with $2$ recoils "upward" at time $t=0$, $4$ recoils "downward" at time $t=T$, and $2$ recoils `upward' at time $t=2T$. The other arm ($a=2$) is kicked with equal and opposite recoils to the first. We take the interferometer beam to be oriented perfectly along the $z$ axis so that no component of the laser recoils are directed in the transverse dimensions, $J_a^{1}[t] = J_a^{2}[t] = 0$. We take the initial center-of-mass (COM) position and velocity of the Gaussian wavepacket to be zero. 
In this case, under only a harmonic potential, the phase shift would be zero owing to the mirror symmetry of the potential and the symmetry with which the two arms of the interferometer are kicked. The small cubic non-linearity in $L_1$ breaks that mirror symmetry and produces a non zero phase shift.

There is an important limitation of the semiclassical approach to computing phase shifts in the presence of external potentials with cubic or higher-order spatial dependence: the computed phase shift depends on when, after the final beamsplitter, it is evaluated.
This is problematic because physical interferometer phase shifts, which are encoded in output port populations, do not change after the final beamsplitter has been applied, owing to the unitarity of time evolution.
\footnote{After the final beamsplitter, the two output ports are completely uncoupled, so that the population in each port is conserved. Any predicted change in output port populations after the beamsplitter therefore violates conservation of population. The fact that the output port populations do not change after the final beamsplitter reflects an important practical feature of matter-wave interferometers: The measured interferometer phase shift does not depend on when after the final beamsplitter these interference fringes are experimentally detected, so that timing precision requirements associated with imaging do not need to be as precise as those associated with the interferometric laser pulses.
}
A corresponding calculation of the interferometer phase shift should, in turn, be independent of the time after the final beamsplitter that it is evaluated. 
Yet in the semiclassical formalism, the two trajectories which interfere to form a single output port can have separation and propagation phases which continue to evolve after the final beamsplitter.
Let us denote the time between the final beamsplitter and the time at which the interferometer phase shift is evaluated as $\tau$.
In the absence of a cubic nonlinearity, the semiclassical formalism correctly predicts a $\tau$-independent phase shift: The $\tau$ dependence of the separation phase exactly cancels with that of the propagation phase. Under a cubic nonlinearity, however, these two contributions no longer cancel, and what results is a nonphysical, $\tau$-dependent phase shift.
While a definite phase shift prediction can be obtained by choosing to evaluate the phase shift at the time of the final beamsplitter ($\tau=0$), as we do in Fig. \ref{fig:split_step_harmonic_potential}, the semiclassical formalism has a fundamental ambiguity: the computed phase shift's dependence on cubic or higher-order nonlinearities is always not well-defined owing to its unphysical dependence on $\tau$.
The diagrammatic approach, on the other hand, correctly predicts the dependence of the phase shift on cubic nonlinearities to be independent of $\tau$.

\subsection{Validating with Numerics in 1D}\label{sec:harmonic_trap_split_step}

\begin{figure}
\includegraphics[width=3in]{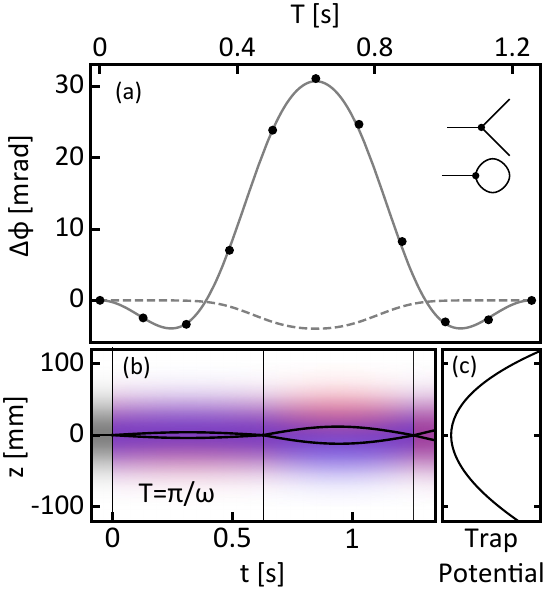}
\caption{\label{fig:split_step_harmonic_potential} Simulation of matter wave interferometer confined to an anharmonic trap in 1D. Simulation parameters (in SI units) are $\hbar/m = 10^{-2}$, $w_{0,z} = 0.0632456$, $\omega = \omega_z = 5$, $v_z = 0$, $v_r = 10^{-2}$, $Q_{zzz} = 10^{2}$,
$ \Delta z = 5\times10^{-4}$, and $\Delta t = 5\times10^{-4}$. The waist $w_{0,z}$ is chosen so that the initial wavepacket is the ground state of the harmonic potential. (a) Numerically computed phase shifts (black dots), along with the semi-classically computed phase (dashed gray line) and diagrammatically computed phase (solid gray line). The two diagrams that contribute to the leading order phase shift in $Q_{zzz}$ are displayed on the right hand side. (b) Classical trajectories of the two interferometer arms (solid black lines) along with the width of the wavepackets associated with both interferometer arms (blue and red).
The trajectories are plotted for the interrogation time for which the phase shift is maximized ($T = \pi/\omega$).
(c) The form of the anharmonic trapping potential, to contextualize the classical trajectories.}
\end{figure}

To validate the diagrammatic calculations, we demonstrate agreement between the diagrammatic and numerical evaluations of the phase shift in a parameter space where quantum corrections are enhanced beyond physically realistic values (parameters indicated in caption of Fig. \ref{fig:split_step_harmonic_potential}). Figure \ref{fig:split_step_harmonic_potential} compares the result of evaluating these phase shifts numerically (black dots), via the semi-classical formalism (dashed gray line), and via the diagrammatic formalism (solid gray line). The semi-classical solution is evaluated at the time immediately following the final beamsplitter, even though technically its value changes as a function of time after the final beamsplitter, as discussed earlier. The two three-vertex diagrams which contribute to the phase shift are included in an inset of panel (a). Panel (b) indicates the semi classical trajectories of the two interferometer arms (black lines), along with the wavepacket size (blue for the upper arm and red for the lower). Panel (c) plots the shape of the potential, which is anharmonic owing to a small cubic dependence which breaks the mirror symmetry of the system about $z=0$.

\subsection{3D Calculation of Phase Shifts in Matter Wave Interferometers Confined in Anharmonic Traps}\label{sec:anharmonic_traps}

Here we apply the analytic expressions which emerge from the diagrammatic formalism to an experimentally realistic case of a matter wave confined to a magnetic trap. \cite{Burke_2009}
While an ideal magnetic trap would be purely harmonic, it can be experimentally challenging to suppress anharmonicities in the trapping potential, and these anharmonicities give rise to quantum corrections to the interferometer phase shift.
Here we compute these corrections using parameters from a realistic trapping potential, taking the magnitude of the anharmonicity from measurements of a real magnetic trap. \cite{Moan2020ControllingTA}
We consider a trap frequency of $\omega = 2 \pi \times 11.2 \text{ Hz}$, a trap anharmonicity of $Q_{zzz} = 9.98\times10^{6} \text{m}^{-1}\text{s}^{-2}$, no initial COM velocity, a recoil velocity of $v_r \approx 5.89 \text{ mm/s}$, and an initial wavepacket waist of $w_0 = 10 \mu \text{m}$. \footnote{As described in Ref.\citenum{Beydler_2024}, an upgraded version of this apparatus has been implemented with reduced anharmonicities}
The initial wavepacket size is based on the observed condensate size in Ref. \citenum{Moan_2019}.

\begin{figure}
\includegraphics[width=3in]{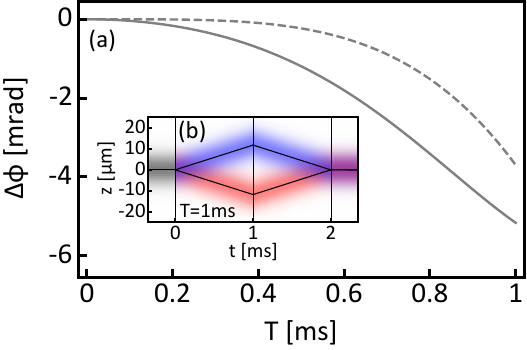}
\caption{\label{fig:real_harmonic_trap} (a) Quantum corrections for a trapped matter wave interferometer with experimentally realistic parameters, computed via the semi-classical method (dashed gray line) and the diagrammatic method (solid gray line). In the semi-classical computation, the phase shift is evaluated immediately after the final beamsplitter.
(b) The corresponding interferometer arms for a $1 \text{ ms}$ interrogation time interferometer sequence. The black lines indicate the classical trajectories and the blue and red density plots indicate the wavepacket waist of the two interferometer arms. The computed interferometer contrast drops to $\approx 73\%$ at $T=1\text{ ms}$ owing to the two arms of the interferometer imperfectly overlapping at the time of the final beamsplitter.
}
\end{figure}

Figure \ref{fig:real_harmonic_trap} shows expected phase shifts under a sequence where the two arms are kicked symmetrically.
In this scenario, the quantum corrections are on the order of the semi-classically computed phase shifts. The leading-order quantum corrections are approximately 1 mrad and might therefore be observable. This example suggests that beyond-semi-classical corrections to the interferometer phase could be important to consider for accurately modeling trapped interferometers.

The numerical value of the condensate size used in this calculation is larger than the harmonic oscillator ground state size, possibly due to repulsive mean-field interactions in the BEC. The diagrammatic calculator does not include interaction effects, which can be important for accurate phase shift predictions in this regime. \cite{thomas_modeling_2022}
In addition, when we refer to the initial ground state wavefunction size, we neglect the anharmonic corrections to the true ground state of the trap and treat the initial wavepacket as Gaussian. We assume the non-Gaussian features of the wavepacket prior to the first beamsplitter pulse are small compared to those generated during free propagation in the interferometer sequence.
Both interaction effects and non-Gaussian initial-state corrections from trap anharmonicities may contribute additional phase shift contributions that are not captured in the present calculation.

\section{\label{sec:phase_shift_from_arbitary_potential} Phase Shift For Potentials with Arbitrary Spatial Dependence}

In this section, we will consider an external potential with a more complicated dependence on $z$. We will use the diagrammatic formalism outlined in Sec. \ref{sec:phase_shifts_with_feynman_diagrams} to compute phase shifts to all orders in $z$, but only first order in $\epsilon$ by summing over all 1-vertex diagrams.

\subsection{\label{sec:calculation_of_phase_response_1D} Calculation of Phase Shift by Summing Over All 1-Vertex Diagrams in 1D}

Consider the Lagrangian

\begin{equation}
L_1 = - m \phi_g[z]
\end{equation}

\noindent
where $\phi_g$ is a gravitational potential with an arbitrary spatial dependence. We will write out the value of a diagram with an arbitrary number of external edges and loops, then sum over the number of external edges and loops. This will amount to summing over the potentially infinite number of one vertex diagrams associated with the potentially infinite number of Taylor series coefficients in the potential $\phi_g[z] = \sum_{n=0}^{\infty}\frac{1}{n!}d_n z^n$. The coefficients of the expansion of $\phi_g$ ($d_n$) are related to the coefficients of the expansion of $L_1$, ($c_n$, Eq. \eqref{eq:diagram_component_values_1D}), by $c_n = -m d_n$.
A diagram with $j$ external edges and $k$ loops will have value $\mathcal{A}[j, k]$, where

\begin{equation}
\centering
\vcenter{\hbox{\includegraphics[width=2.7in]{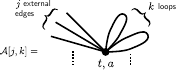}}}
\end{equation}

\noindent
We can write the value of this diagram in terms of the symmetry factor, vertex value, and edge values as $\mathcal{A}[j, k] = (SYM)(VERT)(EDGE)$ with

\begin{equation}
\begin{split}
SYM &= \frac{1}{j!k!2^k}
\\
VERT &= 
\left(-m ~ d_{j+2k}\right)\lambda_{a}^{j+2k} \int_{t_i}^{t_f} dt
\\
EDGE &= \left(G_a^3[t]\right)^j \left(G_{aa}^{33}[t, t]\right)^k
\end{split}
\end{equation}

\noindent
Using Eq. \eqref{eq:varepsilon_definition}, we can write this value as $\mathcal{A}[j, k] = \mathcal{A}_1[j, k] + \mathcal{A}_2[j, k]$ where we explicitly write out the summation over the Latin index $a$, and

\begin{equation}
\begin{split}
\mathcal{A}_a[j, k] = &
\frac{1}{j!k!2^k}
\Big(
(-1)^{a+1}\frac{\hbar}{i}
\Big)^{j + 2k - 1}
(-m \epsilon) d_{j+2k}
\\
& \times \int_{t_i}^{t_f} dt (G_{a}^3[t])^j(G_{aa}^{33}[t,t])^{k}
\end{split}
\end{equation}

\noindent
with $a = 1, 2$.
We can solve for the phase shift by summing over the number of external edges and loops. If we express the phase shift as a Taylor series in $\epsilon$, $\Delta\phi = \Delta\phi^{(0)} + \Delta\phi^{(1)}\epsilon + \mathcal{O}[\epsilon^2]$, then

\begin{equation}
\Delta\phi^{(1)} = \text{Im}\bigg[
\sum_{a=1}^{2}\sum_{j=0}^{\infty}\sum_{k=0}^{\infty}\mathcal{A}_{a}[j, k]
\bigg]
\end{equation}

\noindent
We start by evaluating the sum over the number of external edges $j$

\begin{subequations}\label{eq:sum_over_edges}
\begin{align}
\begin{split}
\sum_{j=0}^{\infty}
\!
\mathcal{A}_{a}[j, k]
\!
&=
\!
-m \epsilon \int_{t_i}^{t_f} \!\!\! dt
\frac{1}{k!2^k}
\Big(
(-1)^{a+1}\frac{\hbar}{i}
\Big)^{2k - 1}
(G_{aa}^{33}[t, t])^k
\\
&
\;\;\;\;\;\;\;\;\;\;\;\;\;\;\;\;\;\;\;\;\;
\times
\sum_{j=0}^{\infty}
\frac{1}{j!}
d_{j+2k}
\Big(
(-1)^{a+1}\frac{\hbar}{i} G_{a}^{3}[t]
\Big)^{j}
\end{split}
\\
\begin{split}
& =
\!
-m \epsilon \int_{t_i}^{t_f} \!\!\! dt
\frac{1}{k!2^k}
\Big(
(-1)^{a+1}\frac{\hbar}{i}
\Big)^{2k - 1}
(G_{aa}^{33}[t,t])^k
\\
&
\;\;\;\;\;\;\;\;\;\;\;\;\;\;\;\;\;\;\;\;\;
\times
\partial_z^{2k}
\phi_g[z]
\Big |_{z= (-1)^{a+1}\frac{\hbar}{i}G_{a}^{3}[t]}
\end{split}
\end{align}
\end{subequations}

\noindent
where to arrive at the second equality, we applied the identity
$
\sum_{j=0}^{\infty}
\frac{1}{j!}
d_{j+2k}
z^j
=
\partial_{z}^{2k}
\sum_{j=0}^{\infty}
\frac{1}{j!}
d_{j}
z^j
= 
\partial_{z}^{2k}
\phi_g[z]
$. Now performing the sum over the number of loops $k$, we have

\begin{subequations}
\begin{align}
\begin{split}
\sum_{k=0}^{\infty} & \sum_{j=0}^{\infty} \mathcal{A}_{a}[j, k]
\\
= & -m \epsilon\Big(
(-1)^{a+1}\frac{\hbar}{i}
\Big)^{-1}
\int_{t_i}^{t_f} dt
\\
&
\times
\sum_{k=0}^{\infty}
\frac{1}{k!}
\Big(
-\frac{1}{2}\hbar^2 G_{aa}^{33}[t,t]
\partial_z^2
\Big)^k
\phi_g[z]
\Big |_{z= (-1)^{a+1}\frac{\hbar}{i} G_{a}^{3}[t]}
\end{split}
\\
\begin{split}
= &
-m \epsilon
(-1)^{a+1}
\frac{i}{\hbar}
\int_{t_i}^{t_f} dt
\\
& 
\times
e^{-\frac{1}{2}\hbar^2 G_{aa}^{33}[t,t] \partial_z^{2}}
\phi_g[z]
\Big |_{z= (-1)^{a+1}\frac{\hbar}{i}G_{a}^{3}[t]}
\end{split}
\\
\begin{split}
= &
(-1)^{a+1}
\frac{-i m \epsilon}{\hbar}
\int_{t_i}^{t_f} dt
\;
\overline{\phi_g}
\Big[i \hbar(-1)^{a} G_{a}^{3}[t], t
\Big]
\end{split}
\end{align}
\end{subequations}

\noindent
with 

\begin{equation}
\begin{split}
\overline{\phi_g}[z, t] = \int_{\infty}^{\infty} dz' & \Big(
\frac{1}{-2\pi \hbar^2 G_{aa}^{33}[t, t]}
\Big)^{1/2}
\\
&\times
e^{\frac{1}{2\hbar^2 G_{aa}^{33}[t, t]}
(z-z')^2
}
\phi_g[z']
\end{split}
\end{equation}

\noindent
where to arrive at the third equality, we used the identity \cite{hirschman2012convolution}

\begin{equation}
e^{b \partial_x^2}\phi_g[x] = \int_{-\infty}^{\infty}dx'\Big(\frac{1}{4\pi b}\Big)^{1/2}e^{-\frac{1}{4b}(x-x')^2}\phi_g[x']
\end{equation}

\noindent
where $b > 0$. This expression is valid whether we include a harmonic component as part of the unperturbed potential or not. If we choose not to include a harmonic component as part of the unperturbed potential, such that 

\begin{equation}
L_0 = \frac{m}{2}\dot{z}^2 - m g z
\end{equation}

\noindent
then referencing the propagators of Sec. \ref{sec:free_potential_diagram_values}, we have

\begin{equation}
G_{11}^{33}[t,t] = G_{22}^{33}[t,t] = 
-\frac{1}{2m\hbar}\Big(\beta_3 t^2 + \frac{1}{\beta_3}\Big) 
\end{equation}

\noindent
and referencing the external edge values of Eq. \eqref{eq:free_potential_external_edge_value}, the first order correction to the phase shift can be written out as

\begin{equation}\label{eq:diagram_phase_shift_proof_mass}
\Delta \phi^{(1)} = 
-\frac{m}{\hbar}
\int_{t_i}^{t_f} dt
\;
\Big(
\overline{\phi_g}
\big[
z_{\text{cl}, 1}^{(0)}[t], t
\big]
-
\overline{\phi_g}
\big[
z_{\text{cl}, 2}^{(0)}[t], t
\big]
\Big)
\end{equation}

\noindent
where

\begin{equation}
\begin{split}
\overline{\phi_g}[z, t] = \int_{\infty}^{\infty} dz' & \Big(
\frac{1}{\pi (\hbar/m) (\beta_3 t^2 + 1/\beta_3)}
\Big)^{1/2}
\\
& \times
e^{-\frac{1}{(\hbar/m)(\beta_3 t^2 + 1/\beta_3)}
(z-z')^2
}
\phi_g[z']
\end{split}
\end{equation}

Each one-vertex diagram has pure imaginary value, resulting in no corrections to the contrast at first order in the potential $\phi_g$. For comparison, following the usual semi-classical formalism, the correction to the phase shift to first order in the potential can be written as \cite{Storey_1994}

\begin{equation}\label{eq:phase_shift_proof_mass_semiclassical}
\Delta \phi^{(1)} = 
-\frac{m}{\hbar}
\int_{t_i}^{t_f} dt
\;
\Big(
\phi_g
\big[
z_{\text{cl},1}^{(0)}[t]
\big]
-
\phi_g
\big[
z_{\text{cl},2}^{(0)}[t]
\big]
\Big)
\end{equation}

\noindent
if the unperturbed classical trajectories have the same position and velocity at the time of the final beamsplitter. 
In Sec. \ref{sec:proof_mass_split_step}, we use Eq. \eqref{eq:diagram_phase_shift_proof_mass} to compute higher order quantum corrections to matter wave interferometers under more interesting spatially dependent potentials, and compare to the results of a numerical simulation in 1D.

The 3D extension of Eq. \eqref{eq:diagram_phase_shift_proof_mass} is derived in detail in Appendix \ref{sec:appendix_summing_1_vertex_diagrams_in_3D} but is conceptually similar to the 1D result. Instead of defining $\overline{\phi}_g$ as a convolution in $z$ between a gaussian function and the classical potential $\phi_g$, the convolution is now carried out over all three spatial dimensions of the system.
As described in Sec. \ref{sec:gravity_gradients}, 
even if the semi-classical trajectories that correspond to the atom wavepacket remain along the $z$-axis, quantum corrections can still depend on the gravitational field strength in transverse directions. Therefore, when performing calculations or making estimates with realistic physical parameters (rather than the simplified parameters used for 1D simulations) it can be important to work with the fully 3D expressions.

\subsection{Agreement With Numerics in 1D}\label{sec:proof_mass_split_step}

Here we validate Eq. \eqref{eq:diagram_phase_shift_proof_mass} by numerically solving for the phase response of a matter wave interferometer to the gravitational potential from a proof mass. We use nonphysical simulation parameters to amplify quantum corrections, as done in Secs. \ref{sec:power_law_potential_split_step} and \ref{sec:harmonic_trap_split_step}. In Sec. \ref{sec:calculation_of_phase_response_3D}, we estimate the phase response using realistic experimental parameters.

\begin{figure}
\includegraphics[width=3in]{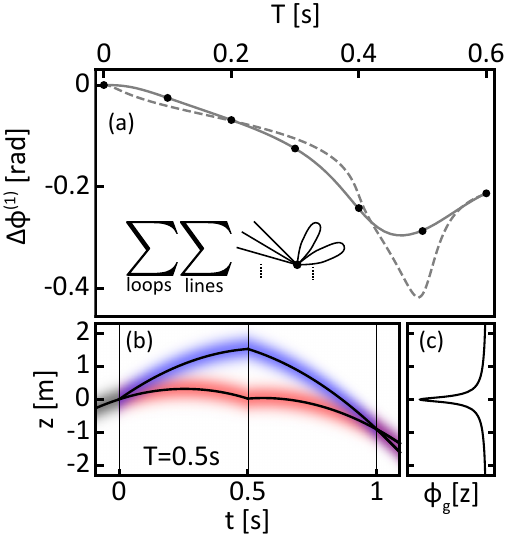}
\caption{\label{fig:proof_mass_split_step} Simulation of phase shift from proof mass potential. (a) The numerically evaluated interferometer phase as a function of the interrogation time $T$ (black points), superimposed with the phase shifts as computed via the semi-classical method (dashed gray line, Eq. \eqref{eq:phase_shift_proof_mass_semiclassical}) and the diagrammatic method (solid gray line, Eq. \eqref{eq:diagram_phase_shift_proof_mass}). The diagrammatic solution comes from evaluating the sum over an infinite number of 1-vertex diagrams, as indicated by the expression in the left hand side. (b) The two arms of the interferometer for a $T=0.5\text{ s}$ sequence. The black lines denote the classical trajectories and the red and blue density plots denote the wavepackets associated with the two interferometer arms. (c) The form of the proof mass potential (Eq. \eqref{eq:1D_proof_mass_potential}), for reference against the arm trajectories. The calculations are performed using the following parameters (in SI units): $w_0 = 0.5$, $\hbar / m = 10^{-2}$, $v_z = 2.5$, $v_r = 3$, $g = 10$, $G M = 10^{-3}$, $R = 7\times10^{-2}$, $\Delta z = 1.25\times10^{-3}$, and 
$\Delta t = 2.5\times10^{-4}$.}
\end{figure}

Consider a system where the gravitational potential $\phi_g[z]$ emerges from a 1D ring shaped proof mass with radius $R$ and total mass $M$ aligned with the interferometer axis $z$, along with a linear gravity term from the Earth, such that the potential contributed by the proof mass has the following form:

\begin{equation}\label{eq:1D_proof_mass_potential}
\phi_g[z] = - \frac{G M}{\sqrt{R^2 + z^2}}
\end{equation}

\noindent
where $G$ is Newton's gravitational constant, and the unperturbed Lagrangian does not have the harmonic component included, $L_0 = m \dot{z}^2/2 - m g z$. This potential has an infinite number of Taylor series coefficients in $z$, so computing phase shifts perturbatively in $z$ could lead to inaccurate predictions. 
In Fig. \ref{fig:proof_mass_split_step}, we display the result of computing the phase shift in an unphysical parameter space where quantum corrections are heightened (see caption for simulation parameters). Panel (a) contains the phase shift as determined by a split step wavepacket simulation (black dots), the semi-classical method (gray dashed line), and the diagrammatic method (gray solid line). The semi-classical phase shift is determined by numerically evaluating the integral over time of Eq. \eqref{eq:phase_shift_proof_mass_semiclassical}, and the diagrammatic phase shift calculation is performed by numerically evaluating the integrals over space and time of Eq. \eqref{eq:diagram_phase_shift_proof_mass}. The inset of that panel indicates the summation that is performed to arrive at the diagrammatically evaluated phase shift. Here, we plot the first order correction in $\epsilon$, which in this case is also the first order correction in Newton's gravitational constant $G$.
$\Delta\phi = \Delta\phi^{(0)} + \Delta\phi^{(1)} G + \mathcal{O}[G^2]$, where $\Delta\phi^{(0)} = -\frac{m}{\hbar}v_r g T^2$. The Mach-Zehnder pulse sequence simulated here corresponds to a source term $J_a$ that has the same form as that of Eq. \eqref{eq:Mach-Zehnder_J}.

The "smearing" effect on the phase response of the interferometer, resulting from incorporating the finite size of the matter wavepacket into the phase shift calculation, suppresses the peak response from the semi-classically computed phase. This suppression highlights how higher order quantum corrections could manifest as sources of systematic error.
Panel (b) presents the semi-classical trajectories (black lines) associated with the two interferometer arms in this parameter space for $T=0.5 \text{ s}$.
The density plot indicates the finite extent of the wavepackets of the two interferometer arms (blue and red).
Panel (c) shows the shape of the external potential $\phi_g[z]$ in relation to the classical trajectories and the size of the wavepackets.

\subsection{\label{sec:calculation_of_phase_response_3D} 3D Calculation of Response of Matter Wave Interferometer to Proof-Mass Potential}

\begin{figure}
\includegraphics[width=3.3in]{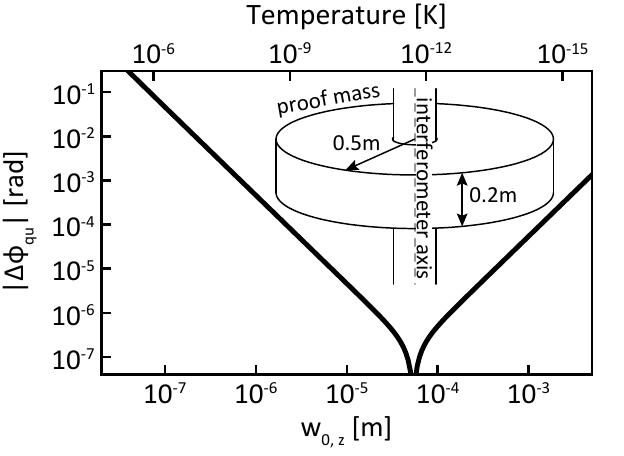}
\caption{\label{fig:real_proof_mass} Estimate of quantum corrections to interferometer phase shift to first order in the proof mass potential for a $\approx 2933 \text{ kg}$ proof mass. It is assumed that the interferometer has a 100 photon recoil momentum splitting, $T \approx 1.40 \text{ s}$, and $v_z \approx 13.35 \text{ m/s}$.
We take the interferometer to operate on the $698 \text{ nm}$ clock transition of $^{87}$Sr. The first order semi-classical contribution from just the proof masses 
is $\approx 240 \text{ rad}$.
As we scan the longitudinal initial wavepacket waist $w_{0,z}$, we keep the  initial transverse waist sizes fixed relative to it by setting $w_{0,x} = w_{0, y} = w_{0,z}/\sqrt{2}$. The temperature is defined relative to the $z$-axis. The relationship between temperature and beam waist is taken to be $w_{0,z} = \hbar / \sqrt{m k_B T_K}$, where $m$ is the mass of the strontium atom, $k_B$ is Boltzmann's constant and $T_K$ is the effective temperature.
A 0.15\;m outer diameter vacuum tube is indicated for reference. 
The inner diameter of the proof mass is $0.2$m and the position of the center of the proof mass is $0.2 \text{ m}$ below the apex of the upper interferometer arm's trajectory along the interferometer axis. The axis of the interferometer, $z$, is indicated in the dashed gray line.
}
\end{figure}

Here we estimate the scale of quantum corrections in a 3D system under realistic experimental parameters associated with the response of an interferometer in the presence of a gravitational potential from a proof mass. We consider a proof mass with nonzero 3D thickness, rather than the 1D ring proof mass of Sec. \ref{sec:proof_mass_split_step}.
Computing higher-order quantum corrections to proof mass potentials using the 3D version of Eq. \eqref{eq:diagram_phase_shift_proof_mass} (see Eq. \eqref{eq:3D_proof_mass_response_convolution}) can be challenging in a physical parameter space because the Gaussian over which proof mass potential is convolved can be narrow relative to the span of the potential along the interferometer axis. For instance, referencing the 1D calculation Eq. \eqref{eq:diagram_phase_shift_proof_mass} with $w_{0,z} = 10 \; \mu \text{m}$, $t = 2.8\text{ s}$, and $\hbar/m = 7.25\times10^{-10} \text{ m}^2/\text{s}$, the standard deviation of this Gaussian is $\sigma = \frac{1}{2}\sqrt{(w_{0,z})^2 + \frac{4\hbar^2t^2}{m^2 (w_{0,z})^2}}\approx 0.2\text{ mm}$, which is small compared to meter-scale span of the proof mass potential.
The semi-classical approximation takes the limit that the Gaussian over which the potential is convolved is a delta function (see Eqs. \eqref{eq:diagram_phase_shift_proof_mass} and \eqref{eq:phase_shift_proof_mass_semiclassical}). Quantum corrections are encoded in the small but finite width of this Gaussian.
It is computationally expensive to reliably evaluate quantum corrections by numerically evaluating the convolution integral owing to the large grid sampling required by the large span of the proof mass potential, and the fine grid resolution required by the narrow Gaussian. A more productive approach is to evaluate Eq. \eqref{eq:diagram_phase_shift_proof_mass} analytically order by order in $\sigma$, then integrate each term numerically in time.
This approach is akin to using Eq. \eqref{eq:sum_over_edges} to evaluate the corrections, where $k=0$ corresponds to the semi-classical case, and $k \geq 1$ terms contain the quantum corrections.
In this parameter space, the contributions for larger values of $k$, which encode higher-order quantum corrections, decrease as $k$ increases.
The extension of these ideas to 3D are worked out in detail in Appendix \ref{sec:details_of_3D_proof_mass_calculation}.
In Fig. \ref{fig:real_proof_mass}, the quantum correction to the interferometer phase shift is displayed as a function of the size of the initial wavepacket waist along the interferometer axis $w_{0,z}$, along with the associated atom cloud temperature along that axis, and a cartoon of the proof mass dimensions used to generate the potential.
The vertical vacuum tube, whose axis of cylindrical symmetry is co-linear with the interferometer axis (the $z$ axis), has a $15$\;cm outer diameter. The proof mass is a hollow cylinder with the density of Tungsten, an inner radius of $0.1 \text{ m}$, an external radius of $0.5 \text{ m}$, and a height of $0.2 \text{ m}$.
In Appendix \ref{sec:details_of_3D_proof_mass_calculation}, the quantum corrections to the phase shift, $\Delta\phi_{\text{qu}}$, are computed by evaluating the terms in the 3D version of Eq. \eqref{eq:sum_over_edges} (Eq. \eqref{eq:proof_mass_actual_calculation_expression}) by taking derivatives of the analytic form of the thick proof mass potential and integrating numerically in time.
For certain values of $w_{0,z}$, the scale of these quantum corrections can be $>$ mrad, which could be measurable.
This result indicates that quantum corrections to the phase shift might need to be considered for gravitational experiments with proof masses, including precision measurements of Newton's gravitational constant $G$ or searches for fifth forces.

\subsection{\label{sec:conditions_for_quantum_corrections_to_be_zero} Conditions for Quantum Corrections to Vanish to First Order in External Potential}

As shown in greater detail in Appendix \ref{sec:appendix_summing_1_vertex_diagrams_in_3D},
if the initial wavepacket has equal waists in all three spatial dimensions (i.e. it is prepared with spherical symmetry), such that $w_{0,x} = w_{0, y} = w_{0, z}$, then we have $\beta_1 = \beta_2 = \beta_3 = \beta$ and $G_{aa}^{\alpha\alpha}[t,t]$ becomes independent of the Latin and Greek indices as follows:

\begin{equation}
G_{aa}^{\alpha\alpha}[t,t] = -\frac{1}{2m\hbar}
\left(
\beta t^2 + \frac{1}{\beta}
\right)
~\forall ~ \alpha = 1,2,3, a = 1,2
\end{equation}

\noindent
Under these conditions, the value of $\Delta\phi^{(1)}$ expressed in 3D can be written as

\begin{equation}\label{eq:first_order_correction_nabla}
\begin{split}
&\Delta\phi^{(1)} = \frac{m}{\hbar} \int_{t_i}^{t_f} \!dt
\\
&
\;\;\;\;\;\;\;\;\;\;\;\;
\sum_{a = 1}^{2}
\left.
(-1)^{a}
\left(
e^{-\frac{\hbar^2}{2}
G_{aa}^{\alpha\alpha}[t,t]
\nabla^2
}
\phi_g[r_1,r_2,r_3]
\right)
\right|_{
r_{\ell} = r_{\text{cl},a}^{\ell}[t]
}
\end{split}
\end{equation}

\noindent
where $r_1 = x$, $r_2 = y$, $r_3 = z$ and 

\begin{equation}
r_{\text{cl},a}^{\ell}[t]
=
\left\{
\begin{matrix}
x_{\text{cl}, a}^{(0)} [t],  & \ell = 1
\\
y_{\text{cl}, a}^{(0)} [t],  & \ell = 2
\\
z_{\text{cl}, a}^{(0)} [t],  & \ell = 3
\end{matrix}
\right.
\end{equation}

\noindent
Taylor expanding the exponential in the integrand of Eq. \eqref{eq:first_order_correction_nabla}, the zeroth order expansion reproduces the semiclassical result the $n^{th}$ order expansion has the factor $\nabla^{2n}\phi_g$ in the integrand. If $\phi_g$ satisfies Laplace's equation $\nabla^{2}\phi_g = 0$, then $\nabla^{2n}\phi_g = 0~\forall n\geq1$, and all quantum corrections to $\Delta\phi^{(1)}$ vanish. In summary, if the initial atom wavefunction is spherically symmetric, and the external potential $\phi_g$ satisfies Laplace's equation, then the first order correction to the interferometer phase shift in $\phi_g$ will have no quantum corrections.

\section{Outlook}\label{sec:outlook}

We have introduced a new formalism for analytically calculating quantum corrections to the traditional semi-classical approximations used to analyze matter wave interferometers. A key result of this approach is a general analytic expression for the phase shift (including quantum corrections) to a matter-wave interferometer to first order in an arbitrary spatially dependent potential (Eq. \eqref{eq:diagram_phase_shift_proof_mass}). We have applied this formalism to examples involving anharmonic traps and gravitational potentials from a local proof mass, and demonstrated that these quantum corrections can be significant for realistic experimental parameters.  It could therefore be important to consider and account for their effects in future measurements.

In future work, it could be productive to investigate the application of this approach to rotating frames of reference, in which Coriolis forces emerge, leading to terms in the Lagrangian that couple position and momentum operators.
Corrections to the interferometer phase that are higher than first order in an external potential with arbitrary spatial dependence can also be evaluated by summing over the infinite number of two-vertex diagrams in the same manner as was done in Sec. \ref{sec:calculation_of_phase_response_1D}.
In this case, an arbitrary two-vertex diagram can be parametrized by five integers: the number of external edges and loops on one vertex, the number of external edges and loops on the other, and the number of internal edges between them. Then, a sum can be performed over the five parameters to arrive at an analytic expression for the phase shift at second order in the arbitrary potential. 
The diagrammatic framework could also be extended to include phase shift corrections arising from multipath interference, \cite{wang_robust_2024} such as those appearing in trapped, multi-loop interferometer gyroscopes, \cite{Krzyzanowska_2023, Hyosub_2022} rather than just the simpler two-arm configurations considered in this paper.
In addition to computing interferometer phase shifts, this formalism could also be applied to computing perturbative corrections to matter wavepackets themselves as they evolve in anharmonic potentials. Furthermore, owing to the structural similarity of the free particle Schrodinger equation and the paraxial wave equation, this formalism could prove useful for generating analytic expressions for the profiles of Gaussian laser beams propagating with small aberrations.

\begin{acknowledgments}

We dedicate this paper to Ernst Maria Rasel in celebration of his 60th birthday.  His groundbreaking work to broadly push the boundaries of atom interferometry, through advances such as pioneering atom interferometry in microgravity and inventing many new atom optics techniques, is a constant source of inspiration. The authors thank Cass Sackett, Boris Dubetsky, and Michael Kagan for valuable discussions. The authors would like to thank the anonymous reviewers for their careful reading of the manuscript and constructive feedback. This work was produced by Fermi Forward Discovery Group, LLC under Contract No. 89243024CSC000002 with the U.S. Department of Energy, Office of Science, Office of High Energy Physics. The United States Government retains and the publisher, by accepting the work for publication, acknowledges that the United States Government retains a non-exclusive, paid-up, irrevocable, world-wide license to publish or reproduce the published form of this work, or allow others to do so, for United States Government purposes. The Department of Energy will provide public access to these results of federally sponsored research in accordance with the DOE Public Access Plan (http://energy.gov/downloads/doe-public-access-plan). This work was supported by the U.S. Department of Energy, Office of Science, National Quantum Information Science Research Centers, Superconducting Quantum Materials and Systems Center (SQMS), under Contract No. 89243024CSC000002.  This work is funded in part by the Gordon and Betty Moore Foundation (Grant GBMF7945), the David and Lucile Packard Foundation (Fellowship for Science and Engineering), and the Office of Naval Research (Grant Number N00014-19-1-2181). JG acknowledges support from a National Science Foundation (NSF) Quantum Information Science and Engineering Network (QISE-NET) Graduate Fellowship, funded by NSF award No. DMR-1747426.
\end{acknowledgments}

\section*{Author Declarations}

\noindent \textbf{Conflict of Interest}

\noindent The authors have no conflicts to disclose.

\vspace{5mm}

\noindent \textbf{Author Contributions}

\vspace{5mm}

\noindent \textbf{Jonah Glick:} Conceptualization (lead), Methodology (lead), Investigation (lead), Validation (lead), Formal Analysis (lead),  Writing -- original draft (lead). Writing -- review \& editing (equal). \textbf{Tim Kovachy:} Supervision (lead), Validation (supporting), Writing -- review \& editing (equal).

\section*{Data Availability Statement}

\noindent The data that support the findings of this study are available from the corresponding author upon reasonable request.

\section*{Conflict of Interest Statement}
\noindent The authors have no conflicts to disclose.

\appendix

\begin{widetext}

\section{Calculating Diagram Vertex Values in 1D}\label{sec:appendix_vertex_values}

The appendices are organized as follows: 
Appendix \ref{sec:appendix_vertex_values} evaluates the vertex and edge values in 1D, used in Eqs. \eqref{eq:diagram_component_values_1D} and \eqref{eq:varepsilon_definition}. Appendix \ref{sec:appendix_edge_values} evaluates the external and internal edge values in 1D when including a harmonic component in the unperturbed Lagrangian $L_0$.
Appendices \ref{sec:appendix_vertex_values_3D} and \ref{sec:appendix_edge_values_3D} derive the corresponding diagram vertex and edge values in 3D, providing detailed derivations of
Eqs. \eqref{eq:diagram_component_values_3D}, \eqref{eq:Gab_free_potential}, \eqref{eq:free_potential_external_edge_value}, and \eqref{eq:3D_internal_edge_values}.
Appendix \ref{sec:appendix_summing_1_vertex_diagrams_in_3D} extends the calculation of Sec. \ref{sec:calculation_of_phase_response_1D} to 3D. Appendix \ref{sec:details_of_3D_proof_mass_calculation} provides detailed calculations for the quantum corrections to the phase shift plotted in Fig. \ref{fig:real_proof_mass}.
Appendix \ref{sec:non_gaussian_initial_conditions} outlines an approach to diagrammatically computing phase shifts for initial wavefunctions with arbitrary position dependence, rather than the Gaussian form assumed elsewhere in this paper.
Appendix \ref{sec:appendix_gaussian_integral_identities} states two Gaussian integral identities used in determining diagram component values in various sections of the paper.

Here we will validate the expression for the vertex value (Eq. \eqref{eq:varepsilon_definition}) by mirroring a method by which vertex values can be computed in QFT. We will take Taylor expansions and functional derivatives to manually compute the value of a tree level (loop-free) 1-vertex diagram, then extract the value associated with the vertex of that diagram. Consider Eq. \eqref{eq:ground_to_ground_amplitude_c}, rewritten here as

\begin{equation}\label{eq:supplement_ground_to_ground_amplitude}
\begin{split}
\langle 0 | 0 \rangle_J &= \exp\left[i\int d^4 x \; \epsilon \mathcal{L}_1\big[\frac{1}{i}\frac{\delta}{\delta J[x]}\big]\right]
Z_0[J]
\end{split}
\end{equation}

\noindent
with

\begin{equation}
Z_0[J] = e^{
\frac{i}{2}
\int d^4 x_1 \int d^4 x_2 \;
J[x_1]G[x_1, x_2]J[x_2]}
\end{equation}

\noindent
One way to determine the value of a vertex for the diagrams which compose the solution to $\langle 0 | 0 \rangle_J$ is to Taylor expand $Z_0$ and the functional in Eq. \eqref{eq:supplement_ground_to_ground_amplitude}, then apply functional derivatives to compose a first order tree level diagram manually.
Here we apply the same analysis to evaluate the vertex values for the diagrams which compose phase shifts in matter wave interferometry. We consider the expression of Eq. \eqref{eq:phase_shift_expression}, re-written here as

\begin{equation}\label{eq:supplement_phase_shift}
C e^{i \Delta \phi} = \exp\bigg[
\int_{t_i}^{t_f} dt ~
\varepsilon
\Big(
\underbrace{
\frac{i}{\hbar}
\delta_{a1}
L_1\big[\frac{\hbar}{i}\frac{\delta}{\delta J_a[t]}\big]
}_{\text{term 1}}
+ 
\underbrace{\frac{-i}{\hbar}
\delta_{a2}
L_1\big[\frac{\hbar}{-i}\frac{\delta}{\delta J_{a}[t]}\big]
\Big)
\bigg]}_{\text{term 2}}
Z_0[J]
\end{equation}

\noindent
with

\begin{equation}
Z_0[J] = e^{
\frac{1}{2}
\int_{t_i}^{t_f}dt_1
\int_{t_i}^{t_f}dt_2
\;
J_b[t_1]
G_{bc}[t_1,t_2]
J_c[t_2]
}
\end{equation}

\noindent
We first consider applying the functional derivatives of "term 1." Consider an $L_1$ of the form $L_1 = \frac{c_n}{n!}z^n$. The contribution to the tree level one-vertex diagram will come from the first order expansion in $\varepsilon$ of the first exponential in Eq. \eqref{eq:supplement_phase_shift}, and the $n^{\text{th}}$ order expansion of $Z_0[J]$:

\begin{equation}
\Bigg(
\epsilon\int_{t_i}^{t_f} dt
\frac{i}{\hbar}
\delta_{a1}
\frac{c_n}{n!} 
\bigg(\frac{\hbar}{i}\frac{\delta}{\delta J_a[t]} \bigg)^n
\Bigg)
\Bigg(
\frac{1}{n!}\bigg(
\frac{1}{2}
\int_{t_i}^{t_f} dt_1 \int_{t_i}^{t_f} dt_2
J_{b}[t_1]G_{bc}[t_1, t_2]J_{c}[t_2]
\bigg)^n
\Bigg)
\end{equation}

\noindent
Now using the identities $\frac{\delta}{\delta J_a[t]}J_b[t_1] = \delta_{ab}\delta[t-t_1]$ and $G_{ab}[t_1,t_2] = G_{ba}[t_2, t_1]$, the result of applying one functional derivative from the left hand side to the functional on the right hand side can be written using the identity

\begin{subequations}\label{eq:long_functional_derivative_example}
\begin{align}
\begin{split}
&
\frac{\delta}{\delta J_a[t]}
\left(
\int_{t_i}^{t_f} dt_1 \int_{t_i}^{t_f} dt_2
J_b[t_1]G_{bc}[t_1, t_2]J_c[t_2]
\right)^n
\\
&
=
\frac{\delta}{\delta J_a[t]}
\int_{t_i}^{t_f}\!\!\!\int_{t_i}^{t_f}
\!
...
\!
\int_{t_i}^{t_f}
d t_1 d t_2 ... d t_{2n-1} d t_{2n}
\left(J_{b_1}[t_1]G_{b_1c_1}[t_1, t_2]J_{c_1}[t_2]\right)
...
\left(J_{b_n}[t_{2n-1}]G_{b_n c_n}[t_{2n-1}, t_{2n}]J_{c_n}[t_{2n}]\right)
\end{split}
\label{eq:long_functional_derivative_example_a}
\\
\begin{split}
&
=
\int_{t_i}^{t_f}\!\!\!\int_{t_i}^{t_f}
\!
...
\!
\int_{t_i}^{t_f}
d t_1 d t_2 ... d t_{2n-1} d t_{2n}
\bigg(
\left(\delta_{a b_1}\delta[t-t_1]G_{b_1c_1}[t_1, t_2]J_{c_1}[t_2]\right)
...
\left(J_{b_n}[t_{2n-1}]G_{b_nc_n}[t_{2n-1}, t_{2n}]J_{c_n}[t_{2n}]\right)
\\
&
\;\;\;\;\;\;\;\;\;\;\;\;\;\;\;\;\;\;\;\;\;\;\;\;\;\;\;\;\;\;\;\;\;\;\;\;\;\;\;\;\;\;\;\;\;\;\;\;\;\;\;\;\;\;\;\;\;\;\;\;\;\;
+\left(J_{b_1}[t_1]G_{b_1 c_1}[t_1, t_2]\delta_{a c_1}\delta[t-t_2]\right)
...
\left(J_{b_n}[t_{2n-1}]G_{b_nc_n}[t_{2n-1}, t_{2n}]J_{c_n}[t_{2n}]\right)
\\
&
\;\;\;\;\;\;\;\;\;\;\;\;\;\;\;\;\;\;\;\;\;\;\;\;\;\;\;\;\;\;\;\;\;\;\;\;\;\;\;\;\;\;\;\;\;\;\;\;\;\;\;\;\;\;\;\;\;\;\;\;\;\;\;\;\;\;\;\;\;\;\;\;\;\;\;\;\;\;
\vdots
\\
&
\;\;\;\;\;\;\;\;\;\;\;\;\;\;\;\;\;\;\;\;\;\;\;\;\;\;\;\;\;\;\;\;\;\;\;\;\;\;\;\;\;\;\;\;\;\;\;\;\;\;\;\;\;\;\;\;\;\;\;\;\;\;
+\left(J_{b_1}[t_1]G_{b_1c_1}[t_1, t_2]J_{c_1}[t_2]\right)
...
\left(\delta_{ab_n} \delta[t - t_{2n-1}]G_{b_nc_n}[t_{2n-1}, t_{2n}]J_{c_n}[t_{2n}]\right)
\\
&
\;\;\;\;\;\;\;\;\;\;\;\;\;\;\;\;\;\;\;\;\;\;\;\;\;\;\;\;\;\;\;\;\;\;\;\;\;\;\;\;\;\;\;\;\;\;\;\;\;\;\;\;\;\;\;\;\;\;\;\;\;\;
+\left(J_{b_1}[t_1]G_{b_1c_1}[t_1, t_2]J_{c_1}[t_2]\right)
...
\left(J_{b_n}[t_{2n-1}]G_{b_nc_n}[t_{2n-1}, t_{2n}]\delta_{ac_n}\delta[t - t_{2n}]\right)
\bigg)
\end{split}
\label{eq:long_functional_derivative_example_b}
\\
\begin{split}
&
=
\int_{t_i}^{t_f}\!\!\!\int_{t_i}^{t_f}
\!
...
\!
\int_{t_i}^{t_f}
d t_2 ... d t_{2n-1} d t_{2n}
\left(\delta_{ab_1}G_{b_1 c_1}[t, t_2]J_{c_1}[t_2]\right)
...
\left(J_{b_n}[t_{2n-1}]G_{b_nc_n}[t_{2n-1}, t_{2n}]J_{c_n}[t_{2n}]\right)
\\
&
\;\;\;\;
+\int_{t_i}^{t_f}\!\!\!\int_{t_i}^{t_f}
\!
...
\!
\int_{t_i}^{t_f}
d t_1 ... d t_{2n-1} d t_{2n}
\left(J_{b_1}[t_1]G_{b_1, c_1}[t_1, t]\delta_{a c_1}\right)
...
\left(J_{b_n}[t_{2n-1}]G_{b_nc_n}[t_{2n-1}, t_{2n}]J_{c_n}[t_{2n}]\right)
\\
&
\;\;\;\;\;\;\;\;\;\;\;\;\;\;\;\;\;\;\;\;\;\;\;\;\;\;\;\;\;\;\;\;\;\;\;\;\;\;\;\;\;\;\;\;\;\;\;\;\;\;\;\;\;\;\;\;\;\;\;\;\;\;\;\;\;\;\;\;\;\;\;\;\;\;\;\;\;\;
\vdots
\\
&
\;\;\;\;
+\int_{t_i}^{t_f}\!\!\!\int_{t_i}^{t_f}
\!
...
\!
\int_{t_i}^{t_f}
d t_1 d t_2 ... d t_{2n}
\left(J_{a_1}[t_1]G_{a_1b_1}[t_1, t_2]J_{b_1}[t_2]\right)
...
\left(\delta_{a b_n}G_{b_n c_n}[t, t_{2n}]J_{c_n}[t_{2n}]\right)
\\
&
\;\;\;\;
+\int_{t_i}^{t_f}\!\!\!\int_{t_i}^{t_f}
\!
...
\!
\int_{t_i}^{t_f}
d t_1 d t_2 ... d t_{2n-1}
\left(J_{a_1}[t_1]G_{a_1b_1}[t_1, t_2]J_{b_1}[t_2]\right)
...
\left(J_{b_n}[t_{2n-1}]G_{b_n c_n}[t_{2n-1}, t]\delta_{ac_n}\right)
\end{split}
\label{eq:long_functional_derivative_example_c}
\\
\begin{split}
&
=
2n
\int_{t_i}^{t_f}\!\!\!\int_{t_i}^{t_f}
\!
...
\!
\int_{t_i}^{t_f}
d t_1 d t_2 ... d t_{2n-1} d t_{2n}
\left(J_{b_1}[t_1]G_{b_1c_1}[t_1, t_2]J_{c_1}[t_2]\right)
...
\left(J_{b_n}[t_{2n-1}]G_{b_n c_n}[t_{2n-1}, t]\delta_{a c_n}\right)
\end{split}
\label{eq:long_functional_derivative_example_d}
\\
\begin{split}
&
=
\left(
\int_{t_i}^{t_f} dt_1
\int_{t_i}^{t_f} dt_2
J_{b}[t_1]G_{bc}[t_1, t_2]J_{c}[t_2]
\right)^{n-1}
2n
\left(
\int_{t_i}^{t_f} dt_3 J_{d}[t_3]G_{de}[t_3, t]
\right)
\end{split}
\label{eq:long_functional_derivative_example_e}
\end{align}
\end{subequations}

as

\begin{equation}
\begin{split}
&
\Bigg(
\int_{t_i}^{t_f} dt
\frac{\epsilon c_n}{n!} 
\Big(\frac{\hbar}{i}\Big)^{n-1}
\delta_{a1}
\bigg(\frac{\delta}{\delta J_a[t]} \bigg)^{n-1}
\Bigg)
\\
&
\;\;\;\;\;\;\;\;\;\;\;\;\;\;\;\;\;\;\;\;\;\;\;\;\;\;\;\;
\;\;\;\;\;\;\;\;\;\;\;\;\;\;\;
\times
\Bigg(
\frac{1}{n!2^n}
\underbrace{
\bigg(
\int_{t_i}^{t_f} dt_1 \int_{t_i}^{t_f} dt_2
J_{b}[t_1]G_{bc}[t_1, t_2]J_{c}[t_2]
\bigg)^{n-1}
}_{\text{factor 1}}
\delta_{ae} 2n
\underbrace{
\bigg(
\int_{t_i}^{t_f}dt_3 J_d[t_3]G_{de}[t_3, t]
\bigg)^1
}_{\text{factor 2}}
\Bigg)
\end{split}
\end{equation}

\noindent
where in Eq. \eqref{eq:long_functional_derivative_example_a}, we write the double integral to the power of $n$ as a $2n$-dimensional integral, and in Eq. \eqref{eq:long_functional_derivative_example_b}, we apply the functional derivative to this $2n$-dimensional integral, which forms $2n$ total integrand terms by the product rule. In Eq. \eqref{eq:long_functional_derivative_example_c}, we integrate over the Dirac delta functions, and in Eq. \eqref{eq:long_functional_derivative_example_d}, we re-order the integration variables and Latin indices of each of the $2n$ terms using the property $G_{ab}[t_1,t_2] = G_{ba}[t_2, t_1]$ to identify that these $2n$ terms as identical and combine them into a single integral with a prefactor $2n$. This prefactor ultimately comes from the $2n$ $J$ functions that are acted on by $\delta/\delta J_a[t]$. In Eq. \eqref{eq:long_functional_derivative_example_e}, we re-express the $(2n-1)$-dimensional integral as a power of a double integral.
Now consider applying another functional derivative to "factor 1" (applying it to "factor 2" would result in a loop in the diagram, and we are aiming to construct just the tree level diagram). This results in

\begin{equation}
\begin{split}
&
\Bigg(
\int_{t_i}^{t_f} dt
\frac{\epsilon c_n}{n!} 
\Big(\frac{\hbar}{i}\Big)^{n-1}
\delta_{a1}
\bigg(\frac{\delta}{\delta J_a[t]} \bigg)^{n-2}
\Bigg)
\\
&
\;\;\;\;\;\;\;\;\;\;\;\;\;\;\;\;\;\;\;\;\;\;\;\;\;\;\;\;
\;\;\;\;\;\;\;\;\;\;\;\;\;\;\;
\times
\Bigg(
\frac{1}{n!2^n}
\underbrace{
\bigg(
\int_{t_i}^{t_f} dt_1 \int_{t_i}^{t_f} dt_2
J_{b}[t_1]G_{bc}[t_1, t_2]J_{c}[t_2]
\bigg)^{n-2}
}_{\text{factor 1}}
\delta_{ae} 2^2n(n-1)
\underbrace{
\bigg(
\int_{t_i}^{t_f}dt_3 J_d[t_3]G_{de}[t_3, t]
\bigg)^2
}_{\text{factor 2}}
\Bigg)
\end{split}
\end{equation}

Then, if we keep applying functional derivatives to "factor 1" until we are out of functional derivatives, we will be left with

\begin{equation}
\Bigg(
\int_{t_i}^{t_f} dt
\frac{\epsilon c_n}{n!} 
\Big(\frac{\hbar}{i}\Big)^{n-1}
\delta_{a1}
\Bigg)
\Bigg(
\frac{1}{n!2^n}
\delta_{ae} 2^nn!
\bigg(
\int_{t_i}^{t_f}dt_3 J_d[t_3]G_{de}[t_3, t]
\bigg)^n
\Bigg)
=
\frac{1}{n!}\epsilon c_n \Big(\frac{\hbar}{i}\Big)^{n-1}\delta_{a1}
\int_{t_i}^{t_f}dt G_a[t]^n
\end{equation}

\noindent
where $G_{a}[t] = \int_{t_i}^{t_f}dt' J_{b}G_{ba}[t', t]$. The symmetry factor of the diagram is $1/n!$ And the value of all the external edges are $G_a[t]^n$, so the vertex value is given by $\epsilon c_n \big(\frac{\hbar}{i}\big)^{n-1}\delta_{a1}
\int_{t_i}^{t_f}dt$. Doing the same analysis with "term 2" in Eq. \eqref{eq:supplement_phase_shift} results in a vertex value of $\epsilon c_n \big(\frac{\hbar}{-i}\big)^{n-1}\delta_{a2}
\int_{t_i}^{t_f}dt$. It happens that even beyond the first order expansion in $\epsilon$, these two separate vertex values can be combined into a single vertex whose value is a sum of the two. The rest of this section will illustrate this concept. Let us first consider these two terms (term 1 and term 2) in Eq. \eqref{eq:supplement_phase_shift} to contribute two distinct vertices. We can draw the vertex associated with term 1 with a filled-in circle, and the vertex associated with term 2 as a hollow circle as follows:

\begin{equation}
\centering
\vcenter{\hbox{\includegraphics[width=4.0in]{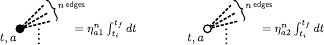}}}
\end{equation}

\noindent
where

\begin{equation}
\eta_{jk}^n = \epsilon \,
c_n
\Big(\frac{\hbar}{i (-1)^{k+1}}\Big)^{n-1} 
\delta_{jk}
\end{equation}

\noindent
Consider a diagram with $m$ vertices indexed by $i=1,2,...m$, where vertex $i$ is an $n_i$-point vertex. Such a diagram will emerge from taking the following functional derivatives of $Z_0[J]$ in Eq. \eqref{eq:supplement_phase_shift},

\begin{equation}
\begin{split}
&
\prod_{i=1}^{m}
\frac{1}{{n_i}!} \int_{t_i}^{t_f} dt_i
\Bigg(
\eta_{a_i 1}^{n_i}
\bigg(
\frac{\delta}{\delta J_{a_i}[t_i]}
\bigg)^{n_i}
+
\eta_{a_i 2}^{n_i}
\bigg(
\frac{\delta}{\delta J_{a_i}[t_i]}
\bigg)^{n_i}
\Bigg)
\\
&
\;\;\;\;\;\;\;\;\;\;\;\;\;\;\;\;\;\;
=
\frac{1}{\prod_{i=1}^{m}{n_i}!}
\bigg(\int_{t_i}^{t_f} dt_1\int_{t_i}^{t_f} dt_2 ... \int_{t_i}^{t_f} dt_m\bigg)
\sum_{b_1, b_2, ... b_m \in \{1,2\}}
\prod_{i=1}^m
\eta^{n_i}_{a_i b_i} 
\bigg(\frac{\delta}{\delta J_{a_i}[t_i]}\bigg)^{n_i}
\end{split}
\end{equation}

\noindent
where the sum is over all the $2^m$ combinations of $b_1, b_2, ..., b_m$, with each taking the values of $1$ or $2$, and vertex $i$ is labeled `$t_i, a_i$' for $i = 1,2, ..., m$. The sum comes from expanding out all the terms in the product over $i$ in the left hand side.
Consider a diagram that comes out of one of the elements in this sum, corresponding to one possible set of values for $b_1, b_2, ..., b_m$. We can express the value of this diagram as

\begin{equation}
\bigg(
\prod_{i=1}^{m}\eta_{a_ib_i}^{n_i}
\bigg)
\mathcal{E}[a_1, a_2, ..., a_m]
\end{equation}

\noindent
where included in the function $\mathcal{E}$ are the edge values, the symmetry factor of the diagram, and the integrals over time associated with the vertices.
The sum over the possible values of $b_1, b_2, ..., b_m$ can we written as 

\begin{equation}\label{eq:sum_over_hollow_and_filled_in_vertices}
\sum_{b_1, b_2, ... b_m \in \{1,2\}}
\bigg(
\prod_{i=1}^{m}\eta_{a_ib_i}^{n_i}
\bigg)
\mathcal{E}[a_1, a_2, ..., a_m]
=
\bigg(
\prod_{i=1}^{m}(\eta^{n_i}_{a_i 1}+\eta^{n_i}_{a_i 2})
\bigg)
\mathcal{E}[a_1, a_2, ..., a_m]
=
\bigg(
\prod_{i=1}^{m}\lambda^{n_i}_{a_i}
\bigg)
\mathcal{E}[a_1, a_2, ..., a_m]
\end{equation}

\noindent
where $\lambda_{a}^n = \eta_{a1}^{n} + \eta_{a2}^{n}$, so that the value of each of the $m$ vertices can be expressed as a sum of the corresponding hollow and filled-in $n_i$-point vertices.
As an example, consider a perturbative potential $L_1 = \frac{c_3}{3!}z^3 + \frac{c_4}{4!}z^4$, and a diagram which contributes to the phase shift at first order in $c_3$ and second order in $c_4$ under this potential. In this case, $m=3$, and we label vertices so that $n_1 = 3, n_2=n_3 = 4$. For notational convenience, we write $a_1 = a, a_2=b, a_3=c$. The $2^3$ different possible values of $b_1, b_2,$ and $b_3$ will produce eight diagrams, whose sum can be written as

\begin{figure}[H]
\centering
\includegraphics[width=5.875in]
{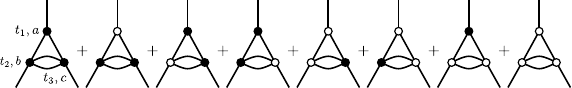}
\end{figure}

\vspace{-0.7cm}

\begin{equation}
\begin{split}
\;\;\;\;\;\;\;\;\;\;\;\;\;\;\;\;\;\;\;\;\;
&
\big(
\eta_{a1}^3\eta_{b1}^4\eta_{c1}^4
+
\eta_{a2}^3\eta_{b1}^4\eta_{c1}^4
+
\eta_{a1}^3\eta_{b2}^4\eta_{c1}^4
+
\eta_{a1}^3\eta_{b1}^4\eta_{c2}^4
+
\eta_{a2}^3\eta_{b2}^4\eta_{c1}^4
+
\eta_{a2}^3\eta_{b1}^4\eta_{c2}^4
+
\eta_{a1}^3\eta_{b2}^4\eta_{c2}^4
+
\eta_{a2}^3\eta_{b2}^4\eta_{c2}^4
\big) \mathcal{E}[a, b, c]
\\
&
\;\;\;\;\;\;\;\;\;\;\;\;\;\;\;
=
(\eta_{a1}^3+\eta_{a2}^3)
(\eta_{b1}^4+\eta_{b2}^4)
(\eta_{c1}^4+\eta_{c2}^4)
\mathcal{E}[a, b, c]
\\
&
\;\;\;\;\;\;\;\;\;\;\;\;\;\;\;
=
\lambda_{a}^{3}\lambda_{b}^{4}\lambda_{c}^{4}
\mathcal{E}[a, b, c]
\end{split}
\end{equation}

\noindent
with

\begin{equation}
\mathcal{E}[a, b, c] = \frac{1}{4}
\int_{t_i}^{t_f} dt_1
\int_{t_i}^{t_f} dt_2
\int_{t_i}^{t_f} dt_3
\big(
G_{a}[t_1]G_{b}[t_2]G_{c}[t_3]
G_{bc}[t_2, t_3]^2 G_{ba}[t_2, t_1]G_{ca}[t_3, t_1]
\big)
\end{equation}

\noindent
where the factor of $1/4$ comes from the symmetry factor of the diagram. The two terms in Eq. \eqref{eq:supplement_phase_shift} (term 1 and term 2), can therefore be associated with a single vertex, rather than two distinct ones.
The vertex value associated with the perturbative term $L_1 = \frac{c_n}{n!}z^n$ can be written as

\begin{equation}
\centering
\vcenter{\hbox{\includegraphics[width=1.75in]{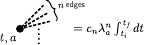}}}
\end{equation}

\noindent
where

\begin{equation}
\lambda_a^{n}= \epsilon \,
c_n \bigg(
\Big(\frac{\hbar}{i}\Big)^{n-1} 
\delta_{a1}
+
\Big(\frac{\hbar}{-i}\Big)^{n-1}
\delta_{a2}
\bigg)
\end{equation}

\noindent
as is expressed in Eq. \eqref{eq:varepsilon_definition}.

\section{Calculating Value of Diagram External and Internal Edges in 1D}\label{sec:appendix_edge_values}

Here we will evaluate in more detail the spatial integrals which are part of Eqs. \eqref{eq:integral_over_K} and \eqref{eq:phase_shift_expression}.
Consider the Lagrangian

\begin{equation}
L = \frac{1}{2}\dot{x}^2 - \frac{1}{2}m \omega^2 x^2 + J[t] x
\end{equation}

\noindent
for which a kernel solution can be written as \cite{feynman1965quantum}

\begin{equation}
\begin{split}
&
K^{(0)}[x_f, t_f ; x_i, t_i ; J[t]] =
\bigg(
\frac{m \omega}{2 \pi i \hbar \sin[\omega (t_f-t_i)]}
\bigg)^{1/2}
\\
&
\;\;\;\;\;\;\;\;\;\;\;\;\;\;\;\;\;\;\;\;
\times \exp \bigg[
\frac{i}{\hbar}
\frac{m \omega}{2 \sin[\omega(t_f-t_i)]}
\Big(
(x_f^2 + x_i^2)\cos[\omega(t_f-t_i)] - 2 x_f x_i
\\
&
\;\;\;\;\;\;\;\;\;\;\;\;\;\;\;\;\;\;\;\;\;\;\;\;\;\;\;\;\;\;\;\;\;\;\;\;\;\;\;\;\;\;\;\;\;\;\;\;\;\;\;\;\;\;\;\;\;\;\;\;\;\;\;\;
+ \frac{2 }{m \omega}\int_{t_i}^{t_f} dt J[t] (x_f \sin[\omega(t-t_i)] + x_i \sin[\omega(t_f-t)])
\\
&
\;\;\;\;\;\;\;\;\;\;\;\;\;\;\;\;\;\;\;\;\;\;\;\;\;\;\;\;\;\;\;\;\;\;\;\;\;\;\;\;\;\;\;\;\;\;\;\;\;\;\;\;\;\;\;\;\;\;\;\;\;\;\;\;
- \frac{1}{m^2 \omega^2}
\int_{t_i}^{t_f}dt_1 \int_{t_i}^{t_f}dt_2 J[t_1]J[t_2]
\begin{cases} 
      \sin[\omega(t_f-t_2)]\sin[\omega(t_1-t_i)] & , t_1 < t_2 \\
      \sin[\omega(t_f-t_1)]\sin[\omega(t_2-t_i)] & , t_1 > t_2
\end{cases}
\Big)
\bigg]
\end{split}
\end{equation}

\noindent
If we defining an initial wavefunction of the following form:

\begin{equation}
\psi_0[x_i, t_i] = \left(
\frac{2}{\pi w_0^2}
\right)^{1/4}
e^{-\frac{1}{w_0^2}x_i^2}
\end{equation}

\noindent
then the unperturbed wavefunction at a later time $t_f$ can be expressed as

\begin{equation}
\psi^{(0)}[x_f, t_f ; J[t]] = 
\int_{-\infty}^{\infty} dx_i
K^{(0)}[x_f, t_f ; x_i, t_i ; J[t]]\psi_0[x_i, t_i]
\end{equation}

\noindent
which can be solved by leveraging Eq. \eqref{eq:first_integral_identity} with

\begin{equation}
\begin{split}
A &= \left(
\frac{m \omega}{2 \pi i \hbar \sin[\omega(t_f-t_i)]}
\right)^{1/2}
\left(\frac{2}{\pi w_0^2}\right)^{1/4}
\\
F^{(0,2)} &= \frac{i m \omega \cos[\omega(t_f-t_i)]}{2\hbar\sin[\omega(t_f-t_i)]}
-\frac{1}{w_0^2}
\\
F^{(0,1)} &= \frac{-i m \omega x_f }{\hbar \sin[\omega(t_f-t_i)]}
\\
F^{(0,0)} &= \frac{i m \omega \cos[\omega(t_f-t_i)]x_f^2}{2\hbar\sin[\omega(t_f-t_i)]}
\\
F^{(1,1)}[t] &= \frac{i \sin[\omega(t_f-t)]}{\hbar \sin[\omega(t_f-t_i)]}
\\
F^{(1,0)}[t] &= \frac{i \sin[\omega(t-t_i)]x_f}{\hbar \sin[\omega(t_f-t_i)]}
\\
F^{(2,0)}[t_1, t_2] &= \frac{-i}{m \omega \hbar \sin[\omega(t_f-t_i)]}
\begin{cases} 
      \sin[\omega(t_f-t_2)]\sin[\omega(t_1-t_i)] & , t_1 < t_2 \\
      \sin[\omega(t_f-t_1)]\sin[\omega(t_2-t_i)] & , t_1 > t_2
\end{cases}
\end{split}
\end{equation}

\noindent
which results in 

\begin{equation}
\psi^{(0)}[x_f, t_f ; J[t]] = A_{\psi}
\exp\bigg[
x_f^2 F_{\psi}^{(0)}
+ x_f \int_{t_i}^{t_f}dt F_{\psi}^{(1)}[t]J[t]
+\frac{1}{2}\int_{t_i}^{t_f} dt_1 \int_{t_i}^{t_f}dt_2
J[t_1]F_{\psi}^{(2)}[t_1, t_2]J[t_2]
\bigg]
\end{equation}

\noindent
where

\[
\begin{split}
A_{\psi} &= \big(\frac{2}{\pi w_0^2}\big)^{1/4}
\big(
\frac{\omega}{\omega \cos[\omega(t_f-t_i)] + i \beta \sin[\omega(t_f-t_i)]}
\big)^{1/2}
\\
F_{\psi}^{(0)} &= -\frac{m \omega}{2\hbar}\frac{i \omega \sin[\omega(t_f-t_i)] + \beta \cos[\omega(t_f-t_i)]}{i \beta \sin[\omega(t_f-t_i)] + \omega \cos[\omega(t_f-t_i)]}
\\
F_{\psi}^{(1)}[t] &= \frac{i}{\hbar}
\Big(
\frac{\sin[\omega(t-t_i)]}{\sin[\omega(t_f-t_i)]}
+
\frac{\omega}{\omega \cos[\omega(t_f-t_i)] + i \beta \sin[\omega(t_f-t_i)]} \frac{\sin[\omega(t_f-t)]}{\sin[\omega(t_f-t_i)]}
\Big)
\\
F_{\psi}^{(2)}[t_1, t_2] &= -\frac{1}{m \hbar}
\Big(
\frac{i}{\omega \sin[\omega(t_f-t_i)]}
\begin{cases} 
      \sin[\omega(t_f-t_2)]\sin[\omega(t_1-t_i)] & , t_1 < t_2 \\
      \sin[\omega(t_f-t_1)]\sin[\omega(t_2-t_i)] & , t_1 > t_2
\end{cases}
\\
&\;\;\;\;+
\frac{1}{\beta \sin[\omega(t_f-t_i)] - i \omega \cos[\omega(t_f-t_i)]}
\frac{\sin[\omega(t_f-t_1)]\sin[\omega(t_f-t_2)]}{\sin[\omega(t_f-t_i)]}
\Big)
\end{split}
\]

\noindent
and $\beta = \frac{2\hbar}{m w_0^2}$. This is the integration done as part of Eq. \eqref{eq:integral_over_K} in the main text. Next, we compute the following integral:

\begin{equation}
\int_{-\infty}^{\infty}dx_f
\psi^{(0)}[x_f, t_f ; J_1[t]](\psi^{(0)}[x_f, t_f ; J_2[t]])^*
\end{equation}

\noindent
using Eq. \eqref{eq:second_integral_identity} with

\begin{equation}
\begin{split}
A &=  |A_{\psi}|^2 \\
F^{(0,2)} &= F_{\psi}^{(0)} + (F_{\psi}^{(0)})^{*} \\
F^{(0,1)} &= F^{(0,0)} = 0 \\
F^{(1,0)}_1[t] &= F^{(1,0)}_2[t] = 0 \\
F^{(1,1)}_1[t] &= F_{\psi}^{(1)}[t] \\
F^{(1,1)}_2[t] &= (F_{\psi}^{(1)}[t])^{*} \\
F_{11}^{(2,0)}[t_1, t_2] &= F_{\psi}^{(2)}[t_1, t_2] \\
F_{22}^{(2,0)}[t_1, t_2] &= (F_{\psi}^{(2)}[t_1, t_2])^{*}
\end{split}
\end{equation}

\noindent
which results in 

\begin{equation}\label{eq:phase_shift_1}
\begin{split}
\int_{-\infty}^{\infty}dx_f
\psi^{(0)}[x_f, t_f ; J_1[t]](\psi^{(0)}[x_f, t_f ; J_2[t]])^* = \exp\Big[
\frac{1}{2}
\int_{t_i}^{t_f} dt_1
\int_{t_i}^{t_f} dt_2
\Big(
&\;\;\;\;J_1[t_1]G_{11}[t_1, t_2]J_1[t_2]\\
&+J_1[t_1]G_{12}[t_1, t_2]J_2[t_2]\\
&+J_2[t_1]G_{21}[t_1, t_2]J_1[t_2]\\
&+J_2[t_1]G_{22}[t_1, t_2]J_2[t_2]
\Big)
\Big]
\end{split}
\end{equation}

\noindent
with

\begin{equation}\label{eq:appendix_G2_values_1D}
\begin{split}
G_{11}[t_1, t_2] &= \frac{1}{2 m \hbar}\big(
-\frac{\beta\sin[\omega(t_1-t_i)]\sin[\omega(t_2-t_i)]}{\omega^2}
-\frac{\cos[\omega(t_1-t_i)]\cos[\omega(t_2-t_i)]}{\beta}
+i\frac{\sin[\omega|t_1-t_2|]}{\omega}
\Big)
\\
G_{12}[t_1, t_2] &= \frac{1}{2 m \hbar}\big(
\frac{\beta\sin[\omega(t_1-t_i)]\sin[\omega(t_2-t_i)]}{\omega^2}
+\frac{\cos[\omega(t_1-t_i)]\cos[\omega(t_2-t_i)]}{\beta}
+i\frac{\sin[\omega(t_1-t_2)]}{\omega}
\big)
\\
G_{21}[t_1, t_2] &= \frac{1}{2 m \hbar}\big(
\frac{\beta\sin[\omega(t_1-t_i)]\sin[\omega(t_2-t_i)]}{\omega^2}
+\frac{\cos[\omega(t_1-t_i)]\cos[\omega(t_2-t_i)]}{\beta}
-i\frac{\sin[\omega(t_1-t_2)]}{\omega}
\big)
\\
G_{22}[t_1, t_2] &= \frac{1}{2 m \hbar}\big(
-\frac{\beta\sin[\omega(t_1-t_i)]\sin[\omega(t_2-t_i)]}{\omega^2}
-\frac{\cos[\omega(t_1-t_i)]\cos[\omega(t_2-t_i)]}{\beta}
-i\frac{\sin[\omega|t_1-t_2|]}{\omega}
\big)
\end{split}
\end{equation}

In Sec. \ref{sec:phase_shift_calculation_background}, we simplify the notation of Eq. \eqref{eq:phase_shift_1} by expressing it as

\begin{equation}
\int_{-\infty}^{\infty}dx_f
\psi^{(0)}[x_f, t_f ; J_1[t]](\psi^{(0)}[x_f, t_f ; J_2[t]])^* = \exp
\left[
\frac{1}{2}
\int_{t_i}^{t_f} dt_1
\int_{t_i}^{t_f} dt_2
J_a[t_1]G_{ab}[t_1,t_2]J_b[t_2]
\right]
\end{equation}

\noindent
where Latin indices can have value 1 or 2 and summation over repeated indices is assumed. This is the integration that is done as part of Eq. \eqref{eq:integral_over_psi} in the main text.
Note that performing the integration this way results in $G_{ab}[t_j, t_k] = G_{ba}[t_k, t_j]$, which is a necessary symmetry of the propagator, akin to the property $G[x_1, x_2] = G[x_2, x_1]$ in QFT.
To arrive at the expression for the propagators $G_{ab}[t_1, t_2]$ in Sec. \ref{sec:harmonic_potential}, we take $t_i = 0$, and to arrive at the propagators of Sec. \ref{sec:free_potential_diagram_values}, we take the limit that $\omega\rightarrow 0$.

Now we compute the value of the external edge $G_a[t] = \int_{t_i}^{t_f}dt' J_{b}[t']G_{b a}[t',t]$, which depends on the form of the effective force $J_a[t]$. In this paper, we consider two different pulse sequences, corresponding to two different choices of $J_a$; A Mach-Zehnder pulse sequence, as in Eq. \eqref{eq:Mach-Zehnder_J}

\begin{equation}
\frac{J_a[t]}{m} = 
-g + 
\begin{cases} 
      v_r\delta[t-0]-v_r\delta[t-T], & a = 1 \\
      v_r\delta[t-T]-v_r\delta[t-2T], & a = 2 \\
\end{cases}
\end{equation}

\noindent
and a sequence where the two arms of the interferometer are kicked symmetrically as in Eq. \eqref{eq:symmetric_J}.

\begin{equation}
\frac{J_a[t]}{m} = 
-g + 
\begin{cases} 
      2v_r\delta[t-0] -4 v_r\delta[t-T] + 2v_r\delta[t-2T] & a = 1 \\
    -2v_r\delta[t-0]+4v_r\delta[t-T] -2v_r\delta[t-2T] & a = 2 \\
\end{cases}
\end{equation}

\noindent
For the Mach Zehnder sequence, the value of the external edge $G_a[t]$ is given by

\begin{equation}\label{eq:appendix_G1_1D_Mach_Zehnder}
\begin{split}
& G_1[t] = \Big(
-\frac{2 i g}{\hbar \omega^2}
\Big)
\Big(
\sin[\omega t / 2]^2
\Big)
+
\Big(
\frac{i v_r}{2 \hbar \omega}
\Big)
\Big(
\sin[\omega |t|] + \sin[\omega(t-T)] - \sin[\omega |t-T|] - \sin[\omega(t-2T)]
\Big)
\\
&
+
\Big(
\frac{i v_z}{2\hbar \omega}
\Big)
\Big(
\sin[\omega t] + \sin[\omega |t|]
\Big)
+
\Big(
\frac{2 v_r}{\hbar \beta}
\Big)
\Big(
\cos[\omega t]\cos[\omega T]\sin[\omega T/2]^2
\Big)
+
\Big(
\frac{2v_r \beta}{\hbar \omega^2}
\Big)
\Big(
\sin[\omega t]\sin[\omega T]\sin[\omega T / 2]^2
\Big)
\\
& G_2[t] = \Big(
\frac{2 i g}{\hbar \omega^2}
\Big)
\Big(
\sin[\omega t / 2]^2
\Big)
+
\Big(
-\frac{i v_r}{2 \hbar \omega}
\Big)
\Big(
\sin[\omega t] - \sin[\omega(t-T)] + \sin[\omega |t-T|] - \sin[\omega|t-2T|]
\Big)
\\
&
+
\Big(
-\frac{i v_z}{2\hbar \omega}
\Big)
\Big(
\sin[\omega t] + \sin[\omega |t|]
\Big)
+
\Big(-
\frac{2 v_r}{\hbar \beta}
\Big)
\Big(
\cos[\omega t]\cos[\omega T]\sin[\omega T/2]^2
\Big)
+
\Big(-
\frac{2v_r \beta}{\hbar \omega^2}
\Big)
\Big(
\sin[\omega t]\sin[\omega T]\sin[\omega T / 2]^2
\Big)
\end{split}
\end{equation}

\noindent
where we have evaluated the integral in Mathematica and expressed the result using absolute value functions.
For the symmetrically kicked sequence, external edges have value

\begin{equation}\label{eq:appendix_G1_1D_symmetrically_kicked}
\begin{split}
&
G_1[t] = \Big(
-\frac{2 i g}{\hbar \omega^2}
\Big)
\Big(
\sin[\omega t / 2]^2
\Big)
\\
&
\;\;
+
\Big(
-\frac{i v_r}{\hbar \omega}
\Big)
\Big(
\sin[\omega t] - \sin[\omega |t|] - 2 \sin[\omega (t-T)] + 2 \sin[\omega |t-T|] + \sin[\omega(t-2T)] - \sin[\omega|t-2T|]
\Big)
\\
&
\;\;
+
\Big(
\frac{i v_z}{2\hbar \omega}
\Big)
\Big(
\sin[\omega t] + \sin[\omega |t|]
\Big)
+
\Big(
\frac{8 v_r}{\hbar \beta}
\Big)
\Big(
\cos[\omega t]\cos[\omega T]\sin[\omega T/2]^2
\Big)
+
\Big(
\frac{8v_r \beta}{\hbar \omega^2}
\Big)
\Big(
\sin[\omega t]\sin[\omega T]\sin[\omega T / 2]^2
\Big)
\\
&
G_2[t] = \Big(
\frac{2 i g}{\hbar \omega^2}
\Big)
\Big(
\sin[\omega t / 2]^2
\Big)
\\
&
\;\;
+
\Big(
-\frac{i v_r}{\hbar \omega}
\Big)
\Big(
\sin[\omega t] - \sin[\omega |t|] - 2 \sin[\omega (t-T)] + 2 \sin[\omega |t-T|] + \sin[\omega(t-2T)] - \sin[\omega|t-2T|]
\Big)
\\
&
\;\;
+
\Big(
-\frac{i v_z}{2\hbar \omega}
\Big)
\Big(
\sin[\omega t] + \sin[\omega |t|]
\Big)
+
\Big(
-\frac{8 v_r}{\hbar \beta}
\Big)
\Big(
\cos[\omega t]\cos[\omega T]\sin[\omega T/2]^2
\Big)
+
\Big(
-\frac{8v_r \beta}{\hbar \omega^2}
\Big)
\Big(
\sin[\omega t]\sin[\omega T]\sin[\omega T / 2]^2
\Big)
\end{split}
\end{equation}

\noindent
For both sequences, in the limit $\omega \rightarrow 0$, the edge values can be expressed in terms of the unperturbed classical trajectory, as written in Eq. \eqref{eq:free_potential_external_edge_value}.

\begin{equation}
\begin{split}
G_1[t] & = \frac{i}{\hbar} x_{\text{cl},1}^{(0)} [t] \\
G_2[t] & = -\frac{i}{\hbar} x_{\text{cl}, 2}^{(0)} [t] \\
\end{split}
\end{equation}

\noindent
where $x_{\text{cl},1}^{(0)} [t]$ is the unperturbed classical trajectory of the upper interferometer arm, and $x_{\text{cl},2}^{(0)} [t]$ is the unperturbed classical trajectory of the lower interferometer arm. For example, for the Mach Zehnder sequence of Eq. \eqref{eq:appendix_G1_1D_Mach_Zehnder}, in the limit that $\omega \rightarrow 0$, we have

\[
\begin{split}
x_{\text{cl}, 1}^{(0)}[t] 
&=
-\frac{1}{2}g t^2 + \frac{1}{2}v_r \left(
T + |t| - |t-T|
\right)
 + \frac{1}{2}v_z\left(
t + |t|
\right)
\\
&=
-\frac{1}{2}g t^2 + (v_z+v_r)\Theta[t-0](t-0)
+(-v_r)\Theta[t-T](t-T)
\\
x_{\text{cl}, 2}^{(0)}[t] 
&=-\frac{1}{2}g t^2
+\frac{1}{2}v_r
\left(
T + |t-T| - |t-2T|
\right)
+\frac{1}{2}v_z
\left(
t + |t|
\right)
\\
&=
-\frac{1}{2}g t^2
+
(v_z) \Theta[t-0](t-0)
+
(v_r)\Theta[t-T](t-T)
+
(-v_r)\Theta[t-2T](t-2T)
\end{split}
\]

\noindent
where $\Theta[t]$ is the Heaviside theta function.

\section{Calculating Diagram Vertex Values in 3D}\label{sec:appendix_vertex_values_3D}

Here we will follow the same steps as Sec. \ref{sec:appendix_vertex_values}, using the 3D rules. The 3D version of Eq. \eqref{eq:phase_shift_expression} can be expressed as

\begin{equation}\label{eq:supplement_phase_shift_3D}
\begin{split}
C e^{i \Delta \phi} = \exp\Bigg[
\int_{t_i}^{t_f} dt ~
\epsilon
\bigg(
&
\underbrace{
\frac{i}{\hbar}
\delta_{a1}
L_1\left[
\frac{\hbar}{i}\delta_{\alpha 1}\frac{\delta}{\delta J_a^{\alpha}[t]},
\frac{\hbar}{i}\delta_{\alpha 2}\frac{\delta}{\delta J_a^{\alpha}[t]},
\frac{\hbar}{i}\delta_{\alpha 3}\frac{\delta}{\delta J_a^{\alpha}[t]}
\right]
}_{\text{term 1}}
\\
&
+ 
\underbrace{\frac{-i}{\hbar}
\delta_{a2}
L_1\left[
\frac{\hbar}{-i}\delta_{\alpha 1}\frac{\delta}{\delta J_a^{\alpha}[t]},
\frac{\hbar}{-i}\delta_{\alpha 2}\frac{\delta}{\delta J_a^{\alpha}[t]},
\frac{\hbar}{-i}\delta_{\alpha 3}\frac{\delta}{\delta J_a^{\alpha}[t]}
\right]
}_{\text{term 2}}
\bigg)
\Bigg]
Z_0[J]
\end{split}
\end{equation}

\noindent
with

\begin{equation}
Z_0[J] = e^{
\frac{1}{2}
\int_{t_i}^{t_f}dt_1
\int_{t_i}^{t_f}dt_2
\;
J_b^{\beta}[t_1]
G_{bc}^{\beta \gamma}[t_1,t_2]
J_c^{\gamma}[t_2]
}
\end{equation}

\noindent
where the Greek indices run from 1 to 3, and index the three cartesian coordinates. 
Consider a perturbing Lagrangian of the following form:

\begin{equation}
L_1 = -m \phi_g[x,y,z]
\end{equation}

\noindent
The 3D Taylor series expansions of $\phi_g[x,y,z]$ and $L_1$ in position coordinates can be written in two different ways as follows:

\begin{equation}\label{eq:appendix_a_tale_of_two_taylor_series}
\begin{aligned}
\begin{split}
L_1[x,y,z] &= 
\sum_{n=0}^{\infty}
\frac{1}{n!}
\Phi_{
\alpha_1 \alpha_2 ... \alpha_n
}
r_{\alpha_1} r_{\alpha_2} ... r_{\alpha_n}
\\
&= 
\sum_{n_x=0}^{\infty}
\sum_{n_y=0}^{\infty}
\sum_{n_z=0}^{\infty}
\frac{1}{n_x!n_y!n_z!}
c_{n_x, n_y, n_z} 
x^{n_x} y^{n_y} z^{n_z}
\end{split}
\end{aligned}
\quad
\begin{aligned}
\begin{split}
\phi_g[x,y,z] &= 
\sum_{n=0}^{\infty}
\frac{1}{n!}
\Gamma^{(n-1)}_{
\alpha_1 \alpha_2 ... \alpha_n
}
r_{\alpha_1} r_{\alpha_2} ... r_{\alpha_n}
\\
&= 
\sum_{n_x=0}^{\infty}
\sum_{n_y=0}^{\infty}
\sum_{n_z=0}^{\infty}
\frac{1}{n_x!n_y!n_z!}
d_{n_x, n_y, n_z} 
x^{n_x} y^{n_y} z^{n_z}
\end{split}
\end{aligned}
\end{equation}

\noindent
where $\Phi_{\alpha_1 \alpha_2 ... \alpha_n} = -m \Gamma^{(n-1)}_{
\alpha_1 \alpha_2 ... \alpha_n
}$, and $c_{n_x, n_y, n_z} = -m d_{n_x, n_y, n_z}$, repeating Greek indices are summed, and $x = r_1, y = r_2, z = r_3$. Here we are adopting the notation of Ref. \citenum{ufrecht2020perturbative} for the coefficients $\Gamma^{(n-1)}$ of the Taylor series expansion of $\phi_g$, where $\Gamma^{(n-1)}$ is defined as the $n^\text{th}$ gravity gradient tensor.
The Taylor series coefficients $\Phi$ and $\Gamma$ are symmetric about an exchange of indices (for example, $\Gamma_{1,1,2} = \Gamma_{1,2,1} = \Gamma_{2,1,1} = d_{2, 1, 0}$, and $\Gamma_{1,2,3} = \Gamma_{1,3,2} = \Gamma_{2,1,3} = 
\Gamma_{2,3,1} = \Gamma_{3,1,2} = \Gamma_{3,2,1} = d_{1, 1, 1}$)
since the position operators associated with the different spatial dimensions commute with one another.
The first equality of Eq. \eqref{eq:appendix_a_tale_of_two_taylor_series} (where $L_1$ is expressed in terms of $\Phi$), leads to simpler expressions for the diagram vertex values than the second equality (where $L_1$ is expressed in terms of $c_{n_x, n_y, n_z}$). 
The second approach to expressing the Taylor series expansion, however, leads to a easier calculation of quantum corrections to atom interferometer phase shifts to first order in $\phi_g[x,y,z]$.
The two approaches to Taylor expanding $\phi_g[x,y,z]$ and $L_1[x,y,z]$ leads to two different approaches to assigning value to diagram elements. The first one is

\begin{equation}
\centering
\vcenter{\hbox{\includegraphics[width=3.0in]{AI_values_3D_v3.pdf}}}
\end{equation}

\noindent
which is indicated in the main text Sec. \ref{sec:phase_shifts_with_feynman_diagrams}. The alternate way to assign value to the diagrams is to label the diagram edges with the cartesian coordinate, and explicitly write out all the permutations over the cartesian coordinates as independent diagrams. In this approach, the various diagram vertices and edges values have values given by

\begin{equation}
\centering
\vcenter{\hbox{\includegraphics[width=4.0in]{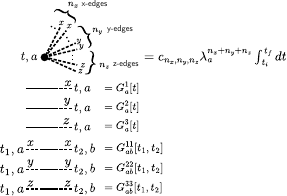}}}
\end{equation}

\noindent
where one must be careful about setting the correct symmetry factor given that the symmetry of the diagram needs to include symmetries associated with the cartesian indexes. The first approach to writing the diagram rules in 3D can be thought of as encoding the summation over all the diagrams, which have the same topology but differently labeled edges, in terms of the sum over Greek indices.
When evaluating individual diagrams in 3D, it is productive to leverage the first form of the Taylor series expansion, so that the total number of diagrams required to compute perturbative corrections is minimized, but when computing general rules for the first order corrections to an arbitrary potential, the second expansion is more computationally convenient. In this section, we will express the external potential as a Taylor series of the first form, and when computing the leading order corrections to a potential with arbitrary spatial dependence in Sec.\ref{sec:appendix_summing_1_vertex_diagrams_in_3D}, we will use the second form.
We will now derive the vertex value for the case of expressing the Taylor series of $L_1[x,y,z]$ in terms of the $\Phi$ coefficients.

Consider the following perturbation to the system Lagrangian

\begin{equation}
L_1[x,y,z] = 
\frac{1}{n!}
\Phi_{\alpha_1 \alpha_2 ... \alpha_n}
r_{\alpha_1} r_{\alpha_2} ... r_{\alpha_n}
\end{equation}

\noindent
As in the 1D case of Sec. \ref{sec:appendix_vertex_values}, the tree level 1-vertex diagram here will come from the first order expansion of the functional, and the $n^{\text{th}}$ order Taylor series expansion of $Z_0$ in Eq. \eqref{eq:supplement_phase_shift_3D}. We will first consider the contribution of just the "term 1" of the functional in Eq. \eqref{eq:supplement_phase_shift_3D}. This Taylor series coefficients can be written as

\begin{equation}
\left(
\!
\epsilon\int_{t_i}^{t_f} dt
\frac{i}{\hbar}
\delta_{a1}
\frac{1}{n!}
\Phi_{\alpha_1 \alpha_2 ... \alpha_n}
\left(
\frac{\hbar}{i}
\right)^n
\!
\frac{\delta}{\delta J_a^{\alpha_n}[t]}
...
\frac{\delta}{\delta J_a^{\alpha_2}[t]}
\frac{\delta}{\delta J_a^{\alpha_1}[t]}
\!
\right)
\!\!
\left(
\frac{1}{n!}\left(
\frac{1}{2}
\int_{t_i}^{t_f} \!\! dt_1 \int_{t_i}^{t_f} \!\! dt_2
\;
J_{b_1}^{\beta_1}[t_1]G_{b_1 b_2}^{\beta_1\beta_2}[t_1, t_2]J_{b_2}^{\beta_2}[t_2]
\right)^n
\right)
\end{equation}

\noindent
where

\begin{equation}
\frac{\delta}{\delta J_a^{\alpha}[t_2]}J_{b}^{\beta}[t_1] = \delta[t_1-t_2]\delta_{ab}\delta_{\alpha\beta}
\end{equation}

\noindent
Applying the first functional derivative (the one with respect to $J_{a}^{\alpha_1}[t]$) gives us

\begin{equation}
\begin{split}
&
\left(
\epsilon\int_{t_i}^{t_f} \!dt
~
\delta_{a1}
\frac{1}{n!}
\Phi_{\alpha_1 \alpha_2 ... \alpha_n}
\left(
\frac{\hbar}{i}
\right)^{n-1}
\frac{\delta}{\delta J_a^{\alpha_n}[t]}
...
\frac{\delta}{\delta J_a^{\alpha_2}[t]}
\right)
\\
&
\;\;\;\;\;\;\;\;\;\;\;\;\;\;\;\;\;\;\;\;
\times
\left(
\frac{1}{n!2^n}
\underbrace{
\left(
\int_{t_i}^{t_f} \!\! dt_1 \int_{t_i}^{t_f} \!\! dt_2
\;
J_{b_1}^{\beta_1}[t_1]G_{b_1 b_2}^{\beta_1\beta_2}[t_1, t_2]J_{b_2}^{\beta_2}[t_2]
\right)^{n-1}
}_{\text{factor 1}}
2n
\underbrace{
\left(
\int_{t_i}^{t_f}\!dt_3
~
J_{b_3}^{\beta_3}[t_3]
G_{b_3 a}^{\beta_3 \alpha_1}[t_3, t]
\right)
}_{\text{factor 2}}
\right)
\end{split}
\end{equation}

\noindent
Then defining

\begin{equation}
G_a^{\alpha}[t]=
\int_{t_i}^{t_f}\!dt'
~
J_{b}^{\beta}[t']
G_{b a}^{\beta \alpha}[t', t]
\end{equation}

\noindent
we can simplify the notation of "factor 2", and write the result of the application of all $n$ functional derivatives (where we only apply the functional derivatives to the terms inside of "factor 1" rather than "factor 2" so as to build only the corresponding tree-level diagram) as

\begin{equation}
\left(
\epsilon\int_{t_i}^{t_f} \!dt
~
\delta_{a1}
\frac{1}{n!}
\Phi_{\alpha_1 \alpha_2 ... \alpha_n}
\left(
\frac{\hbar}{i}
\right)^{n-1}
\right)
\left(
G_{a}^{\alpha_1}[t]
G_{a}^{\alpha_2}[t]
...
G_{a}^{\alpha_n}[t]
\right)
\end{equation}

\noindent
including the contributions of "term 2" in the functional of Eq. \eqref{eq:supplement_phase_shift_3D}, the full tree level diagram can be written as

\begin{equation}
\underbrace{\frac{1}{n!}}_{\text{SYM}}
\underbrace{
\Phi_{\alpha_1 \alpha_2 ... \alpha_n}
\lambda_{a}^{n}
\int_{t_i}^{t_f} \!dt
}_{\text{VERTEX}}
~
\underbrace{
G_{a}^{\alpha_1}[t]
G_{a}^{\alpha_2}[t]
...
G_{a}^{\alpha_n}[t]
}_{\text{EDGE}}
\end{equation}

\noindent
with

\begin{equation}
\lambda_a^{n}= \epsilon \,
\bigg(
\Big(\frac{\hbar}{i}\Big)^{n-1} 
\delta_{a1}
+
\Big(\frac{\hbar}{-i}\Big)^{n-1}
\delta_{a2}
\bigg)
\end{equation}

\section{Calculating Value of Diagram External and Internal Edges in 3D}\label{sec:appendix_edge_values_3D}

In this paper, we will make the simplifying assumption that the unperturbed Lagrangian is separable, as stated in Sec. \ref{sec:phase_shifts_with_feynman_diagrams}

\begin{equation}\label{eq:separable_3D_L0}
L_0[\dot{x},x, \dot{y},y, \dot{z}, z] = L_x[\dot{x},x] + L_y[\dot{y},y] + L_z[\dot{z}, z]
\end{equation}

\noindent
If $L_0$ contains only kinetic terms and a potential that is at most quadratic in the coordinates (including possible mixed terms such as $xy$, $xz$, or $yz$), then a linear change of coordinates can always be performed that brings $L_0$ into the separable form of Eq. \eqref{eq:separable_3D_L0}.
The analysis of Sec \ref{sec:appendix_edge_values} can be applied to each of the three cartesian dimensions individually, where the 2 point correlation functions G can be expressed as 

\begin{equation}
G_{ab}^{\alpha\beta}[t_1,t_2] =
G^{11}_{ab}[t_1,t_2]\delta_{\alpha 1}\delta_{\beta 1}
+G^{22}_{ab}[t_1,t_2]\delta_{\alpha 2}\delta_{\beta 2}
+G^{33}_{ab}[t_1,t_2]\delta_{\alpha 3}\delta_{\beta 3}
\end{equation}

\noindent
where the values of $G_{ab}^{\alpha\alpha}[t_1, t_2]$ are equal to the value of the 1D $G_{ab}[t_1, t_2]$ of Eq. \eqref{eq:appendix_G2_values_1D} but with $\beta = \beta_{\alpha}, \omega = \omega_\alpha$ to account for the fact that the trap stiffness and initial wavepacket sizes could be different in the different cartesian dimensions. To write this out explicitly, if we have an unperturbed Lagrangian of the following form:

\begin{equation}
L_0 = \frac{m}{2}\left(
\dot{x}^2 + \dot{y}^2 + \dot{z}^2
\right)
-
\frac{m}{2}\left(
\omega_x^2 x^2 + \omega_y^2 y^2 + \omega_z^2 z^2
\right)
\end{equation}

\noindent
then the values of $G_{ab}^{\alpha\alpha}[t_1, t_2]$ are given by

\begin{equation}
\begin{split}
G_{11}^{\alpha\alpha}[t_1, t_2] &= \frac{1}{2 m \hbar}\left(
-\frac{\beta_{\alpha}\sin[\omega_{\alpha}(t_1-t_i)]\sin[\omega_{\alpha}(t_2-t_i)]}{\omega_{\alpha}^2}
-\frac{\cos[\omega_{\alpha}(t_1-t_i)]\cos[\omega_{\alpha}(t_2-t_i)]}{\beta_{\alpha}}
+i\frac{\sin[\omega_{\alpha}|t_1-t_2|]}{\omega_{\alpha}}
\right)
\\
G_{12}^{\alpha\alpha}[t_1, t_2] &= \frac{1}{2 m \hbar}\left(
\frac{\beta_{\alpha}\sin[\omega_{\alpha}(t_1-t_i)]\sin[\omega_{\alpha}(t_2-t_i)]}{\omega_{\alpha}^2}
+\frac{\cos[\omega_{\alpha}(t_1-t_i)]\cos[\omega_{\alpha}(t_2-t_i)]}{\beta_{\alpha}}
+i\frac{\sin[\omega_{\alpha}(t_1-t_2)]}{\omega_{\alpha}}
\right)
\\
G_{21}^{\alpha\alpha}[t_1, t_2] &= \frac{1}{2 m \hbar}\left(
\frac{\beta_{\alpha}\sin[\omega_{\alpha}(t_1-t_i)]\sin[\omega_{\alpha}(t_2-t_i)]}{\omega_{\alpha}^2}
+\frac{\cos[\omega_{\alpha}(t_1-t_i)]\cos[\omega_{\alpha}(t_2-t_i)]}{\beta_{\alpha}}
-i\frac{\sin[\omega_{\alpha}(t_1-t_2)]}{\omega_{\alpha}}
\right)
\\
G_{22}^{\alpha\alpha}[t_1, t_2] &= \frac{1}{2 m \hbar}\left(
-\frac{\beta_{\alpha}\sin[\omega_{\alpha}(t_1-t_i)]\sin[\omega_{\alpha}(t_2-t_i)]}{\omega_{\alpha}^2}
-\frac{\cos[\omega_{\alpha}(t_1-t_i)]\cos[\omega_{\alpha}(t_2-t_i)]}{\beta_{\alpha}}
-i\frac{\sin[\omega_{\alpha}|t_1-t_2|]}{\omega_{\alpha}}
\right)
\end{split}
\end{equation}

\noindent
where $\alpha = 1,2,3$ corresponding to the three cartesian dimensions, and $\omega_x = \omega_1, \omega_y = \omega_2, \omega_z = \omega_3$. It happens that this function can be written more compactly in terms of Pauli matrices in the subspace indexed by the Latin indices.

\begin{equation}
\begin{split}
\mathbf{G}^{\alpha\alpha}[t_1, t_2] =
\frac{1}{2m\hbar}
\bigg(
&
\left(
\boldsymbol{\sigma}_1 - \boldsymbol{1}
\right)
\left(
\frac{\beta_{\alpha}\sin[\omega_{\alpha}(t_1-t_i)]\sin[\omega_{\alpha}(t_2-t_i)]}{\omega_{\alpha}^2}
+\frac{\cos[\omega_{\alpha}(t_1-t_i)]\cos[\omega_{\alpha}(t_2-t_i)]}{\beta_{\alpha}}
\right)
\\
&
\;\;\;\;\;\;
+i
\boldsymbol{\sigma}_3
\frac{\sin[\omega_{\alpha}|t_1-t_2|]}{\omega_{\alpha}}
-\boldsymbol{\sigma}_2
\frac{\sin[\omega_{\alpha}(t_1-t_2)]}{\omega_{\alpha}}
\bigg)
\end{split}
\end{equation}

\noindent
where the Pauli matrices are

\begin{equation}
\boldsymbol{\sigma}_1 = \begin{pmatrix}
0 & 1 \\
1 & 0
\end{pmatrix}
,~
\boldsymbol{\sigma}_2 = \begin{pmatrix}
0 & -i \\
i & 0
\end{pmatrix}
,~
\boldsymbol{\sigma}_3 = \begin{pmatrix}
1 & 0 \\
0 & -1
\end{pmatrix}
,~
\boldsymbol{1} = \begin{pmatrix}
1 & 0 \\
0 & 1
\end{pmatrix}
\end{equation}

\noindent
Then we have $G^{12}_{jk}[t_1, t_2] = G^{21}_{jk}[t_1, t_2] = G^{13}_{jk}[t_1, t_2] = G^{31}_{jk}[t_1, t_2] = G^{23}_{jk}[t_1, t_2] = G^{32}_{jk}[t_1, t_2] = 0$, since there are no cross terms for the position dependent coordinates in the unperturbed Lagrangian $L_0$.
The values of the $G_a^{\alpha}[t]$ can be written 

\begin{equation}
G_a^{\alpha}[t] = \int_{t_i}^{t_f} dt' J_{a'}^{\alpha'}[t']G_{a' a}^{\alpha' \alpha}[t', t]
\end{equation}

\noindent
For an interferometer beam oriented along $z$, $G_a^{3}[t]$ has the same form as the 1D expressions of Eq. \eqref{eq:appendix_G1_1D_Mach_Zehnder} for a Mach-Zehnder interferometer sequence, and of Eq. \eqref{eq:appendix_G1_1D_symmetrically_kicked} for the symmetrically kicked interferometer sequence, where $\beta = \beta_{3}$ and $\omega = \omega_{3}$. Since the interferometer beam is oriented only along $z$, we have $J_a^1[t] = J_a^2[t] = 0$ so that $G_a^{1}[t] = G_a^{2}[t] = 0$. When we include nonzero initial kinematics (either nonzero initial transverse positions $x_0$ or $y_0$ or initial transverse velocities $v_x$ or $v_y$), $G_{a}^1[t]$ and $G_{a}^2[t]$ can become nonzero. In the limit that $\omega_x \rightarrow 0$ and $\omega_y \rightarrow 0$,

\begin{equation}
\begin{split}
G_a^1[t] &= (-1)^{a+1}\frac{i}{\hbar} x_{\text{cl},a}^{(0)} [t]
\\
G_a^2[t] &= (-1)^{a+1}\frac{i}{\hbar} y_{\text{cl},a}^{(0)} [t]
\end{split}
\end{equation}

\noindent
where if the interferometer beam is oriented along the $z$-axis, then

\begin{equation}
\begin{split}
x_{\text{cl},a}^{(0)} [t] = x_0 + v_x (t-t_i)
\\
y_{\text{cl},a}^{(0)} [t] = y_0 + v_y (t-t_i)
\end{split}
\end{equation}

\section{Summing Over All One-Vertex Diagrams in 3D}\label{sec:appendix_summing_1_vertex_diagrams_in_3D}

Consider a perturbative Lagrangian of the following form:

\begin{equation}
L_1[x, y, z] = -m \phi_g[x, y, z]
\end{equation}

\noindent
where $\phi_g$ can be written in terms of its Taylor series coefficients as 

\begin{equation}
\phi_g[x, y, z] =
\sum_{n_x=0}^{\infty}
\sum_{n_y=0}^{\infty}
\sum_{n_z=0}^{\infty}
\frac{1}{n_x! n_y! n_z!} d_{n_x, n_y, n_z}
x^{n_x}y^{n_y}z^{n_z}
\end{equation}

In this paper, we will consider a gravitational external potential $\phi_g[\mathbf{r}]$, although the response to a general potential $V[\mathbf{r}]$ (originating from non-gravitational sources like inhomogeneous B fields or optical potentials) could also be computed by setting $\phi_g[x,y,z] = V[x,y,z]/m$.
We will make the same simplifying assumption that has been made throughout the paper that the unperturbed Lagrangian $L_0$ can be written separably in terms of an $x$-, $y$-, and $z$-dependent components, $L_0[x, \dot{x}, y, \dot{y}, z, \dot{z}] = L_{0,x}[x, \dot{x}] + L_{0,y}[y, \dot{y}] + L_{0,z}[z, \dot{z}]$. 
A diagram with $j_1$ x external edges, $j_2$ y external edges, $j_3$ z external edges, $k_1$ x-loops, $k_2$ y-loops, and $k_3$ z-loops has value $\mathcal{A}[j_1, j_2, j_3, k_1, k_2, k_3]$, where

\begin{equation}
\centering
\vcenter{\hbox{\includegraphics[width=5.0in]{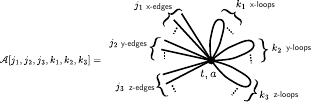}}}
\end{equation}

\noindent
The value of this diagram can be written in terms of the symmetry factor ($SYM$), the value of the vertex ($VERT$), and the value of all the edges ($EDGE$), $\mathcal{A}[j_1, j_2, j_3, k_1, k_2, k_3] = (SYM)(VERT)(EDGE)$ with

\begin{equation}
\begin{split}
SYM &= 
\frac{1}{j_1!j_2!j_3!
k_1! 2^{k_1}
k_2! 2^{k_2}
k_3! 2^{k_3}
}
\\
VERT &= 
\lambda_{a}^{n}
\left(
-
m ~
d_{j_1 + 2 k_1, j_2 + 2 k_2, j_3 + 2 k_3}
\right)
\int_{t_i}^{t_f} \!dt
\\
EDGE &= 
\left(G_a^{1}[t]\right)^{j_1}
\left(G_a^{2}[t]\right)^{j_2}
\left(G_a^{3}[t]\right)^{j_3}
\left(G_{aa}^{11}[t,t]\right)^{k_1}
\left(G_{aa}^{22}[t,t]\right)^{k_2}
\left(G_{aa}^{33}[t,t]\right)^{k_3}
\end{split}
\end{equation}

\noindent
with

\begin{equation}
\lambda_{a}^{n} = 
\epsilon
\left(
\left(\frac{\hbar}{i}\right)^{n-1}
\delta_{a1}
+
\left(\frac{\hbar}{-i}\right)^{n-1}
\delta_{a2}
\right)
\end{equation}

\noindent
and $n = j_1 + j_2 + j_3 + 2k_1 + 2k_2 + 2k_3$. 
Now let us write the summation over the Latin index $a$ explicitly as

\begin{equation}
\mathcal{A}[j_1, j_2, j_3, k_1, k_2, k_3] = 
\mathcal{A}_1[j_1, j_2, j_3, k_1, k_2, k_3]
+
\mathcal{A}_2[j_1, j_2, j_3, k_1, k_2, k_3]
\end{equation}

\noindent
with

\begin{equation}
\begin{split}
\mathcal{A}_{a}[j_1, j_2, j_3, k_1, k_2, k_3] = &
\frac{1}{j_1!j_2!j_3!
k_1! 2^{k_1}
k_2! 2^{k_2}
k_3! 2^{k_3}
}
\epsilon
\left(\frac{\hbar}{i (-1)^{a+1}}\right)^{j_1+j_2+j_3 + 2k_1 + 2k_2 + 2k_3 - 1}
\left(
-
m ~
d_{j_1 + 2 k_1, j_2 + 2 k_2, j_3 + 2 k_3}
\right)
\\
&
\;\;\;\;\;\;\;\;\;\;\;\;
\times
\int_{t_i}^{t_f} \!dt
\left(G_{a}^{1}[t]\right)^{j_1}
\left(G_{a}^{2}[t]\right)^{j_2}
\left(G_{a}^{3}[t]\right)^{j_3}
\left(G_{aa}^{11}[t,t]\right)^{k_1}
\left(G_{aa}^{22}[t,t]\right)^{k_2}
\left(G_{aa}^{33}[t,t]\right)^{k_3}
\end{split}
\end{equation}

we can write this more compactly as

\begin{subequations}
\begin{align}
\begin{split}
&\mathcal{A}_{a}[j_1, j_2, j_3, k_1, k_2, k_3]
\\
&
\;\;\;\;\;\;\;\;\;\;\;\;\;\;\;\;
=
\frac{1}{
\prod_{\ell=1}^{3}
j_{\ell}!k_{\ell}!2^{k_{\ell}}
}
\epsilon
\left(\frac{\hbar}{i (-1)^{a+1}}\right)^{
-1+
\sum_{\ell=1}^{3} j_\ell + 2k_{\ell}
}
\left(
-
m ~
d_{j_1 + 2 k_1, j_2 + 2 k_2, j_3 + 2 k_3}
\right)
\int_{t_i}^{t_f}dt
\prod_{\ell=1}^{3}
\left(G_{a}^{\ell}[t]\right)^{j_\ell}
\left(G_{aa}^{\ell\ell}[t,t]\right)^{k_\ell}
\end{split}
\\
\begin{split}
&
\;\;\;\;\;\;\;\;\;\;\;\;\;\;\;\;
=
\epsilon
\left(\frac{\hbar}{i (-1)^{a+1}}\right)^{-1}
\left(
-
m ~
d_{j_1 + 2 k_1, j_2 + 2 k_2, j_3 + 2 k_3}
\right)
\int_{t_i}^{t_f} \!dt
\prod_{\ell=1}^{3}
\frac{1}{j_{\ell}!}
\left(
\frac{\hbar}{i (-1)^{a+1}}
G_{a}^{\ell}[t]
\right)^{j_\ell}
\frac{1}{k_\ell! 2^{k_\ell}}
\left(
-\hbar^2
G_{aa}^{\ell\ell}[t,t]
\right)^{k_{\ell}}
\end{split}
\end{align}
\end{subequations}

The sum over the number of external edges can be written as

\begin{equation}
\begin{split}
&
\left(
\prod_{\ell=1}^{3}
\sum_{j_\ell=0}^{\infty}
\right)
\mathcal{A}_{a}[j_1, j_2, j_3, k_1, k_2, k_3] = 
\\
&
\;\;\;\;\;\;\;\;\;\;\;\;
\epsilon
\left(\frac{\hbar}{i (-1)^{a+1}}\right)^{-1}
\left(
-
m
\right)
\int_{t_i}^{t_f} \!dt
\left(
\prod_{\ell=1}^{3}
\frac{1}{k_\ell! 2^{k_\ell}}
\left(
-\hbar^2
G_{aa}^{\ell\ell}[t,t]
\right)^{k_\ell}
\sum_{j_\ell=0}^{\infty}
\frac{1}{j_{\ell}!}
\left(
\frac{\hbar}{i (-1)^{a+1}}
G_a^{\ell}[t]
\right)^{j_\ell}
\right)
d_{j_1 + 2 k_1, j_2 + 2 k_2, j_3 + 2 k_3}
\end{split}
\end{equation}

Then we can use the identity

\begin{equation}
\begin{split}
\left(
\prod_{\ell=1}^{3}
\sum_{j_\ell=0}^{\infty}
\frac{1}{j_{\ell}!}
\left(r_{\ell}\right)^{j_{\ell}}
\right)
d_{j_1 + 2 k_1, j_2 + 2 k_2, j_3 + 2 k_3}
& =
\left(
\prod_{\ell=1}^{3}
\partial_{\ell}^{2k_{\ell}}
\sum_{j_\ell=0}^{\infty}
\frac{1}{j_{\ell}!}
\left(r_{\ell}\right)^{j_{\ell}}
\right)
d_{j_1, j_2, j_3}
\\
& = 
\left(
\prod_{\ell=1}^{3}
\partial_{\ell}^{2k_{\ell}}
\right)
\phi_g[r_1,r_2,r_3]
\end{split}
\end{equation}

where

\begin{equation}
\begin{split}
&r_1 = x, r_2 = y, r_3 = z
\\
&\partial_1 = \partial_x,
\partial_2 = \partial_y,
\partial_3 = \partial_z
\end{split}
\end{equation}

to write

\begin{equation}
\begin{split}
&
\left(
\prod_{\ell=1}^{3}
\sum_{j_\ell=0}^{\infty}
\right)
\mathcal{A}_{a}[j_1, j_2, j_3, k_1, k_2, k_3] = 
\\
&
\;\;\;\;\;\;\;\;\;\;\;\;\;\;\;\;\;\;\;\;
\left.
\epsilon
\left(\frac{\hbar}{i (-1)^{a+1}}\right)^{-1}
\left(
-
m
\right)
\int_{t_i}^{t_f} \!dt
\left(
\prod_{\ell=1}^{3}
\frac{1}{k_\ell! 2^{k_\ell}}
\left(
-\hbar^2
G_{aa}^{\ell\ell}[t,t]
\right)^{k_\ell}
\partial_{\ell}^{2k_{\ell}}
\right)
\phi_g[r_1,r_2,r_3]
\right|_{
r_{\ell} = \frac{\hbar}{i (-1)^{a+1}}
G_{a}^{\ell}[t]
}
\end{split}
\end{equation}

now summing over the loops as well, we have

\begin{subequations}
\begin{align}
\begin{split}
&
\left(
\prod_{\ell=1}^{3}
\left(
\sum_{j_\ell=0}^{\infty}
\sum_{k_\ell=0}^{\infty}
\right)
\right)
\mathcal{A}_{a}[j_1, j_2, j_3, k_1, k_2, k_3]
\\
&
\;\;\;\;\;\;\;\;\;\;\;\;\;\;\;\;\;\;
= 
\left.
\epsilon
\left(\frac{\hbar}{i (-1)^{a+1}}\right)^{-1}
\left(
-
m
\right)
\int_{t_i}^{t_f} \!dt
\prod_{\ell=1}^{3}
\left(
\sum_{k_\ell=0}^{\infty}
\frac{1}{k_\ell!}
\left(
-\frac{\hbar^2}{2}
G_{aa}^{\ell\ell}[t,t]
\partial_{\ell}^{2}
\right)^{k_\ell}
\right)
\phi_g[r_1,r_2,r_3]
\right|_{
r_{\ell} = \frac{\hbar}{i (-1)^{a+1}}
G_{a}^{\ell}[t]
}
\end{split}
\\
\begin{split}
&
\;\;\;\;\;\;\;\;\;\;\;\;\;\;\;\;\;\;
=
\left.
\epsilon
\left(\frac{\hbar}{i (-1)^{a+1}}\right)^{-1}
\left(
-
m
\right)
\int_{t_i}^{t_f} \!dt
\left(
\prod_{\ell=1}^{3}
e^{-\frac{\hbar^2}{2}G_{aa}^{\ell\ell}[t,t]\partial_{\ell}^2}
\right)
\phi_g[r_1,r_2,r_3]
\right|_{
r_{\ell} = \frac{\hbar}{i (-1)^{a+1}}
G_\nu^{\ell}[t]
}
\end{split}
\\
\begin{split}
&
\;\;\;\;\;\;\;\;\;\;\;\;\;\;\;\;\;\;
=
\left.
\epsilon
\left(\frac{\hbar}{i (-1)^{a+1}}\right)^{-1}
\left(
-
m
\right)
\int_{t_i}^{t_f} \!dt
\left(
e^{-\frac{\hbar^2}{2}
\sum_{\ell=1}^{3}
G_{aa}^{\ell\ell}[t,t]\partial_{\ell}^2}
\right)
\phi_g[r_1,r_2,r_3]
\right|_{
r_{\ell} = \frac{\hbar}{i (-1)^{a+1}}
G_{a}^{\ell}[t]
}
\end{split}
\end{align}
\end{subequations}

\noindent
So that the leading order correction to the interferometer phase shift from this arbitrary gravitational potential can be written as

\begin{equation}\label{eq:3D_proof_mass_first_order_correction}
\Delta\phi^{(1)} = 
\text{Im}\left[
\sum_{a = 1}^{2}
\left.
\left(\frac{\hbar}{i (-1)^{a+1}}\right)^{-1}
\left(
-
m
\right)
\int_{t_i}^{t_f} \!dt
\left(
e^{-\frac{\hbar^2}{2}
\sum_{\ell=1}^{3}
G_{aa}^{\ell\ell}[t,t]\partial_{\ell}^2}
\right)
\phi_g[r_1,r_2,r_3]
\right|_{
r_{\ell} = \frac{\hbar}{i (-1)^{a+1}}
G_{a}^{\ell}[t]
}
\right]
\end{equation}

\noindent
There are two ways to proceed with the evaluation of Eq. \eqref{eq:3D_proof_mass_first_order_correction}. One is to express the result as a convolution integral in the $x$, $y$, and $z$ coordinates, and the second is to Taylor expand the exponential inside of the integrand and compute the time integral order by order in that expansion by taking higher and higher order derivatives of the potential $\phi_g[r_1, r_2, r_3]$. We will find that when computing the response to proof mass potentials, this second approach will be more productive.
We can express Eq. \eqref{eq:3D_proof_mass_first_order_correction} as a convolution integral as

\begin{equation}
\Delta\phi^{(1)} = 
\text{Im}
\left[
-
i
\frac{m}{\hbar}
\int_{t_i}^{t_f} \!dt
~
\left(
\overline{\phi}_g
\left[
\frac{\hbar}{i}
G_1^{1}[t]
,
\frac{\hbar}{i}
G_1^{2}[t]
,
\frac{\hbar}{i}
G_1^{3}[t]
\right]
- 
\overline{\phi}_g
\left[
\frac{\hbar}{-i}
G_2^{1}[t]
,
\frac{\hbar}{-i}
G_2^{2}[t]
,
\frac{\hbar}{-i}
G_2^{3}[t]
\right]
\right)
\right]
\end{equation}

\noindent
where

\begin{equation}
\begin{split}
\overline{\phi}_g[x,y,z] 
=
&
\int_{-\infty}^{\infty} \!\!\!\!\!\!
dx'
\int_{-\infty}^{\infty} \!\!\!\!\!\!
dy'
\int_{-\infty}^{\infty} \!\!\!\!\!
dz'
\left(
\frac{1}{4\pi \left(
-\frac{\hbar^2}{2}G_{aa}^{11}[t,t]
\right) }
\right)^{1/2}
\left(
\frac{1}{4\pi \left(
-\frac{\hbar^2}{2}G_{aa}^{22}[t,t]
\right) }
\right)^{1/2}
\left(
\frac{1}{4\pi \left(
-\frac{\hbar^2}{2}G_{aa}^{33}[t,t]
\right) }
\right)^{1/2}
\\
&
\;\;\;\;\;\;\;\;\;\;\;\;\;\;\;\;\;\;\;\;\;\;\;\;\;\;\;\;\;\;\;\;\;\;\;\;\;\;\;\;\;\;
\times
e^{-\frac{
\left(x - x'\right)^2
}{4\left(
-\frac{\hbar^2}{2}G_{aa}^{11}[t,t]
\right)}}
e^{-\frac{
\left(y - y'\right)^2
}{4\left(
-\frac{\hbar^2}{2}G_{aa}^{22}[t,t]
\right)}}
e^{-\frac{
\left(z - z'\right)^2
}{4\left(
-\frac{\hbar^2}{2}G_{aa}^{33}[t,t]
\right)}}
\phi_g[x', y', z']
\end{split}
\end{equation}

\noindent
where, including the harmonic components as part of the unperturbed Lagrangian $L_0$, we have

\begin{equation}
G_{11}^{\ell\ell}[t,t] = G_{22}^{\ell\ell}[t,t] = 
-
\frac{1}{2m\hbar}
\left(
\frac{1}{
\beta_{\ell}}
\cos\left[\omega_{\ell}\left(t-t_i\right)\right]^2
+
\frac{\beta_{\ell}}{\omega_{\ell}^2}
\sin\left[\omega_{\ell}\left(t-t_i\right)\right]^2
\right)
\end{equation}

\noindent
where $\beta_{\ell} = \frac{2\hbar}{m (w_{0,\ell})^2}$. When the initial wavepacket waist is the ground state of the harmonic oscillator, we have

\begin{equation}
\begin{split}
(w_{0,\ell})^2 &= \frac{2 \hbar}{m \omega_{\ell}}
\\
\beta_{\ell} &= \omega_{\ell}
\end{split}
\end{equation}

\noindent
in which case the functions $G_{aa}^{\ell\ell}[t,t]$ become time-independent

\begin{equation}
G_{11}^{\ell\ell}[t,t] = G_{22}^{\ell\ell}[t,t] = 
-
\frac{1}{2m\hbar}
\frac{1}{\omega_{\ell}}
\end{equation}

\noindent
When the harmonic components are not included as part of $L_0$, the $G_a^{\alpha}[t]$ functions take the form of

\begin{equation}
\begin{aligned}
G_a^1[t] & = (-1)^{a+1}\frac{i}{\hbar} x_{\text{cl}, a}^{(0)} [t]
\end{aligned}
\quad
\begin{aligned}
G_a^2[t] & = (-1)^{a+1}\frac{i}{\hbar} y_{\text{cl}, a}^{(0)} [t]
\end{aligned}
\quad
\begin{aligned}
G_a^3[t] & = (-1)^{a+1}\frac{i}{\hbar} z_{\text{cl}, a}^{(0)} [t]
\end{aligned}
\end{equation}

and the $G_{aa}^{\alpha \alpha}[t, t]$ functions have the following form:

\begin{equation}
G_{11}^{\alpha \alpha}[t_1, t_2] = G_{22}^{\alpha \alpha}[t_1, t_2] = 
-\frac{1}{2m\hbar}
\left(
\beta_{\alpha} t^2 + \frac{1}{\beta_{\alpha}}
\right)
\end{equation}

\noindent
in which case, the first order correction to the interferometer phase can be expressed as

\begin{equation}\label{eq:3D_proof_mass_response_convolution}
\Delta\phi^{(1)} = 
-
\frac{m}{\hbar}
\int_{t_i}^{t_f} \!dt
~
\left(
\overline{\phi}_g
\left[
x_{\text{cl},1}^{(0)} [t]
,
y_{\text{cl},1}^{(0)} [t]
,
z_{\text{cl},1}^{(0)} [t]
\right]
- 
\overline{\phi}_g
\left[
x_{\text{cl},2}^{(0)} [t]
,
y_{\text{cl},2}^{(0)} [t]
,
z_{\text{cl},2}^{(0)} [t]
\right]
\right)
\end{equation}

\noindent
with

\begin{equation}
\begin{split}
\overline{\phi}_g[x,y,z] 
=
&
\int_{-\infty}^{\infty} \!\!\!\!\!\!
dx'
\int_{-\infty}^{\infty} \!\!\!\!\!\!
dy'
\int_{-\infty}^{\infty} \!\!\!\!\!
dz'
\left(
\frac{1}{\pi \frac{\hbar}{m} \left(\beta_{1} t^2 + 1/\beta_1
\right) }
\right)^{1/2}
\left(
\frac{1}{\pi \frac{\hbar}{m} \left(\beta_{2} t^2 + 1/\beta_2
\right) }
\right)^{1/2}
\left(
\frac{1}{\pi \frac{\hbar}{m} \left(\beta_{3} t^2 + 1/\beta_3
\right) }
\right)^{1/2}
\\
&
\;\;\;\;\;\;\;\;\;\;\;\;\;\;\;\;\;\;\;\;\;\;\;\;\;\;\;\;\;\;\;\;\;\;\;\;\;\;\;\;\;\;
\times
e^{-\frac{
\left(x - x'\right)^2
}{\frac{\hbar}{m} \left(\beta_{1} t^2 + 1/\beta_1
\right)}}
e^{-\frac{
\left(y - y'\right)^2
}{\frac{\hbar}{m} \left(\beta_{2} t^2 + 1/\beta_2
\right)}}
e^{-\frac{
\left(z - z'\right)^2
}{\frac{\hbar}{m} \left(\beta_{3} t^2 + 1/\beta_3
\right)}}
\phi_g[x', y', z']
\end{split}
\end{equation}

The alternative way of computing the right hand side of Eq. \eqref{eq:3D_proof_mass_first_order_correction} is to Taylor expand the exponential in the integrand. This is the approach we will take to making predictions for quantum corrections to proof mass potentials (see Sec. \ref{sec:details_of_3D_proof_mass_calculation}), but this approach is also useful for seeing the fact that when the wavepacket is prepared with spherical symmetry, the quantum corrections vanish if the external potential satisfies Laplace's equation.
 Consider a system for which

\begin{equation}
G_{aa}^{11}[t,t] = G_{aa}^{22}[t,t] = G_{aa}^{33}[t,t]
\end{equation}

\noindent
Such a system could be one in which the initial wavefunction is spherically symmetric ($w_{0,x} = w_{0,y} = w_{0,z}$).
Let us also consider the case that the harmonic term is not included as part of the unperturbed potential.
In such a case, the integrand on the right hand side of Eq. \eqref{eq:3D_proof_mass_first_order_correction} can be written as

\begin{subequations}
\begin{align}
\begin{split}
&
\left.
\left(
e^{-\frac{\hbar^2}{2}
\sum_{\ell=1}^{3}
G_{aa}^{\ell\ell}[t,t]\partial_{\ell}^2}
\right)
\phi_g[r_1,r_2,r_3]
\right|_{
r_{\ell} = \frac{\hbar}{i (-1)^{a+1}}
G_{a}^{\ell}[t]
}
=
\left.
\left(
e^{-\frac{\hbar^2}{2}
G_{aa}^{11}[t,t]
\sum_{\ell=1}^{3}\partial_{\ell}^2}
\right)
\phi_g[r_1,r_2,r_3]
\right|_{
r_{\ell} = \frac{\hbar}{i (-1)^{a+1}}
G_{a}^{\ell}[t]
}
\end{split}
\\
\begin{split}
&
\;\;\;\;\;\;\;\;\;\;\;\;\;\;\;\;\;\;\;\;\;\;\;\;\;\;\;\;\;\;\;\;\;\;\;\;\;\;\;\;\;\;\;\;\;\;\;\;\;\;\;\;\;\;\;\;\;\;\;\;\;\;\;\;\;\;\;\;\;\;\;\;\;\;\;\;\;\;
=
\left.
\left(
e^{-\frac{\hbar^2}{2}
G_{aa}^{11}[t,t]
\nabla^2
}
\right)
\phi_g[r_1,r_2,r_3]
\right|_{
r_{\ell} = \frac{\hbar}{i (-1)^{a+1}}
G_{a}^{\ell}[t]
}
\end{split}
\\
\begin{split}
&
\;\;\;\;\;\;\;\;\;\;\;\;\;\;\;\;
\;\;\;\;\;\;\;
=
\left.
\left(
1 
-
\frac{\hbar^2}{2}
G_{aa}^{11}[t,t]
\nabla^2
+
\frac{1}{2}
\left(-
\frac{\hbar^2}{2}
G_{aa}^{11}[t,t]
\nabla^2\right)^2
+ ...
\right)
\phi_g[r_1,r_2,r_3]
\right|_{
r_{\ell} = \frac{\hbar}{i (-1)^{a+1}}
G_{a}^{\ell}[t]
}
\end{split}
\end{align}
\end{subequations}

\noindent
where in the last line, we Taylor expand the exponentiated Laplace operator $\nabla^2$.
The zeroth-order term yields the semiclassical approximation, and the higher-order terms (proportional to increasing powers of $\hbar$) represent increasingly higher-order quantum corrections.
When $\nabla^2 \phi_g = 0$ (i.e. when the external potential satisfies Laplace's equation, as is the case for gravitational potentials in regions free of mass) every term beyond zeroth order in this expansion vanishes. Therefore, to first order in $\phi_g$, there are no quantum corrections to the interferometer phase shift.

\section{Details of Calculation of Response to Proof Mass Potential in 3D}\label{sec:details_of_3D_proof_mass_calculation}

It is difficult to express the solution $\phi_g[x,y,z]$ to Poisson's equation for the "hollow cylinder" proof mass geometry indicated in Sec. \ref{sec:calculation_of_phase_response_3D}, in closed form for all spatial dimensions $x$, $y$, and $z$. Solving just along $z$, through ($\phi_g[0, 0, z]$), is more straightforward. The general solution to Poisson's equation for a gravitational potential is given by

\begin{equation}
\phi_g[\mathbf{r}] =
- G
\int d^3 \mathbf{r}' \frac{\rho[\mathbf{r}']}{|\mathbf{r} - \mathbf{r}'|}
\end{equation}

\noindent
where $\rho[\mathbf{r}]$ is the density of the proof mass and $G$ is Newton's gravitational constant. Expressed in cylindrical coordinates, for a hollow cylinder proof mass potential with perfectly homogeneous mass density $\rho_g$, we have

\begin{equation}
\phi_g[0, 0, z] =
- G
\int_{0}^{2\pi}d\phi'
\int_{a}^{b}ds'
\int_{-L}^{L}dz'
s'
\frac{\rho_g}{\sqrt{(s')^2 + (z-z')^2}}
\end{equation}

\noindent
where $z$ is the axial coordinate, $s$ is the radial coordinate, $\phi$ is the azimuthal coordinate, and the origin of this coordinate system is the center-of-mass of the proof mass.
The cylinder length is $2L$, the inner radius is $a$, and the outer radius is $b$, and the axis of cylindrical symmetry of the cylinder is collinear with the interferometer axis $z$. Evaluating these integrals gives us

\begin{equation}\label{eq:3D_proof_mass_potential_along_z}
\begin{split}
\phi_g[0,0,z] = & 
-G \pi \rho_g \bigg(
\left(L - z\right)
\left(
\sqrt{b^2 + (L - z)^2} - \sqrt{
a^2 + (L - z)^2}
\right)
+ \left(L + z\right)
\left(
\sqrt{b^2 + (L + z)^2} - \sqrt{a^2 + (L + z)^2}
\right)
\\
&
+ b^2 \left(
\log\left[(z + L) + \sqrt{b^2 + (L + z)^2}\right]
-
\log\left[(z - L) + \sqrt{b^2 + (L - z)^2}\right]
\right)
\\
&
- 
a^2 \left(
\log\left[(z + L) + \sqrt{a^2 + (L + z)^2}\right] - 
\log\left[(z - L) + \sqrt{a^2 + (L - z)^2}\right]
\right)
\bigg)
\end{split}
\end{equation}

\noindent
Assuming the gravitational potential is even in the radial coordinate $r^2 = x^2 + y^2$, we can write the general solution to $\phi_g[x,y,z]$ perturbatively in $r$ as

\begin{equation}\label{eq:expansion_of_phig_in_r}
\phi_g[x,y,z] = \phi_g^{(0)}[z] + \phi_g^{(2)}[z](x^2+y^2)
+\phi_g^{(4)}[z](x^2+y^2)^2
+\phi_g^{(6)}[z](x^2+y^2)^3
+ ...
\end{equation}

\noindent
where the $\phi_g^{(n)}[z]$ terms are coefficients, with $\phi_g^{(0)}[z] \equiv \phi_g[0,0,z]$. 
It happens that in order for $\phi_g$ to satisfy the Laplace equation $\nabla^2\phi_g = 0$ (for the mass-free regions), the functions $\phi^{(n)}[z]$ must satisfy

\begin{equation}
\phi_g^{(n)}[z] = -\frac{1}{n^2}\partial_z^2\phi_g^{(n-2)}[z]
\end{equation}

\noindent
So that given our solution to $\phi_g^{(0)}[z]$ (Eq. \eqref{eq:3D_proof_mass_potential_along_z}), we can build up solutions to $\phi_g^{(n)}[z]$ up from $n=0$. Solving the recursive relationship above in order to express $\phi_g^{(n)}[z]$ in terms of $\phi_g^{(0)}[z]$, we have

\begin{equation}\label{eq:supplement_coefficients_of_phi_g}
\phi_g^{(2m)}[z] = (-1)^m\frac{1}{4^m(m!)^2}\partial_z^{2m}\phi_g^{(0)}[z]
\end{equation}

\noindent
Looking at the expansion of $\phi_g[x,y,z]$ in Eq. \eqref{eq:expansion_of_phig_in_r}, we can isolate the value of $\phi_g^{(n)}[z]$ for constant $n$ by taking derivatives in the radial coordinate and setting the radial coordinate to zero.

\begin{equation}
(\partial_{\perp}^2)^{n_{\perp}}\phi_g[x,y,z]\Big|_{x=y=0} = 4^{n_\perp}(n_{\perp}!)^2 \phi_g^{(2n_{\perp})}[z]
\end{equation}

\noindent
where $\partial_{\perp}^2 = \partial_x^2 + \partial_y^2$. We can express the right hand side in terms of $\phi_g^{(0)}[z]$ using Eq. \eqref{eq:supplement_coefficients_of_phi_g} as

\begin{equation}\label{eq:3D_potential_radial_coefficients_in_terms_of_phi0g}
(\partial_{\perp}^2)^{n_{\perp}}\phi_g[x,y,z]\Big|_{x=y=0} = 
(-1)^{n_{\perp}}
\partial_z^{2n_{\perp}}\phi_g^{(0)}[z]
\end{equation}

\noindent
Now consider a case where the unperturbed Lagrangian does not include harmonic terms, and that the semiclassical trajectories associated with the wavefunctions of both arms stay along the interferometer ($z$) axis, such that the initial kinematics about the transverse dimensions are zero ($x_0 = y_0 = v_x = v_y = 0$). Let us also make the assumption that the initial wavepacket waist about the transverse coordinates are equal to one another. In this case, we have

\begin{equation}
\begin{split}
w_{0,x} = w_{0,y} &= w_{0, \perp}
\\
\beta_x = \beta_y = \beta_{\perp} &= \frac{2\hbar}{m (w_{0, \perp})^2}
\\
G_{ab}^{11}[t_1,t_2] = G_{ab}^{22}[t_1,t_2] &= G_{ab}^{\perp}[t_1,t_2]
\\
G_{11}^{\perp}[t,t] = G_{22}^{\perp}[t,t] &=
-\frac{1}{2m\hbar}
\left(
\beta_{\perp} t^2 + \frac{1}{\beta_{\perp}}
\right)
\\
G_{11}^{33}[t,t] = G_{22}^{33}[t,t] &=
-\frac{1}{2m\hbar}
\left(
\beta_{3} t^2 + \frac{1}{\beta_{3}}
\right)
\end{split}
\end{equation}

\noindent
with $\beta_3 = 2\hbar / \left(m (w_{0,z})^2\right)$.
We will assume that there are no semiclassical kinematics in the perpendicular dimensions, so that the external edges have the value:

\begin{equation}
\begin{split}
G_{a}^{1}[t] &= G_{a}^{2}[t] = 0
\\
G_{a}^{3}[t] &= \frac{i}{\hbar}(-1)^{a+1}
z_{\text{cl},a}^{(0)}[t]
\end{split}
\end{equation}

\noindent
in this case, Eq. \eqref{eq:3D_proof_mass_first_order_correction} becomes

\begin{equation}
\Delta\phi^{(1)}
=
\sum_{a=1}^{2}
\left(
\frac{\hbar}{(-1)^{a + 1}}
\right)^{-1}
\int_{t_i}^{t_f}dt
\;\;
e^{-\frac{\hbar^2}{2} G_{aa}^{\perp}[t,t] \partial_{\perp}^{2}}
e^{-\frac{\hbar^2}{2}G_{aa}^{33}[t,t] \partial_{z}^{2}}
\phi_g[x, y, z]
\Bigg|_{
x = y = 0, \; z = z_{\text{cl}, a}^{(0)}[t]
}
\end{equation}

\noindent
Taylor expanding the exponentials gives us

\begin{equation}
\begin{split}
&
\Delta\phi^{(1)}
=
\\
&
\;\;\;\;\;
\sum_{a=1}^{2}
\left(
\frac{\hbar}{(-1)^{a + 1}}
\right)^{-1}
\!\!
\int_{t_i}^{t_f}dt
\sum_{n_{\perp}=0}^{\infty}
\sum_{n_{z}=0}^{\infty}
\frac{1}{n_{\perp}!}
\frac{1}{n_{z}!}
\left(
\!
-\frac{\hbar^2}{2} G_{aa}^{\perp}[t,t]
\right)^{n_{\perp}}
\!
\left(
\!
-\frac{\hbar^2}{2} G_{aa}^{33}[t,t]
\right)^{n_{z}}
\partial_{\perp}^{2n_{\perp}}\partial_{z}^{2 n_z}
\phi_g[x, y, z]
\Bigg|_{
x = y = 0, \; z = z_{\text{cl}, a}^{(0)}[t]
}
\end{split}
\end{equation}

\noindent
We can express this in terms of derivatives applied to $\phi^{(0)}[z]$ using Eq. \eqref{eq:3D_potential_radial_coefficients_in_terms_of_phi0g}

\begin{equation}
\partial_{\perp}^{2n_{\perp}}\partial_{z}^{2 n_z}
\phi_g[x, y, z]
\Bigg|_{
x = y = 0, \; z = z_{\text{cl}, a}^{(0)}[t]
}
=
\partial_{z}^{2 n_z}
\left(
\partial_{\perp}^{2n_{\perp}}
\phi_g[x, y, z]
\bigg|_{x = y = 0}
\right)
\bigg|_{
z = z_{\text{cl}, a}^{(0)}[t]
}
=
(-1)^{n_{\perp}}
\partial_z^{2(n_{\perp} + n_z)}\phi_g^{(0)}[z]
\bigg|_{
z = z_{\text{cl}, a}^{(0)}[t]
}
\end{equation}

\noindent
Our first order correction to the interferometer phase shift can therefore be expressed purely in terms of derivatives of $\phi_g^{(0)}[z]$ as

\begin{equation}\label{eq:proof_mass_actual_calculation_expression}
\begin{split}
&
\Delta\phi^{(1)}
=
\\
&
\;\;\;\;\;\;
\sum_{a=1}^{2}
\left(
\frac{\hbar}{(-1)^{a+1}}
\right)^{-1}
\!\!\!
\int_{t_i}^{t_f}dt
\!
\sum_{n_{\perp}=0}^{\infty}
\sum_{n_{z}=0}^{\infty}
\frac{1}{n_{\perp}!}
\frac{1}{n_{z}!}
\left(-\frac{\hbar^2}{2} G_{aa}^{\perp}[t,t]\right)^{n_{\perp}}
\left(-\frac{\hbar^2}{2} G_{aa}^{33}[t,t]\right)^{n_{z}}
(-1)^{n_{\perp}}
\partial_z^{2(n_{\perp} + n_z)}\phi_g^{(0)}[z]
\bigg|_{
z = z_{\text{cl}, a}^{(0)}[t]
}
\end{split}
\end{equation}

\noindent
For the experimental parameter of the 3D proof mass calculation, it happens that higher and higher orders of $\hbar$ contribute more weakly (see Fig. \ref{fig:real_proof_mass_supplement}). As part of the computation presented in Sec. \ref{sec:calculation_of_phase_response_3D}, we include orders up to $n_{\perp} + n_{z} = 3$ (i.e. summing over all values of $n_z$ and $n_{\perp}$ for which $n_{\perp} + n_{z} = 3$), where $n_{\perp} = n_{z} = 0$ is the semiclassical solution. The relative contribution for each of these orders for the parameters used to produce Fig. \ref{fig:real_proof_mass} are indicated in Fig. \ref{fig:real_proof_mass_supplement}.

\begin{figure}
\includegraphics[width=3.3in]{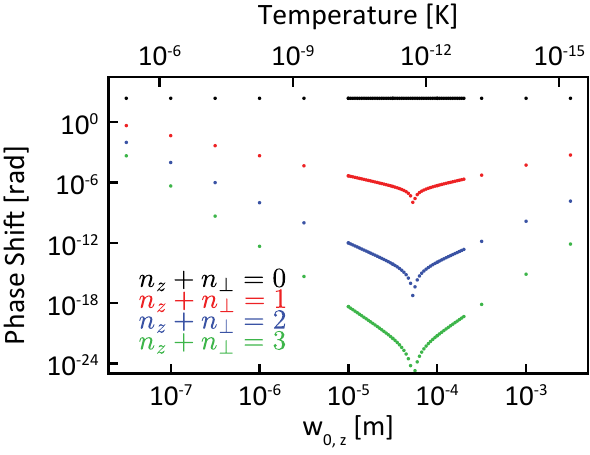}
\caption{\label{fig:real_proof_mass_supplement}
The absolute value of semiclassical and quantum corrections to the atom interferometer phase shift for the interferometer sequence described in Sec. \ref{sec:calculation_of_phase_response_3D}. The semiclassical solutions are written black, the leading order quantum contributions are in red, the second order corrections are in blue, and the third order corrections are in green. The quantum corrections indicated in Fig. \ref{fig:real_proof_mass} are dominated by the leading order quantum corrections (red).
For each value of $n_z$, $n_{\perp}$, and $w_{0,z}$, we numerically evaluate the integral over time associated with Eq. \eqref{eq:proof_mass_actual_calculation_expression}. We take the radial waist $w_{0, \perp}$ to scale with the longitudinal waist $w_{0, z}$ as $w_{\perp} = w_{0,z}/\sqrt{2}$.
}
\end{figure}

\section{Extension to Arbitrary Initial Wavefunctions}\label{sec:non_gaussian_initial_conditions}

The diagrammatic formalism relies on the form of Eq. \eqref{eq:phase_shift_expression_c}, which requires all spatial integrals to have Gaussian functions as integrands. Throughout this paper, we have assumed a Gaussian initial wavepacket, which in 1D takes the following form:

\begin{equation}
\psi_{0}[x_i, t_i] = \left(\frac{2}{\pi w_0^2}\right)^{1/4}\exp\left[
-\frac{1}{w_0^2}x_i^2
\right]
\end{equation}

\noindent
It can be beneficial to consider non-Gaussian initial wavepackets, which could arise from anharmonicity in the trapping potential used to prepare the matter wavefunction before interferometry. However, non-Gaussian initial wavefunctions make the spatial integrands (i.e. the integrals associated with Eqs. \eqref{eq:integral_over_K_a} and \eqref{eq:integral_over_psi}) non-Gaussian, preventing the diagrammatic formalism, as presented in Sec. \ref{sec:phase_shifts_with_feynman_diagrams} from functioning.

One approach to accommodating non-Gaussian initial wavefunctions within the diagrammatic formalism is to evaluate the kernel $K$ perturbatively instead of the full wavefunction $\psi$, then perform the spatial integrals over a non-Gaussian initial wavefunction (Eqs. \eqref{eq:integral_over_K_a} and \eqref{eq:integral_over_psi} for non-Gaussian $\psi_0$) numerically. 
However, this approach yields no analytic predictions and becomes computationally expensive for precise phase shift predictions.

Here, we outline an approach to extend the diagrammatic formalism to non-Gaussian initial conditions by decomposing the initial wavefunction in the harmonic oscillator eigenfunction basis and expressing the Hermite polynomials via a Gaussian generating function. This produces functions that retain the essential form of Eq. \eqref{eq:phase_shift_expression_c} without requiring any spatial integrals to be numerically evaluated. In this approach, the initial wavepacket can be expressed as 

\begin{equation}
\psi_0[x_i, t_i] = \sum_{n=0}^{\infty}a_n \psi_n^{\text{HO}}[x_i]
\end{equation}

\noindent
where $a_n$ are the expansion coefficients in the harmonic oscillator basis that construct the initial wavepacket $\psi_0$, and the superscript "HO" denotes that these are harmonic oscillator ground states, which take the following form:

\begin{equation}
\psi_n^{\text{HO}}[x_i] = \frac{1}{\sqrt{2^n n!}}
\left(\frac{2}{\pi w_0^2}\right)^{1/4} e^{-\frac{1}{w_0^2}x_i^2}
H_n\left[\sqrt{\frac{2}{w_0^2}} x_i\right]
\end{equation}

\noindent
where $H_n[x]$ are Hermite polynomials, and the waist $w_0$ is related to the harmonic oscillator trap frequency by $w_0^2 = \frac{2\hbar}{m \omega}$. These functions form a complete basis set, so any initial wavefunction $\psi_0$ can be constructed from $\psi_n^{\text{HO}}$ through a suitable choice of coefficients $a_n$. The waist $w_0$ is arbitrary: any initial wavefunction can be represented for any choice of $w_0$, which only affects the corresponding values of $a_n$. Hermite polynomials can be expressed as derivatives with respect to a dummy parameter $s$ in a Gaussian generating function as follows:

\begin{equation}
H_n[x] = \left[
\partial_s^n
e^{2 x s - s^2}
\right]_{s=0}
\end{equation}

\noindent
so that the harmonic oscillator ground states can be expressed in a purely Gaussian form in $x_i$ as

\begin{equation}
\psi_n^{\text{HO}}[x_i, t_i] = \frac{1}{\sqrt{2^n n!}} \left[\partial_{s}^{n} \psi_s[x_i, t_i ; s]\right]_{s=0}
\end{equation}

\noindent
where

\begin{equation}
\psi_s[x_i, t_i ; s] = \left(\frac{2}{\pi w_0^2}\right)^{1/4}
\exp\left[
-\frac{1}{w_0^2}x_i^2 + 2\sqrt{\frac{2}{w_0^2}} s x_i
- s^2
\right]
\end{equation}

\noindent
In this form, the arbitrary initial wavefunction can be expressed in terms of $a_n$ and derivatives with respect to $s$ as

\begin{equation}
\psi_0[x_i, t_i] = \sum_{n=0}^{\infty}a_n \frac{1}{\sqrt{2^n n!}} \left[\partial_{s}^{n} \psi_s[x_i, t_i ; s]\right]_{s=0}
\end{equation}

The key trick is that when evaluating spatial integrals, we take derivatives with respect to the dummy parameter $s$ outside the integrand. For example, consider the integral for the wavepacket in the unperturbed ($L_1 = 0$) limit

\begin{equation}
\begin{split}
\psi^{(0)}[x_f, t_f ; J[t]] &= 
\int_{-\infty}^{\infty} dx_i ~ 
K^{(0)}\left[x_f, t_f ; x_i, t_i ; J[t]\right] ~ \psi_0[x_i, t_i]
\\
&=
\int_{-\infty}^{\infty} dx_i ~ 
K^{(0)}\left[x_f, t_f ; x_i, t_i ; J[t]\right]
\left(
\sum_{n=0}^{\infty}a_n \frac{1}{\sqrt{2^n n!}} \left[\partial_{s}^{n} \psi_s[x_i, t_i ; s]\right]_{s=0}
\right)
\\
& = 
\sum_{n=0}^{\infty}a_n 
\frac{1}{\sqrt{2^n n!}}
\left[\partial_s^{n} 
\int_{-\infty}^{\infty} dx_i ~ 
K^{(0)}\left[x_f, t_f ; x_i, t_i ; J[t]\right]
\psi_s[x_i, t_i ; s]
\right]_{s=0}
\\
& = \sum_{n=0}^{\infty}a_n 
\frac{1}{\sqrt{2^n n!}}
\left[\partial_s^{n}  \psi_s^{(0)}[x_f, t_f ; s ; J[t]]\right]_{s=0}
\end{split}
\end{equation}

\noindent
where

\begin{equation}\label{eq:integral_over_K0_w_s}
\psi_s^{(0)}[x_f, t_f ; s ; J[t]] = 
\int_{-\infty}^{\infty} dx_i ~ 
K^{(0)}[x_f, t_f ; x_i, t_i ; J[t]] ~ \psi_s[x_i, t_i ; s]
\end{equation}

\noindent
is a Gaussian integral. For simplicity, we will evaluate this integral in the limit of no harmonic components in $L_0$. In this case the unperturbed kernel solution takes the following form:

\begin{equation}
\begin{split}
K^{(0)}[x_f, t_f ; x_i, t_i ; J[t]] = 
\left(
\frac{m}{2 \pi i \hbar (t_f-t_i)}
\right)^{1/2}
\exp \bigg[
\frac{i}{\hbar}
\frac{m}{2 (t_f-t_i)}
\bigg( &
(x_f - x_i)^2
\\
& + \frac{2 }{m}\int_{t_i}^{t_f} dt J[t] \left(x_f (t-t_i) + x_i (t_f-t)\right)
\\
&
- \frac{1}{m^2}
\int_{t_i}^{t_f}dt_1 \int_{t_i}^{t_f}dt_2 J[t_1]J[t_2]
\begin{cases} 
      (t_f-t_2)(t_1-t_i) & , t_1 < t_2 \\
      (t_f-t_1)(t_2-t_i) & , t_1 > t_2
\end{cases}
\bigg)
\bigg]
\end{split}
\end{equation}

\noindent
so that Eq. \eqref{eq:integral_over_K0_w_s} can be evaluated using Eq. \eqref{eq:first_integral_identity} with

\begin{equation}
\begin{split}
A &= \left(
\frac{m}{2 \pi i \hbar (t_f-t_i)}
\right)^{1/2}
\left(\frac{2}{\pi w_0^2}\right)^{1/4}
\\
F^{(0,2)} &= \frac{i m}{2\hbar(t_f-t_i)}
-\frac{1}{w_0^2}
\\
F^{(0,1)} &= \frac{-i m}{\hbar (t_f-t_i)}x_f + 2 \sqrt{\frac{2}{w_0^2}}s
\\
F^{(0,0)} &= \frac{i m}{2\hbar(t_f-t_i)}x_f^2 - s^2
\\
F^{(1,1)}[t] &= \frac{i (t_f-t)}{\hbar (t_f-t_i)}
\\
F^{(1,0)}[t] &= \frac{i (t-t_i)}{\hbar (t_f-t_i)}x_f
\\
F^{(2,0)}[t_1, t_2] &= \frac{-i}{m \hbar  (t_f-t_i)}
\begin{cases} 
      (t_f-t_2)(t_1-t_i) & , t_1 < t_2 \\
      (t_f-t_1)(t_2-t_i) & , t_1 > t_2
\end{cases}
\end{split}
\end{equation}

\noindent
so that we arrive at

\begin{equation}
\begin{split}
\psi_s^{(0)}[x_f, t_f ; s ; J[t]] = A_{s}
\exp\bigg[
&
x_f^2 F_{s}^{(0, 2)}
+x_f s F_{s}^{(0, 1)}
+ s^2 F_{s}^{(0, 0)}
\\
&
\;\;\;\;\;\;
\;\;\;
+ \int_{t_i}^{t_f}dt \left(x_f F_{s}^{(1,1)}[t] +s  F_s^{(1,0)}[t]\right)J[t]
+\frac{1}{2}\int_{t_i}^{t_f} dt_1 \int_{t_i}^{t_f}dt_2
J[t_1]F_{s}^{(2,0)}[t_1, t_2]J[t_2]
\bigg]
\end{split}
\end{equation}

\noindent
with

\begin{equation}
\begin{split}
A_{s} &= \left(\frac{2}{\pi w_0^2}\right)^{1/4}
\left(
\frac{1}{1 + i \beta (t_f-t_i)}
\right)^{1/2}
\\
F_{s}^{(0,2)} &= -\frac{m \beta}{2\hbar \left(1 + i \beta (t_f-t_i)\right)}
\\
F_{s}^{(0,1)} &= \frac{\sqrt{2} m w_0 \beta}{\hbar \left(1 + i \beta (t_f-t_i)\right)}
\\
F_{s}^{(0,0)} &= \left(1 - \frac{2}{1 + i \beta (t_f-t_i)}\right)
\\
F_{s}^{(1,1)}[t] &= \frac{i - \beta (t-t_i)}{\hbar\left(1 + i \beta(t_f-t_i)\right)}
\\
F_{s}^{(1,0)}[t] &= \frac{w_0 \beta \sqrt{2}(t-t_f)}{\hbar\left(1 + i \beta (t_f-t_i)\right)}
\\
F_{s}^{(2,0)}[t_1, t_2] &= -\frac{i}{m \hbar (t_f-t_i)}
\left(
\left(
\begin{cases} 
      (t_f-t_2)(t_1-t_i) & , t_1 < t_2 \\
      (t_f-t_1)(t_2-t_i) & , t_1 > t_2
\end{cases}
\right)
+
\frac{(t_1-t_f)(t_2-t_f)}{\left(1 + i \beta (t_f-t_i)\right)}
\right)
\end{split}
\end{equation}

\noindent
where $\beta = \frac{2\hbar}{m w_0^2}$. Now we evaluate the overlap integral of the wavefunctions of the two unperturbed arms of the interferometer as follows:

\begin{equation}
\begin{split}
&
\int_{-\infty}^{\infty}dx_f
~
\left(\psi^{(0)}[x_f, t_f ; J_1[t]]\right)
\left(\psi^{(0)}[x_f, t_f ; J_2[t]]\right)^*
\\
&=
\int_{-\infty}^{\infty}dx_f
\left(
\sum_{n_1=0}^{\infty}a_{n_1}
\frac{1}{\sqrt{2^{n_1} n_1!}}
\left[\partial_{s_1}^{n_1} \psi_{s_1}^{(0)}[x_f, t_f ; s_1 ; J_1[t]]\right]_{s_1=0}
\right)
\left(
\sum_{n_2=0}^{\infty}a_{n_2}
\frac{1}{\sqrt{2^{n_2} n_2!}}
\left[\partial_{s_2}^{n_2} \psi_{s_2}^{(0)}[x_f, t_f ; s_2 ; J_2[t]]\right]_{s_2=0}
\right)^*
\\
& = 
\sum_{n_1=0}^{\infty}
\sum_{n_2=0}^{\infty}
a_{n_1}\left(a_{n_2}\right)^{*}
\frac{1}{\sqrt{2^{n_1} n_1!}}
\frac{1}{\sqrt{2^{n_2} n_2!}}
\left[
\partial_{s_1}^{n_1}
\partial_{s_2}^{n_2}
\int_{-\infty}^{\infty}dx_f
\left(\psi_{s_1}^{(0)}[x_f, t_f ; s_1 ; J_1[t]]\right)
\left(\psi_{s_2}^{(0)}[x_f, t_f ; s_2 ; J_2[t]]\right)^*
\right]_{s_1=s_2=0}
\end{split}
\end{equation}

\noindent
where the integral over $x_f$ can be evaluated using Eq. \eqref{eq:second_integral_identity} with

\begin{equation}
\begin{split}
A &=  \left(\frac{2}{\pi w_0^2}\right)^{2/4}
\left(\frac{1}{1 + \beta^2(t_f-t_i)^2}\right)^{2/4}
\\
F^{(0,2)} &= -\frac{m \beta}{\hbar \left(1 + \beta^2(t_f-t_i)^2\right)}
\\
F^{(0,1)} &=
\frac{\sqrt{2}m w_0 \beta}{\hbar\left(1 + i \beta (t_f-t_i)\right)}s_1
+
\frac{\sqrt{2}m w_0 \beta}{\hbar\left(1 - i \beta (t_f-t_i)\right)}s_2
\\
F^{(0,0)} &= \left(1 - \frac{2}{1 + i \beta (t_f-t_i)}\right)s_1^2 + \left(1 - \frac{2}{1 - i \beta (t_f-t_i)}\right)s_2^2
\\
F^{(1,1)}_1[t] &= \frac{i - \beta(t-t_i)}{\hbar(1 + i \beta(t_f-t_i))}
\\
F^{(1,0)}_1[t] &= \frac{w_0 \beta \sqrt{2}(t-t_f)}{\hbar(1 + i\beta (t_f-t_i))}s_1
\\
\\
F^{(1,1)}_2[t] &= \frac{-i - \beta(t-t_i)}{\hbar(1 - i \beta(t_f-t_i))}
\\
F^{(1,0)}_2[t] &= \frac{w_0 \beta \sqrt{2}(t-t_f)}{\hbar(1 - i\beta (t_f-t_i))}s_2
\\
F_{11}^{(2,0)}[t_1, t_2] &=  -\frac{i}{m \hbar (t_f-t_i)}
\left(
\left(
\begin{cases} 
      (t_f-t_2)(t_1-t_i) & , t_1 < t_2 \\
      (t_f-t_1)(t_2-t_i) & , t_1 > t_2
\end{cases}
\right)
+
\frac{(t_1-t_f)(t_2-t_f)}{\left(1 + i \beta (t_f-t_i)\right)}
\right)
\\
F_{22}^{(2,0)}[t_1, t_2] &=  -\frac{-i}{m \hbar (t_f-t_i)}
\left(
\left(
\begin{cases} 
      (t_f-t_2)(t_1-t_i) & , t_1 < t_2 \\
      (t_f-t_1)(t_2-t_i) & , t_1 > t_2
\end{cases}
\right)
+
\frac{(t_1-t_f)(t_2-t_f)}{\left(1 - i \beta (t_f-t_i)\right)}
\right)
\end{split}
\end{equation}

\noindent
which results in

\begin{equation}
\begin{split}
&
\int_{-\infty}^{\infty}dx_f
\left(\psi_{s_1}^{(0)}[x_f, t_f ; s_1 ; J_1[t]]\right)
\left(\psi_{s_2}^{(0)}[x_f, t_f ; s_2 ; J_2[t]]\right)^* 
\\
&
\;\;\;\;\;\;\;\;\;\;\;\;\;\;\;\;\;\;\;\;\;\;\;\;\;\;\;\;
= 
\exp\left[
2 s_1 s_2 
+
\int_{t_i}^{t_f}dt ~ \widetilde{G}_{a}[t ; s_1, s_2]J_a[t] + 
\frac{1}{2}
\int_{t_i}^{t_f}dt_1\int_{t_i}^{t_f}dt_2
J_a[t_1]G_{ab}[t_1, t_2]J_b[t_2]
\right]
\end{split}
\end{equation}

\noindent
where

\begin{equation}
\widetilde{G}_{a}[t ; s_1, s_2] = (-1)^{a+1} \frac{w_0}{\sqrt{2}\hbar}\left(
\left(
s_1 - s_2
\right)
\left(
t-t_i
\right)
\beta
+
i \left(s_1+s_2\right)
\right)
\end{equation}

\noindent
and $G_{ab}[t_1, t_2]$ has the same value as in Eq. \eqref{eq:Gab_free_potential}, written here in 1D, and with full $t_i$ dependence as 

\begin{equation}
\begin{split}
G_{11}[t_j, t_k] &= \frac{1}{2m\hbar}\Big(-\beta \left(t_j-t_i\right) \left(t_k-t_i\right)  - \frac{1}{\beta} + i |t_j-t_k|\Big) \\
G_{12}[t_j, t_k] &= \frac{1}{2m\hbar}\Big( \beta \left(t_j-t_i\right) \left(t_k-t_i\right) + \frac{1}{\beta} + i (t_j-t_k) \Big) \\
G_{21}[t_j, t_k] &= \frac{1}{2m\hbar}\Big( \beta \left(t_j-t_i\right) \left(t_k-t_i\right) + \frac{1}{\beta} - i (t_j-t_k) \Big) \\
G_{22}[t_j, t_k] &= \frac{1}{2m\hbar}\Big( -\beta \left(t_j-t_i\right) \left(t_k-t_i\right) - \frac{1}{\beta} - i |t_j - t_k| \Big)
\end{split}
\end{equation}

Therefore, the interferometer contrast and phase shift can be written, including the perturabtive $L_1$ contribution as 

\begin{equation}\label{eq:full_phase_shift_arbitrary_initial_condition}
\begin{split}
C e^{i \Delta\phi}
&= 
\sum_{n_1=0}^{\infty}
\sum_{n_2=0}^{\infty}
a_{n_1}\left(a_{n_2}\right)^{*}
\frac{1}{\sqrt{2^{n_1} n_1!}}
\frac{1}{\sqrt{2^{n_2} n_2!}}
\\
&
\;\;\;\;\;\;\;\;\;\;\;\;\;\;
\times
\left[
\partial_{s_1}^{n_1}
\partial_{s_2}^{n_2}
e^{2s_1s_2}
e^{
\int_{t_i}^{t_f} dt'
\epsilon
\Big(
\frac{i}{\hbar}
\delta_{a1}
L_1\big[\frac{\hbar}{i}\frac{\delta}{\delta J_a[t']}\big]
+ \frac{-i}{\hbar}
\delta_{a2}
L_1\big[\frac{\hbar}{-i}\frac{\delta}{\delta J_{a}[t']}\big]
\Big)
}
e^{\int_{t_i}^{t_f}dt ~ \widetilde{G}_{b}[t ; s_1, s_2]J_b[t] + 
\frac{1}{2}
\int_{t_i}^{t_f}dt_1\int_{t_i}^{t_f}dt_2
J_b[t_1]G_{bc}[t_1, t_2]J_c[t_2]}
\right]_{s_1=s_2=0}
\\
&= 
\sum_{n_1=0}^{\infty}
\sum_{n_2=0}^{\infty}
a_{n_1}\left(a_{n_2}\right)^{*}
\frac{1}{\sqrt{2^{n_1} n_1!}}
\frac{1}{\sqrt{2^{n_2} n_2!}}
\left[
\partial_{s_1}^{n_1}
\partial_{s_2}^{n_2}
e^{2s_1s_2}
e^{
\text{sum of connected diagrams}
}
\right]_{s_1=s_2=0}
\end{split}
\end{equation}

\noindent
where the unperturbed diagram has value

\begin{equation}
\int_{t_i}^{t_f}dt ~ \widetilde{G}_{a}[t ; s_1, s_2]J_a[t] + 
\frac{1}{2}
\int_{t_i}^{t_f}dt_1\int_{t_i}^{t_f}dt_2
J_a[t_1]G_{ab}[t_1, t_2]J_b[t_2]
\end{equation}

\noindent
and the diagrams associated with the perturbative expansion in $\epsilon$ follow the same rules of Eq. \eqref{eq:diagram_component_values_1D}, but where the external edge value has picked up a dependence on $s_1$ and $s_2$

\begin{equation}
G_a[t] = \int_{t_i}^{t_f} dt' G_{a a'}[t, t']J_{a'}[t'] + \widetilde{G}_{a}[t; s_1, s_2]
\end{equation}

In this approach, one evaluates the sum of connected diagrams as described in the main text (Sec. \ref{sec:phase_shifts_with_feynman_diagrams}), but instead of directly relating the imaginary and real parts of the diagram value to the interferometer phase shift and contrast, one computes the derivatives with respect to $s_1$ and $s_2$ and takes the absolute value and argument of the double sum in Eq. \eqref{eq:full_phase_shift_arbitrary_initial_condition} to obtain corrections to the interferometer contrast and phase shift, respectively, for arbitrary initial wavefunctions.
Symbolic differentiation with respect to $s_1$ and $s_2$ is far less computationally costly than integrating numerically over continuous spatial coordinates, and crucially yields analytic corrections to the interferometer response.
This approach also scales to 3D and can accommodate a harmonic component in $L_0$.

\section{Two Gaussian Integral Identities}\label{sec:appendix_gaussian_integral_identities}

To compute the value of diagram edges, it will be important to leverage two Gaussian integral identities, denoted as Eqs. \eqref{eq:first_integral_identity} and \eqref{eq:second_integral_identity}. Both of these integral identities are ultimately just extensions of the identity
$\int_{-\infty}^{\infty}dx e^{-a x^2 + b x} = \sqrt{\pi/a}\;e^{b^2/(4a)}
$ for $\text{Re}[a] > 0$.
The first integral identity is

\begin{equation}\label{eq:first_integral_identity}
\begin{split}
& I_1[x] = A \exp \bigg[
x^2 F^{(0,2)} + x F^{(0,1)} + F^{(0,0)}
+\int_{t_i}^{t_f}dt (x F^{(1,1)}[t] + F^{(1,0)}[t])J[t]
+ \frac{1}{2}\int_{t_i}^{t_f}dt_1 \int_{t_i}^{t_f}dt_2 J[t_1]F^{(2,0)}[t_1, t_2]J[t_2]
\bigg]
\\
&\int_{-\infty}^{\infty}dx ~ I_1[x] = \overline{A} 
\exp\bigg[\overline{F}^{(0)}
+ \int_{t_i}^{t_f} dt \overline{F}^{(1)}[t]J[t]
+ \frac{1}{2} \int_{t_i}^{t_f}dt_1 \int_{t_i}^{t_f}dt_2 J[t_1]\overline{F}^{(2)}[t_1, t_2]J[t_2]
\bigg]
\end{split}
\end{equation}

\noindent
where

\begin{equation}
\begin{split}
\overline{A} &= A\sqrt{\frac{\pi}{-F^{(0,2)}}} \\ 
\overline{F}^{(0)} &= F^{(0,0)} - \frac{(F^{(0,1)})^2}{4 F^{(0,2)}} \\
\overline{F}^{(1)}[t] &= F^{(1,0)}[t] - \frac{F^{(0,1)}F^{(1,1)}[t]}{2 F^{(0,2)}} \\
\overline{F}^{(2)}[t_1, t_2] &= F^{(2,0)}[t_1, t_2] - \frac{F^{(1,1)}[t_1]F^{(1,1)}[t_2]}{2 F^{(0,2)}}
\end{split}
\end{equation}

\noindent
and $F^{(j, k)}[t_1, t_2]$ denotes a function associated with a term in the exponent of the integrand which is $j^{\text{th}}$ order in the integration variable $x$ and $k^{\text{th}}$ order in the function $J[t]$. The second integral identity concerns integrating over an expression which consists of two different source terms $J_1$ and $J_2$,

\begin{equation}\label{eq:second_integral_identity}
\begin{split}
& I_2[x] = A \exp \bigg[
x^2 F^{(0,2)} + x F^{(0,1)} + F^{(0,0)}
+ \int_{t_i}^{t_f}dt \Big(
(x F^{(1,1)}_1[t] + F^{(1,0)}_1[t])J_1[t]
+ (x F^{(1,1)}_2[t] + F^{(1,0)}_2[t])J_2[t]
\Big)
\\
& 
\;\;\;\;\;\;\;\;\;\;\;\;\;\;\;\;\;\;\;\;\;\;\;\;\;\;\;\;
+ \frac{1}{2}\int_{t_i}^{t_f}dt_1 \int_{t_i}^{t_f}dt_2 (J_1[t_1]F^{(2,0)}_{11}[t_1, t_2]J_1[t_2] + J_2[t_1]F^{(2,0)}_{22}[t_1, t_2]J_2[t_2])
\bigg]
\\
& \int_{-\infty}^{\infty}dx ~ I_2[x] = \overline{A} 
\exp\bigg[
\overline{F}^{(0)}
+ \int_{t_i}^{t_f} dt (\overline{F}^{(1)}_1[t]J_1[t] + \overline{F}^{(1)}_2[t]J_2[t])
\\
&
+ \frac{1}{2} \int_{t_i}^{t_f}dt_1  \int_{t_i}^{t_f} dt_2 
( 
J_1[t_1]\overline{F}^{(2)}_{11}[t_1, t_2]J_1[t_2]
+J_1[t_1]\overline{F}^{(2)}_{12}[t_1, t_2]J_2[t_2]
+J_2[t_1]\overline{F}^{(2)}_{21}[t_1, t_2]J_1[t_2]
+J_2[t_1]\overline{F}^{(2)}_{22}[t_1, t_2]J_2[t_2]
)
\bigg]
\end{split}
\end{equation}

\noindent
where

\begin{equation}
\begin{split}
\overline{A} &= A\sqrt{\frac{\pi}{-F^{(0,2)}}} \\ 
\overline{F}^{(0)} &= F^{(0,0)} - \frac{(F^{(0,1)})^2}{4 F^{(0,2)}} \\
\overline{F}^{(1)}_j[t] &= F^{(1,0)}_j[t] - \frac{F^{(0,1)}F^{(1,1)}_j[t]}{2 F^{(0,2)}} \;\;\; , \;\;\;\; j ={1,2} \\
\overline{F}^{(2)}_{jk}[t_1, t_2] &= \delta_{jk} F^{(2,0)}_{jj}[t_1, t_2]- \frac{F^{(1,1)}_{j}[t_1]F^{(1,1)}_{k}[t_2]}{2 F^{(0,2)}}
\;\;\;\; , \;\;\;\;j,k = 1, 2
\end{split}
\end{equation}

\noindent
We use both of these expressions to compute the value of the internal and external edges associated with a diagram in Appendix \ref{sec:appendix_edge_values}.

\end{widetext}

\bibliography{bibliography}

\end{document}